\documentclass[twocolumn,dvipsnames]{aastex631}

\usepackage{amsmath, amssymb}
\usepackage{multirow}
\usepackage{bm}
\usepackage{enumerate}
\usepackage{soul}
\usepackage{bm}

\newcommand{\uaaffil}{Steward Observatory, University of Arizona, 933 N Cherry Ave, Tucson, AZ 85721, USA}

\newcommand{\beagle}{\texttt{BEAGLE}}

\newcommand{\Lya}{Ly$\alpha$}
\newcommand{\Msun}{M$_\odot$}
\newcommand{\Muv}{\ensuremath{M_\textsc{uv}}}

\submitjournal{ApJ}

\shorttitle{JADES: the $z \gtrsim 9$ galaxy UV luminosity function}
\shortauthors{Whitler et al.}

\graphicspath{{figs/}}

\begin{document}

\title{The $z \gtrsim 9$ galaxy UV luminosity function from the JWST Advanced Deep Extragalactic Survey: insights into early galaxy evolution and reionization}

\correspondingauthor{Lily Whitler}
\email{lwhitler@arizona.edu}

\author[0000-0003-1432-7744]{Lily Whitler}
\affiliation{\uaaffil}

\author{Daniel P. Stark}
\affiliation{Department of Astronomy, University of California, Berkeley, Berkeley, CA 94720, USA}

\author[0000-0001-8426-1141]{Michael W. Topping}
\affiliation{\uaaffil}

\author[0000-0002-4271-0364]{Brant Robertson}
\affiliation{Department of Astronomy and Astrophysics University of California, Santa Cruz, 1156 High Street, Santa Cruz CA 96054, USA}

\author[0000-0002-7893-6170]{Marcia Rieke}
\affiliation{\uaaffil}

\author[0000-0003-4565-8239]{Kevin N.\ Hainline}
\affiliation{\uaaffil}

\author[0000-0003-4564-2771]{Ryan Endsley}
\affiliation{Department of Astronomy, University of Texas, Austin, TX 78712, USA}

\author[0000-0002-2178-5471]{Zuyi Chen}
\affiliation{\uaaffil}

\author[0000-0003-0215-1104]{William M.\ Baker}
\affiliation{DARK, Niels Bohr Institute, University of Copenhagen, Jagtvej 128, DK-2200 Copenhagen, Denmark}

\author[0000-0003-0883-2226]{Rachana Bhatawdekar}
\affiliation{European Space Agency (ESA), European Space Astronomy Centre (ESAC), Camino Bajo del Castillo s/n, 28692 Villanueva de la Cañada, Madrid, Spain}

\author[0000-0002-8651-9879]{Andrew J.\ Bunker}
\affiliation{Department of Physics, University of Oxford, Denys Wilkinson Building, Keble Road, Oxford OX1 3RH, UK}

\author[0000-0002-6719-380X]{Stefano Carniani}
\affiliation{Scuola Normale Superiore, Piazza dei Cavalieri 7, I-56126 Pisa, Italy}

\author[0000-0003-3458-2275]{St\'ephane Charlot}
\affiliation{Sorbonne Universit\'e, CNRS, UMR 7095, Institut d'Astrophysique de Paris, 98 bis bd Arago, 75014 Paris, France}

\author[0000-0002-7636-0534]{Jacopo Chevallard}
\affiliation{Department of Physics, University of Oxford, Denys Wilkinson Building, Keble Road, Oxford OX1 3RH, UK}

\author[0000-0002-9551-0534]{Emma Curtis-Lake}
\affiliation{Centre for Astrophysics Research, Department of Physics, Astronomy and Mathematics, University of Hertfordshire, Hatfield AL10 9AB, UK}

\author[0000-0003-1344-9475]{Eiichi Egami}
\affiliation{\uaaffil}

\author[0000-0002-2929-3121]{Daniel J.\ Eisenstein}
\affiliation{Center for Astrophysics $|$ Harvard \& Smithsonian, 60 Garden St., Cambridge MA 02138 USA}

\author[0000-0003-4337-6211]{Jakob M.\ Helton}
\affiliation{\uaaffil}

\author[0000-0001-7673-2257]{Zhiyuan Ji}
\affiliation{\uaaffil}

\author[0000-0002-9280-7594]{Benjamin D.\ Johnson}
\affiliation{Center for Astrophysics $|$ Harvard \& Smithsonian, 60 Garden St., Cambridge MA 02138 USA}

\author[0000-0003-4528-5639]{Pablo G. P\'erez-Gonz\'alez}
\affiliation{Centro de Astrobiolog\'ia (CAB), CSIC–INTA, Cra. de Ajalvir Km.~4, 28850- Torrej\'on de Ardoz, Madrid, Spain}

\author[0000-0002-5104-8245]{Pierluigi Rinaldi}
\affiliation{\uaaffil}

\author[0000-0002-8224-4505]{Sandro Tacchella}
\affiliation{Kavli Institute for Cosmology, University of Cambridge, Madingley Road, Cambridge, CB3 0HA, UK}
\affiliation{Cavendish Laboratory, University of Cambridge, 19 JJ Thomson Avenue, Cambridge, CB3 0HE, UK}

\author[0000-0003-2919-7495]{Christina C.\ Williams}
\affiliation{NSF National Optical-Infrared Astronomy Research Laboratory, 950 North Cherry Avenue, Tucson, AZ 85719, USA}

\author[0000-0001-9262-9997]{Christopher N.\ A.\ Willmer}
\affiliation{\uaaffil}

\author[0000-0002-4201-7367]{Chris Willott}
\affiliation{NRC Herzberg, 5071 West Saanich Rd, Victoria, BC V9E 2E7, Canada}

\author[0000-0002-7595-121X]{Joris Witstok}
\affiliation{Cosmic Dawn Center (DAWN), Copenhagen, Denmark}
\affiliation{Niels Bohr Institute, University of Copenhagen, Jagtvej 128, DK-2200, Copenhagen, Denmark}

\begin{abstract}
The high-redshift UV luminosity function provides important insights into the evolution of early galaxies. JWST has revealed an unexpectedly large population of bright ($\Muv\lesssim-20$) galaxies at $z\gtrsim10$, implying fundamental changes in the star forming properties of galaxies at increasingly early times. However, constraining the fainter population ($\Muv\gtrsim-18$) has been more challenging. In this work, we present the $z\gtrsim9$ UV luminosity function from the JWST Advanced Deep Extragalactic Survey. We calculate the UV luminosity function from several hundred $z\gtrsim9$ galaxy candidates that reach UV luminosities of $\Muv\sim-17$ in redshift bins of $z\sim8.5-12$ (309 candidates) and $z\sim12-16$ (63 candidates). We search for candidates at $z\sim16-22.5$ and find none. We also estimate the $z\sim14-16$ luminosity function from the $z\geq14$ subset of the $z\sim12-16$ sample. Consistent with other measurements, we find an excess of bright galaxies that is in tension with many theoretical models, especially at $z\gtrsim12$. However, we also find high number densities at $-18\lesssim\Muv\lesssim-17$, suggesting that there is a larger population of faint galaxies than expected, as well as bright ones. From our parametric fits for the luminosity function, we find steep faint end slopes of $-2.5\lesssim\alpha\lesssim-2.3$, suggesting a large population of faint ($\Muv\gtrsim-17$) galaxies. Combined, the high normalization and steep faint end slope of the luminosity function could imply that the reionization process is appreciably underway as early as $z=10$.
\end{abstract}


\section{Introduction} \label{sec:intro}

The formation and evolution of galaxies within the first few hundred million years after the Big Bang played an essential part in shaping the evolution of the Universe. The emergence of these early galaxies marks the transition from a nearly homogeneous Universe composed primarily of neutral hydrogen to one filled with galaxies hosted in large scale dark matter overdensities, all embedded in an ionized intergalactic medium (IGM). Moreover, there is substantial evidence that galaxies were not only present for this key phase transition of the Universe from neutral to ionized hydrogen -- called cosmic reionization -- but that they were capable of providing most of the photons responsible for reionizing the Universe \citep[for a recent review, see][]{robertson2022}. Thus, observing galaxies at these early times provides crucial insights into the early phases of star formation and the process of hydrogen reionization.

The identification and characterization of galaxies within a billion years after the Big Bang was first enabled by near-infrared ($\sim 1 - 2$\,$\mu$m) observations with the Hubble Space Telescope (HST) and ground-based facilities. Together, these data enabled thousands of star-forming galaxies at $z \gtrsim 6$ ($\sim 900$\,Myr after the Big Bang) and tens at $z \sim 9 - 10$ ($\sim 500$\,Myr after the Big Bang) to be identified (e.g. \citealt{stanway2003, bunker2004, bunker2010, ellis2013, mclure2013, finkelstein2015, finkelstein2022_hst, mcleod2016, bouwens2021}; also see \citealt{stark2016} for a review) spanning more than four orders of magnitude in rest-frame ultraviolet (UV) continuum luminosities. This allowed the high-redshift rest-UV luminosity function to be measured at luminosities as bright as $\Muv \sim -24$ and as faint as $\Muv \sim -13$ \citep[e.g.][]{bouwens2011, bouwens2015_uvlf, bouwens2021, bowler2014, bowler2015, bowler2020, finkelstein2015, finkelstein2022_hst, mcleod2016, livermore2017, oesch2018, rojas-ruiz2020}. At the faint end of the luminosity function, primarily accessible with space-based observations, these studies revealed steep faint end slopes of $\alpha \lesssim -2$ (implying that faint, $\Muv \gtrsim -18$ objects dominated galaxy number counts during reionization), while ground-based studies found evidence of an increase in the number of very bright ($\Muv \lesssim -22$) galaxies at $z \gtrsim 7$ compared to lower redshifts (i.e. a change in the shape of the bright end of the luminosity function, suggesting a redshift evolution of the physical processes impacting the UV luminosity of galaxies). However, the limited near-infrared wavelength coverage of HST made it extremely challenging to robustly perform rest-UV photometric selections at $z > 9$.

With JWST \citep{gardner2006, gardner2023}, our understanding of galaxies within the first $\sim 500$\,Myr after the Big Bang has changed significantly over the last few years. Photometric selections using early JWST/Near Infrared Camera \citep[NIRCam;][]{rieke2005, rieke2023_nircam} imaging quickly identified large numbers of $z \gtrsim 10$ galaxy candidates \citep[e.g.][]{naidu2022, castellano2022, adams2023, donnan2023, harikane2023, finkelstein2023_ceers, whitler2023_ceers, robertson2023_jades_highz, hainline2024_highz}. Subsequent spectroscopic observations with the JWST/Near Infrared Spectrograph \citep[NIRSpec;][]{jakobsen2022, ferruit2022} have now unequivocally confirmed the existence of many $z \geq 10$ galaxies up to redshifts as high as $z = 14.2$ \citep[e.g.][]{curtis-lake2023, fujimoto2023, arrabalharo2023_ddt, wang2023, castellano2024_ghz2, deugenio2024_carbon, carniani2024, napolitano2024}, many of which are bright in the rest-frame UV \citep[$\Muv \lesssim -20$; e.g.][]{finkelstein2022_maisie, castellano2024_ghz2, carniani2024, napolitano2024}. Given the rarity of similarly bright, early galaxies predicted by existing theoretical galaxy evolution models \citep[e.g.][]{yung2019, behroozi2020, rosdahl2022, wilkins2023}, the discovery of these bright galaxies in moderately small survey areas tentatively suggested that such systems were unexpectedly abundant in the early Universe.

Over time and with the maturation of JWST, insights into the $z \gtrsim 10$ galaxy population as a whole have become accessible. Consistent with the large number of bright galaxies implied by individual discoveries, JWST measurements of the rest-UV luminosity function at $z \sim 10 - 13$ have typically found only a slow decline in the number densities of $\Muv \lesssim -20$ galaxies (e.g. \citealt{adams2023, bouwens2023, castellano2023_lf, donnan2023, donnan2024, harikane2023, harikane2024a, harikane2024b, mcleod2024, finkelstein2024}, though \citealt{willott2024} found a faster evolution that could be attributed to field-to-field variance). This slow evolutionary trend is a significant departure from many theoretical expectations established prior to JWST, prompting many potential explanations to be explored theoretically. For example, the efficiency of star formation may be higher than previously predicted at high redshift \citep[e.g.][]{dekel2023, li2024, harikane2023, ceverino2024, feldmann2025}. Star formation may be extremely stochastic or ``bursty'' at early times \citep[e.g.][]{mason2023, mirocha2023, shen2023, sun2023, kravtsov2024, gelli2024}. The stellar initial mass function may be more top-heavy in the early Universe (e.g. \citealt{harikane2023, yung2024}, though c.f. \citealt{cueto2024}). Dust attenuation may decrease at high redshift \citep[e.g.][]{ferrara2023, fiore2023}. Or, there may be a significant population of active galactic nuclei \citep[e.g.][]{hegde2024}. In particular, bursty star formation has been shown to significantly alleviate the tension between models and observations \citep[e.g.][]{sun2023, kravtsov2024, gelli2024}, which is also consistent with the spectral energy distributions (SEDs) of high-redshift galaxies \citep[e.g.][]{tacchella2023, looser2023, looser2024, endsley2023, endsley2024_jades, witten2024, boyett2024}. However, it is challenging for any single physical mechanism to fully reconcile observations and theoretical predictions and it is likely that a large population of bright galaxies is sustained by a combination of several factors.

Moreover, while much attention has been dedicated to observing and understanding the bright galaxy population, the complete picture of galaxies at all luminosities has been more challenging to characterize. It is now generally agreed that the abundance of $\Muv \lesssim -20$ galaxies declines relatively slowly at $z \gtrsim 10$ at least until $z \sim 13$, but these objects do not represent the majority of the overall galaxy population. The steep faint end slopes of $z \gtrsim 8$ UV luminosity functions imply that faint, $\Muv \gtrsim -18$ galaxies dominate galaxy number counts and the total cosmic UV luminosity density at high redshift. Studying only the bright galaxy population provides an incomplete perspective on galaxies in the early Universe; identifying and characterizing fainter galaxies despite the observational challenges is necessary for a holistic understanding of the evolution of high-redshift galaxies and how galaxies contribute to the process of cosmic reionization.

In this work, we aim to place new, direct constraints on the faint end ($\Muv \gtrsim -18$) of the UV luminosity function at $z \gtrsim 9$ using deep JWST/NIRCam imaging taken as part of the JWST Advanced Deep Extragalactic Survey \citep[JADES;][]{eisenstein2023_jades, rieke2023_jades}. We take advantage of three dropout filters at $\sim 1 - 2$\,$\mu$m to photometrically select galaxy candidates that lie at $z \sim 9 - 17$ and that span nearly a factor of $\sim 100$ in far-UV continuum ($\lambda_\text{rest} = 1500$\,\AA) luminosity. We can select bright, $\Muv \lesssim -20$ candidates over the full area of the JADES imaging we use in this work ($\sim 160$\,arcmin$^{2}$), while $\Muv \sim -17$ candidates are identified in the $\sim 12$\,arcmin$^{2}$ of the imaging that reaches depths of $m_\textsc{ab} \sim 31$. This depth is comparable to the JADES Origins Field (JOF) in which the $z \sim 12 - 14$ UV luminosity function has been independently measured \citep{robertson2024_jof}, as well as the Next Generation Deep Extragalactic Exploratory Public survey \citep{leung2023} and the NIRCam parallel to the MIRI Deep Imaging Survey \citep{perez-gonzalez2023}. Using these deep data from JADES, we can directly constrain the slope of the faint end of the luminosity function, investigate implications for early star formation processes in the context of both the bright and faint galaxy population, and examine the role of galaxies in reionizing the Universe.

This paper is organized as follows. We describe our imaging data, reduction techniques, and photometric measurements in Section\ \ref{sec:data}. We then present our selection methods in Section\ \ref{sec:selection} and discuss the general observed properties of our samples in Section\ \ref{sec:sample_overview}. In Section\ \ref{sec:uvlf}, we present our measurement of the rest-UV luminosity function in two bins of redshift spanning $z \sim 9 - 12$ and $z \sim 12 - 16$, an upper limit on the UV luminosity function at $z \gtrsim 16$, and the corresponding redshift evolution of the cosmic UV luminosity density. We then place our results in the context of theoretical galaxy formation models and discuss implications for the reionization timeline in Section\ \ref{sec:discussion}. Finally, we summarize our key results and conclusions in Section\ \ref{sec:summary}.

Throughout this work, we assume a standard flat $\Lambda$CDM cosmology with $h = 0.7$, $\Omega_M = 0.3$, and $\Omega_\Lambda = 0.7$ and all magnitudes are given in the AB system \citep{oke1983}. Unless stated otherwise, all physical lengths are comoving and reported values and uncertainties correspond to the marginalized median, 16$^\text{th}$, and 84$^\text{th}$ percentiles.

\section{Data and Photometry} \label{sec:data}

\begin{deluxetable*}{cccccccccc}
    \tablecaption{\label{tab:depths}The aperture corrected $5\sigma$ depths in AB magnitudes of the NIRCam images used in this work, measured in 0.2\arcsec{} diameter circular apertures (the same apertures as were used for color selection) in source-masked regions in the field. We report depths only for the filters used individually in this work, though additional filters were included in the stacked SNR image used for detection where available. We report the depths in AB magnitudes in each subregion as described in Section\ \ref{subsec:photometry}, which are defined by the exposure time in F200W, and over the entire area of the imaging we use in this work.}
    \tablehead{\colhead{Subregion} & \colhead{Area  [arcmin$^2$]} & \colhead{F090W} & \colhead{F115W} & \colhead{F150W} & \colhead{F200W} & \colhead{F277W} & \colhead{F356W} & \colhead{F410M} & \colhead{F444W}}
\startdata
$< 5000$ s & 33.7 & 28.4 & 28.8 & 28.7 & 28.7 & 29.3 & 29.0 & 27.9 & 29.1 \\
$5000 - 12\,000$ s & 91.9 & 29.2 & 29.5 & 29.5 & 29.6 & 30.0 & 29.9 & 29.3 & 29.6 \\
$12\,000 - 40\,000$ s & 25.8 & 30.1 & 30.3 & 30.3 & 30.4 & 30.7 & 30.7 & 30.1 & 30.4 \\
$> 40\,000$ s & 12.0 & 30.4 & 30.7 & 30.7 & 30.7 & 31.0 & 31.0 & 30.4 & 30.7 \\
\hline
All & 163.5 & 28.9 & 29.3 & 29.3 & 29.3 & 29.8 & 29.6 & 28.6 & 29.5 \\
\enddata
\end{deluxetable*}

In this work, we use deep near-infrared JWST imaging obtained as part of the JADES program \citep{eisenstein2023_jades}, combined with archival HST/Advanced Camera for Surveys (ACS) optical imaging from the Hubble Legacy Fields (HLF) project \citep{illingworth2016, whitaker2019}. We refer to \citet{eisenstein2023_jades} and \citet{rieke2023_jades} for a full description of the JADES design and dataset but provide a brief summary of the JADES NIRCam data here.

The JADES NIRCam imaging consists of observations in both the North and South fields of the Great Observatories Origins Deep Survey \citep[GOODS-N and GOODS-S;][]{giavalisco2004} under program IDs 1180 and 1181 (PI Eisenstein), 1210 and 1286 (PI Luetzgendorf), and 1287 (PI Isaak). JADES observed both fields in eight NIRCam filters over the entire footprint (F090W, F115W, F150W, F200W, F277W, F356W, F410M, and F444W), with the addition of F070W and/or F335M in some regions of the footprint. Additionally, subregions of the JADES NIRCam footprint overlap with imaging taken by other JWST programs, some of which we use when constructing the data products used in this work. Specifically, we include the $\sim 9$\,arcmin$^2$ of F162M, F182M, F210M, F250M, F300M, and F335M imaging in the JADES GOODS-S footprint taken by program ID 3215 (PIs Eisenstein and Maiolino), the area of which comprises the JOF \citep{eisenstein2023_jof}. There is also $\sim 10$\,arcmin$^2$ of F182M, F210M, F430M, F460M, and F480M imaging in GOODS-S from the JWST Extragalactic Medium-band Survey \citep[JEMS, program ID 1963; PIs Williams, Tacchella, and Maseda;][]{williams2023_jems} and $\sim 70$\,arcmin$^2$ in both GOODS-N and GOODS-S of F182M, F210M, and F444W observations from the First Reionization Epoch Spectroscopically Complete Observations \citep[FRESCO, program ID 1895;][]{oesch2023_fresco} survey. Data from program 3215, JEMS, and FRESCO are co-reduced with JADES observations and included in the final NIRCam mosaics and photometric catalogs. We refer to \citet{rieke2023_jades, eisenstein2023_jades, eisenstein2023_jof}, and Tacchella et al. (in prep.) for more detailed descriptions of the NIRCam image reduction and Robertson et al. (in prep.) for the detection and photometry methods, but provide a summary below.

\subsection{Images} \label{subsec:images}

We process the JADES, JEMS, FRESCO, and program 3215 NIRCam data with version 1.11.4 of the JWST Science Calibration Pipeline (\texttt{jwst}) and Calibration Reference Data System pipeline mapping \texttt{jwst\textunderscore1130.pmap}, with some custom steps for correction of imaging artifacts, astrometric alignment, and background subtraction. We run \texttt{jwst} Stage 1 to perform detector-level corrections and ramp fitting, largely with the default parameters \citep[with the exception of identifying and correcting ``snowball'' artifacts from cosmic rays, for which we use custom parameters;][]{robertson2024_jof}. We then run \texttt{jwst} Stage 2 with the default parameters using sky flats provided by the Space Telescope Science Institute for the short wavelength (SW) filters, F250M, and F300M, and custom super-sky flats for all other long wavelength (LW) filters.

After the completion of Stage 2, we fit and subtract several common additive features seen in NIRCam data \citep{rigby2023}: $1 / f$ noise, artifacts from scattered light\footnote{\url{https://jwst-docs.stsci.edu/known-issues-with-jwst-data/nircam-known-issues/nircam-scattered-light-artifacts}} \citep{rigby2023}, and a large-scale background. We also perform a custom astrometric alignment to HST images that were registered to Gaia DR2 (\citealt{gaia-collaboration2018}; G. Brammer, private communication) using a modified version of the \texttt{jwst TweakReg} package. Finally, for each filter, we combine the individual calibrated exposures into a single mosaic using the default parameters of \texttt{jwst} Stage 3. We set the pixel scale to 0.03\arcsec\,pixel$^{-1}$ and choose a drizzle parameter of \texttt{pixfrac = 1} for both the SW and LW images.

\subsection{Detection and Photometry} \label{subsec:photometry}

\begin{figure*}
    \includegraphics[width=\textwidth]{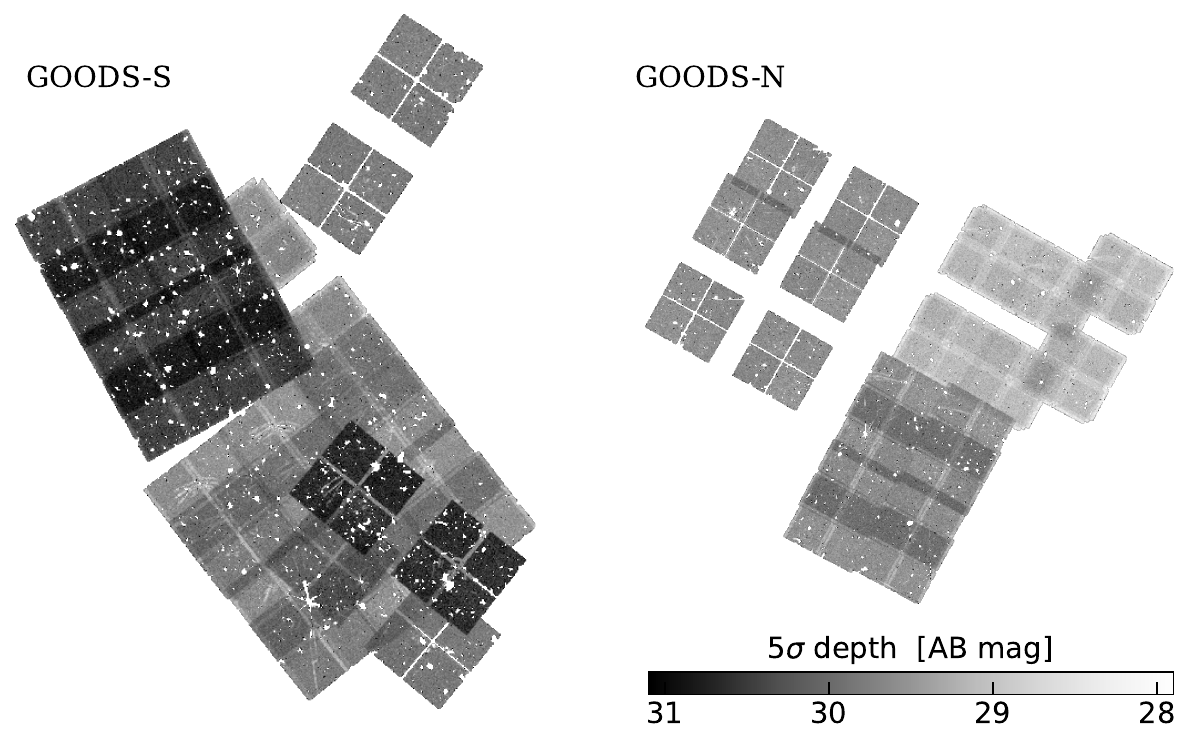}
    \caption{\label{fig:field_image}The F200W footprint of the imaging used in this work. The colorbar shows the $5\sigma$ depth of F200W measured as described in Section\ \ref{subsec:photometry}, where darker grey corresponds to deeper depths. The shallowest region with F200W exposure time of $< 5000$\,s primarily falls in GOODS-N and has typical $5\sigma$ depths of $m_\textsc{AB} \sim 29$\,mag, while the deepest, $>40\,000$\,s region lies exclusively in GOODS-S and reaches $m_\textsc{AB} \sim 30.8$.}
\end{figure*}

We run source detection on a signal-to-noise ratio (SNR) image that includes all of the available LW NIRCam mosaics in a given region, at their original resolution (i.e. without point spread function, PSF, homogenization). In GOODS-S, these filters are F250M, F277W, F300M, F335M, F356W, F410M, F430M, F444W, F460M, and F480M. In GOODS-N, the detection image includes F277W, F335M, F356W, F410M, and F444W. To produce the SNR stack, we first median filter the error images of individual filters to remove small-scale features introduced by incomplete masking in the \texttt{jwst} pipeline, then construct the signal and noise images as the inverse variance weighted stack of the science and median filtered error channels for the relevant filters. The final SNR image on which we perform detection is the ratio of these stacks.

To construct the segmentation map, we select initial regions of interest by applying a significance threshold of $\text{S/N} > 1.5$, then refine the segmentations with a series of custom computational morphology algorithms inspired by \texttt{NoiseChisel} \citep{akhlaghi2015}. Stars and diffraction spikes are masked, the segmentation map is deblended using a logarithmic scaling of the F200W image, then further refined by applying a high-pass filter to the outer regions of large segmentations to identify nearby satellites. Finally, the segmentations of already identified objects are masked out of the detection image and a final search for faint objects that may have previously been missed is performed.

We next use the \texttt{photutils} \citep{bradley2023_photutils} package to do custom forced aperture photometry on the objects detected as described above. Object centroids are computed as the windowed positions used by \texttt{Source Extractor} \citep{bertin1996} and photometry is measured in several circular apertures ($r = 0.1\arcsec, 0.15\arcsec, 0.25\arcsec, 0.3\arcsec, 0.35\arcsec, 0.5\arcsec$) and \citet{kron1980} elliptical apertures with Kron parameters of $k = 1.2$ and $k = 2.5$. Kron apertures are determined using the inverse variance weighted signal image stack (the numerator of the detection image) with area limited to be less than twice that of an object's segmentation area.

To calculate aperture corrections, we produce model point spread functions (mPSFs) following the methods of \citet{ji2023}, wherein the mPSF is measured from mosaics of \texttt{WebbPSF} models that are constructed using the same exposure pattern as the real observations. The aperture correction for a circular aperture of a given size is measured from the mPSF and applied to every source in the catalog in the same way, while the Kron aperture corrections are calculated on a source-by-source basis by placing the appropriate elliptical aperture on the mPSF. Aperture corrections for HST bands are measured from empirical PSFs constructed using visually inspected stars in the field.

Finally, photometric uncertainties are calculated as the quadrature sum of the Poisson noise from a given source and sky noise measured from apertures randomly placed on blank areas of the mosaics. As the depth of the image varies significantly over the entire JADES area, we calculate the contributions of sky noise as a function of both exposure time and aperture size. We refer to section 4.2 of \citet{rieke2023_jades} for a detailed description of the methods we use to measure sky noise, but in brief, we place 100,000 apertures in object-free regions across the entire imaging area for a range of aperture sizes, then sort the apertures into bins of exposure time. We then calculate the root-mean-square (RMS) of electron counts for every aperture in a given bin of exposure time, producing RMS values as a function of aperture size at fixed exposure time, and fit a power law scaling relation to the relationship between RMS and aperture size. Finally, for a given object and aperture size, we use the scaling relation appropriate for the exposure time at its location to determine the sky noise contribution to the total uncertainty. For the Kron photometry, the circularized radius of the elliptical aperture is used to determine the random aperture contribution to the uncertainty.

\bigskip

\begin{deluxetable*}{ccl}
    \tablecaption{\label{tab:selection_criteria}The S/N and color criteria adopted to select dropout candidates as described in Section\ \ref{sec:selection}.}
    \tablehead{\colhead{Dropout Sample} & \colhead{Redshift} & \colhead{Criteria}}
\startdata
\multirow{6}{*}{F115W} & \multirow{6}{*}{$z \sim 8.5 - 12$} &
$\text{S/N} > 5$ in at least four of F150W, F200W, F277W, F356W, F410M, and F444W \\
& & $\text{S/N} > 2$ in all of F150W, F200W, F277W, F356W, and F444W \\
& & ($\text{S/N} < 2$ in F090W) or ($\text{F115W} - \text{F150W} > 3$) \\
& & $\text{F115W} - \text{F150W} > 1.3$ \\
& & $\text{F150W} - \text{F277W} < 1.0$ \\
& & $\text{F115W} - \text{F150W} > \text{F150W} - \text{F277W} + 1.3$ \\
& & $(\texttt{FLAG\textunderscore BN}\tablenotemark{a} < 2$) or ($\text{F115W} - \text{F150W} > 3$) \\ \hline
\multirow{6}{*}{F150W} & \multirow{6}{*}{$z \sim 12 - 16$} &
$\text{S/N} > 5$ in at least three of F200W, F277W, F356W, F410M, and F444W \\
& & $\text{S/N} > 2$ in all of F200W, F277W, F356W, and F444W \\
& & ($\text{S/N} < 2$ in F090W and F115W) or (($\text{S/N} < 2$ in one of F090W and F115W) \\
& & \qquad and ($\text{F150W} - \text{F200W} > 3$)) \\
& & $\text{F150W} - \text{F200W} > 1.3$ \\
& & $\text{F200W} - \text{F356W} < 1.0$ \\
& & $\text{F150W} - \text{F200W} > \text{F200W} - \text{F356W} + 1.3$ \\
& & ($\texttt{FLAG\textunderscore BN}\tablenotemark{a} < 2$) or ($\text{F150W} - \text{F200W} > 3$) \\ \hline
\multirow{6}{*}{F200W} & \multirow{6}{*}{$z \sim 16 - 22.5$} &
$\text{S/N} > 5$ in at least three of F277W, F356W, F410M, and F444W \\
& & $\text{S/N} > 2$ in all of F277W, F356W, and F444W \\
& & ($\text{S/N} < 2$ in F090W, F115W, and F150W) or (($\text{S/N} < 2$ in two of F090W, F115W, \\
& & \qquad and F150W) and ($\text{F200W} - \text{F277W} > 3$)) \\
& & $\text{F200W} - \text{F277W} > 1.3$ \\
& & $\text{F277W} - \text{F444W} < 1.0$ \\
& & $\text{F200W} - \text{F277W} > \text{F277W} - \text{F444W} + 1.3$ \\
& & ($\texttt{FLAG\textunderscore BN}\tablenotemark{a} < 2$) or ($\text{F200W} - \text{F277W} > 3$) \\
\enddata
\tablenotetext{a}{The $\texttt{FLAG\textunderscore BN}$ flag in the JADES catalog denotes the flux ratio between a given object and its brightest neighbor within the bounding box defined by the object's segmentation plus 0.3\arcsec (10 pixels in the JADES mosaics) on all sides. $\texttt{FLAG\textunderscore BN} = 2$ signifies a neighbor $10\times$ brighter than the object under consideration and larger values correspond to brighter neighbors.}
\end{deluxetable*}

The full JADES NIRCam imaging, combined with that from JEMS, FRESCO, and program 3215, comprises a rich dataset, sometimes with up to 14 bands of NIRCam imaging in the same area \citep{eisenstein2023_jof}. However, due to the multiple observational programs targeting different areas with different filters for different amounts of time, the imaging is also very complex with heterogeneous filter coverage and depths over the full area. The image used for source detection incorporates all of the LW images available at a given location (Section\ \ref{subsec:photometry}), but for simplicity, we otherwise restrict our analysis to the eight NIRCam filters (F090W, F115W, F150W, F200W, F277W, F356W, F410M, and F444W) in which JADES obtained its primary imaging \citep{eisenstein2023_jades}. Correspondingly, we only select candidates and calculate the UV luminosity function in the area covered by all eight of these filters, a total of $\sim 163$\,arcmin$^2$ ($\sim 83$\,arcmin$^2$ in GOODS-S and $\sim 80$\,arcmin$^2$ in GOODS-N).

In Table\ \ref{tab:depths}, we report the $5\sigma$ depths for the NIRCam images in this limited area. To measure depths, we mask sources out of the image using the segmentation map, then calculate aperture corrected photometry in blank regions of the field using the $r = 0.1\arcsec$ circular apertures (named CIRC1 in the JADES photometric catalog) that we use for sample selection (Section\ \ref{sec:selection}). We note that the exposure times over the full JADES area can vary by more than a factor of 16 in F200W, leading to upwards of a factor of four difference in depth between the shallowest and deepest regions of the images ($\Delta m \geq 1.5$). We illustrate this variation in depth in Figure\ \ref{fig:field_image}, where darker grey indicates deeper imaging. Thus, though we do not explicitly subdivide the full JADES field into subregions while calculating the UV luminosity function, we report the depths in four regimes of F200W exposure time to approximately describe the area of the JADES imaging that constrains the faint end of the luminosity function. The deepest ($> 40\,000$\,s in F200W) $12$\,arcmin$^2$ area is typically $\sim 2$\,mag deeper than the shallowest ($< 5000$\,s) $\sim 12$\,arcmin$^2$ subregion, reaching typical $5\sigma$ depths of $m \sim 30.7$.

\section{Sample selection} \label{sec:selection}

\subsection{Photometric Selection Criteria} \label{subsec:photometric_selection}

With the photometric catalogs in hand, we now turn to selecting high-redshift galaxy candidates. At $z \sim 8.5 - 22.5$, the \Lya{} break redshifts progressively through the NIRCam F115W ($z \sim 8.5 - 12$), F150W ($z \sim 12 - 16$), and F200W ($z \sim 16 - 22.5$) filter. We search for high-redshift galaxy candidates primarily through three sets of color criteria, one for each of these three dropout filters, then apply secondary selection criteria to further clean the samples. We enumerate the exact color cuts in Table\ \ref{tab:selection_criteria} but provide a general description below. In brief, our selections are designed to require that high-redshift galaxy candidates satisfy the following:

\begin{enumerate}
    \item have a red dropout color ($\text{F115W} - \text{F150W}$, $\text{F150W} - \text{F200W}$, and $\text{F200W} - \text{F277W}$ greater than 1.3 for the F115W, F150W, and F200W dropout samples, respectively),
    \item are robustly detected in the filters expected to probe the rest-frame UV and rest-frame optical ($\text{S/N} > 2$ in all of these filters and $\text{S/N} > 5$ in more than half of them),
    \item are not detected (signal-to-noise of $\text{S/N} < 2$) in any of the NIRCam filters blueward of the filter expected to contain the \Lya{} break,
    \item do not have an extremely red color in filters expected to probe the rest-frame UV ($\text{F150W} - \text{F277W}$, $\text{F200W} - \text{F356W}$, and $\text{F277W} - \text{F44W}$ less than 1.0 for the F115W, F150W, and F200W dropout samples, respectively), and
    \item have an increasingly strong \Lya{} break the redder the rest-UV colors are ($\text{dropout color} > \text{rest--UV color} + 1.3$).
\end{enumerate}

We apply these color selection criteria to the photometric catalog constructed as described in Section\ \ref{subsec:photometry}. We acknowledge that the F115W and F150W dropouts that we search for in this work are expected to be detected in F150W and/or F200W as well as in the filters included in the SNR image we use for detection. However, due to the large number of filters already included in our detection image (especially in GOODS-S), we do not expect only one or two filters to add significant signal to the existing SNR image. We also expect that objects that are not detected in the LW SNR image will have relatively low S/N in the single-band images for the LW filters, thus making it less likely for them to pass criterion 2 of our color selection. We caution that this may bias our selection against objects with sufficiently blue spectral slopes that are only detected in F150W and/or F200W, but we emphasize that we perform source injection and recovery simulations to account for this effect when calculating the UV luminosity function.

We use photometry measured in $r = 0.1\arcsec$ circular apertures (i.e. CIRC1) for sample selection, as these apertures have less background noise than larger apertures and most high-redshift galaxies are expected to be comparable in size to the $r = 0.1\arcsec$ ($r \sim 0.45$\;proper\;kpc at $z = 9$) CIRC1 aperture \citep[e.g.][]{shibuya2015, yang2022, ono2023, ono2024, morishita2024}. We note that we use photometry measured on native resolution mosaics, which may result in bluer colors for our objects of interest than PSF matched images, as the size of the PSF increases with increasing wavelength and our aperture corrections assume point sources (but our objects are likely to be at least marginally resolved). To assess the impact of this effect, we perform our source injection and recovery tests on the same native resolution mosaics (see Section\ \ref{subsec:completeness}) and find only a mild systematic bias in photometric colors at wavelengths longer than the Ly$\alpha$ break (at most $\sim 0.1$\,mag). We observe little, if any, bias in the dropout colors, which we attribute to only the red filter having significant amounts of flux that may introduce bias. At longer wavelengths, both filters are expected to be detected and therefore both may introduce bias in the color (e.g. one filter may be underestimated and one may be overestimated).

\begin{deluxetable*}{cll}
    \tablecaption{\label{tab:beagle_photoz}Priors adopted for our \beagle\ SED models used to infer photometric redshifts (see Section\ \ref{sec:selection}).}
    \tablehead{\colhead{Parameter} & \colhead{Description} & \colhead{Prior}}
\startdata
$z$ & Redshift & Uniform, $z = 0 - 25$ \\
$M_*$ & Formed stellar mass & Uniform in log, $5 \leq \log_{10}(M_* / M_\odot) \leq 12$ \\
$Z_*$ & Stellar metallicity & Uniform in log, $-2.2 \leq \log_{10}(Z_* / Z_\odot) \leq -0.3$ \\
$\tau_\textsc{v}$ & Diffuse dust optical depth in the $V$-band & Uniform in log, $-3 \leq \log_{10}(\tau_\textsc{v}) \leq 0.7$ \\
$U$ & Ionization parameter & Uniform in log, $-4 \leq \log(U) \leq -1$ \\
$t_\text{max}$ & Maximum stellar age (i.e. onset of star formation) & Uniform in log, $7.35 \leq \log_{10}(t_\text{max} / \text{yr}) \leq \log_{10}(t_\text{univ}(z_\text{phot}) / \text{yr})$ \\
$\tau$ & Delayed exponential $e$-folding time & Uniform in log, $6 \leq \log_{10}(\tau / \text{yr}) \leq 10.5$ \\
$t_\text{recent}$ & Duration of recent constant component & Uniform in log, $6 \leq \log_{10}(t_\text{recent} / \text{yr}) \leq 7.3$ \\
$\text{sSFR}_\text{recent}$ & sSFR of recent constant component & Uniform in log, $-14 \leq \log_{10}(\text{sSFR}_\text{recent} / \text{yr}^{-1}) \leq -6$ \\
\enddata
\end{deluxetable*}

If the CIRC1 flux for an object has $\text{S/N} < 1$ in the dropout filter for each selection, we set the flux to the $1\sigma$ flux error before applying the selection criteria in Table\ \ref{tab:selection_criteria}. Additionally, for each of the F150W and F200W dropout candidates, we require that the object is not selected by any of the lower redshift selections (which may happen in the case of partial dropouts). For example, if an F150W dropout is also selected by the F115W dropout selection, we consider the object an F115W dropout and exclude it from the F150W dropout sample.

Balmer breaks and strong rest-optical emission lines at low redshift can mimic the rest-UV colors of high-redshift galaxies, especially near the flux limit of a survey. To minimize this interloper population, we remove objects with any close\footnote{A close neighbor is defined as a source within the bounding box set by the primary object's segmentation plus 0.3\arcsec{} (10 pixels in the JADES mosaics) on all sides.} neighbors that are $> 10\times$ brighter than the object itself, as these sources have a higher probability than more isolated objects do of being star clusters or satellites associated with a bright, low redshift galaxy. However, we note that some legitimate high-redshift galaxies will fall near bright neighbors in projection and be rejected \citep[e.g.][]{hainline2024_highz}, as initially occurred for JADES-GS-z14-0 \citep{robertson2024_jof, carniani2024}. To recover at least some of these systems, we allow objects that have a very red ($> 3$\,mag) dropout color and satisfy all other selection criteria even if they have a close, bright neighbor.

Similarly, photometric scatter may cause true high-redshift galaxies to be removed from our samples if they are formally detected in a filter expected to be blueward of the \Lya{} break. Thus, we make the same exception and allow objects that have a $> 3$\,mag dropout color but are formally detected at $\text{S/N} \geq 2$ in one of the filters blueward of the \Lya{} break. We do not place any specific requirements about which filter can be allowed to have a formal detection, but we do not allow objects that have $\text{S/N} \geq 2$ in more than one of the blue filters. Together, these exceptions for a bright neighbor and a formal detection in a blue filter apply to four candidates in all of the dropout samples.

After applying these initial selection criteria, we obtain samples of 417 F115W dropouts, 124 F150W dropouts, and 90 F200W dropouts.

\bigskip

We further clean our samples using redshift probability distributions obtained by fitting the CIRC1 spectral energy distributions (SEDs) with the BayEsian Analysis of GaLaxy sEds \citep[\beagle;][]{chevallard2016} code. \beagle{} self-consistently models both stellar and nebular emission and based on an updated version of the \citet{bruzual2003} stellar population synthesis models \citep{vidal-garcia2017} and the nebular line and continuum models of \citet{gutkin2016}. We assume a \citet{chabrier2003} stellar initial mass function with mass range $0.1 - 300$\,\Msun, a Small Magellanic Cloud dust extinction curve \citep{pei1992}, and the IGM attenuation model of \citet{inoue2014}. For the star formation history (SFH), we adopt a two component model consisting of a delayed exponential at early times and a constant component at recent times that is completely decoupled from the early time delayed exponential. We place a uniform prior on redshift ranging from $z_\text{phot} = 0 - 25$ and log-uniform priors on all other free parameters; see Table\ \ref{tab:beagle_photoz}.

Using the redshift posterior probability distributions from these \beagle{} models, we require that candidates have a $> 50$\,percent probability of being at high redshift to be included in our samples. That is, we require $\mathcal{P}(z \geq z_\text{lim}) \equiv \int_{z_\text{lim}}^{25} p(z) \text{d}z > 0.5$, where $z = 25$ is the upper limit on redshift that we have placed on our \beagle{} models and $z_\text{lim} = 8, 11, \text{ and } 15$ for the F115W, F150W, and F200W dropout samples, respectively. We note that, though we initially adopt a probability requirement of $\mathcal{P}(z \geq z_\text{lim}) > 0.5$, we demonstrate in Section\ \ref{sec:sample_overview} that a large fraction of our samples have significantly higher probabilities of being at high redshift. This photometric redshift probability requirement removes another $\sim 20 - 30$\,percent of the original color selected samples. Additionally, out of the 23 F115W dropout candidates and four F150W dropout candidates that have spectroscopic redshifts, we identify two F115W dropouts that are spectroscopically confirmed to be at low redshift and remove them from the sample. After these redshift cuts, our samples comprise 325 F115W dropouts, 85 F150W dropouts, and 75 F200W dropouts.

\begin{figure}
    \includegraphics[width=\columnwidth]{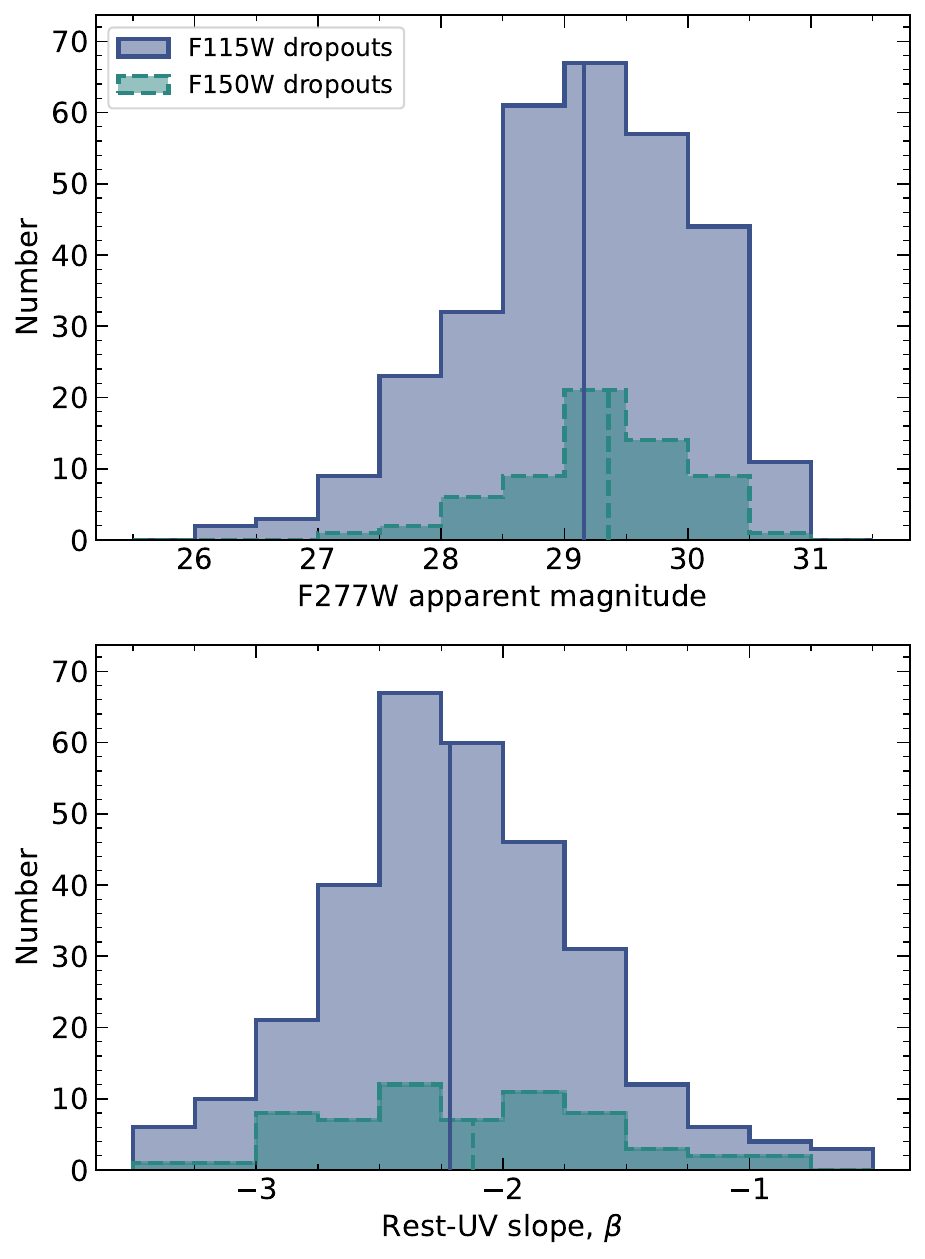}
    \caption{\label{fig:property_histograms}The distributions of observed F277W apparent magnitudes and rest-UV slopes, $\beta$, of our dropout samples. To measure UV slopes, we fit a power law of the form $f_\lambda \propto \lambda^\beta$ to three wide filters expected to probe the rest-frame UV; see Section\ \ref{sec:sample_overview} for details. We show the F115W dropout sample in blue with a solid outline and the F150W dropout sample in dark green with a dashed outline. The median values for the F115W and F150W dropout samples are shown as vertical lines. Our samples are relatively faint, with median F277W apparent magnitudes of $m_\text{F277W} = 29.2$ and $m_\text{F277W} = 29.4$ for the F115W and F150W dropouts, respectively. Consistent with expectations for high-redshift galaxies \citep[e.g.][]{cullen2023, austin2024, topping2024}, we also find generally blue UV continuum slopes with median $\beta = -2.2$ for the F115W dropouts and $\beta = -2.1$ for the F150W dropouts.}
\end{figure}

After applying all of the algorithmic selection criteria, we verify that none of the candidates have been previously identified as transients \citep{decoursey2023_gs, decoursey2023_gn, decoursey2023_3215} or brown dwarf candidates \citep{hainline2024_bds}. Finally, the SEDs and postage stamps of all candidates are visually inspected by authors LW, DPS, MWT, KNH, RE, and ZC to remove spurious detections and imaging artifacts (e.g. diffraction spikes, hot pixels near detector edges, cross-talk between NIRCam amplifiers\footnote{\url{https://www.stsci.edu/files/live/sites/www/files/home/jwst/documentation/technical-documents/_documents/JWST-STScI-004361.pdf}}). All remaining F200W dropout candidates are removed by this visual inspection. During visual inspection, we also identify several F115W dropouts with overlapping Kron apertures (all of which fall within $\sim 0.7$\arcsec of their neighbor, corresponding to $\sim 3$\,kpc at $z = 10$). For these groups of objects with overlapping apertures, we define a single multi-component source by creating an elliptical aperture that contains all of the relevant objects. After these steps, our final samples consist of 309 F115W dropouts, 63 F150W dropouts, and zero F200W dropouts.

\begin{deluxetable*}{cl}
    \tablecaption{\label{tab:properties}A description of the columns of our source catalog table, available online at \url{https://github.com/lwhitler/jades_highz_uvlf}.}
    \tablehead{\colhead{Column Name} & Description}
\startdata
\texttt{ID} & JADES ID \\
\texttt{RA}, \texttt{Dec} & Right ascension and declination in degrees \\
\texttt{f\textunderscore F277W}, \texttt{f\textunderscore F277W\textunderscore err} & Median and error of the flux in F277W in $r = 0.1\arcsec$ circular apertures (CIRC1) in nJy \\
\texttt{beta}, \texttt{beta\textunderscore err} & Median and error of the rest-frame UV continuum slope \\
\texttt{zphot}, \texttt{zphot\textunderscore lowerr}, \texttt{zphot\textunderscore uperr} & Median and errors of the photometric redshift from \beagle{} SED models \\
\texttt{zspec}, \texttt{zspec\textunderscore ref} & If available, the spectroscopic redshift and corresponding literature reference(s) \\
\texttt{p\textunderscore highz} & Probability of being at high redshift from \beagle{} SED models \\
\texttt{MUV}, \texttt{MUV\textunderscore lowerr}, \texttt{MUV\textunderscore uperr} & Median and errors of the absolute UV magnitude in AB magnitudes\\
\texttt{Sample} & The dropout sample in which the source is included \\
\enddata
\end{deluxetable*}

\begin{figure*}
    \centering
    \includegraphics[width=\textwidth]{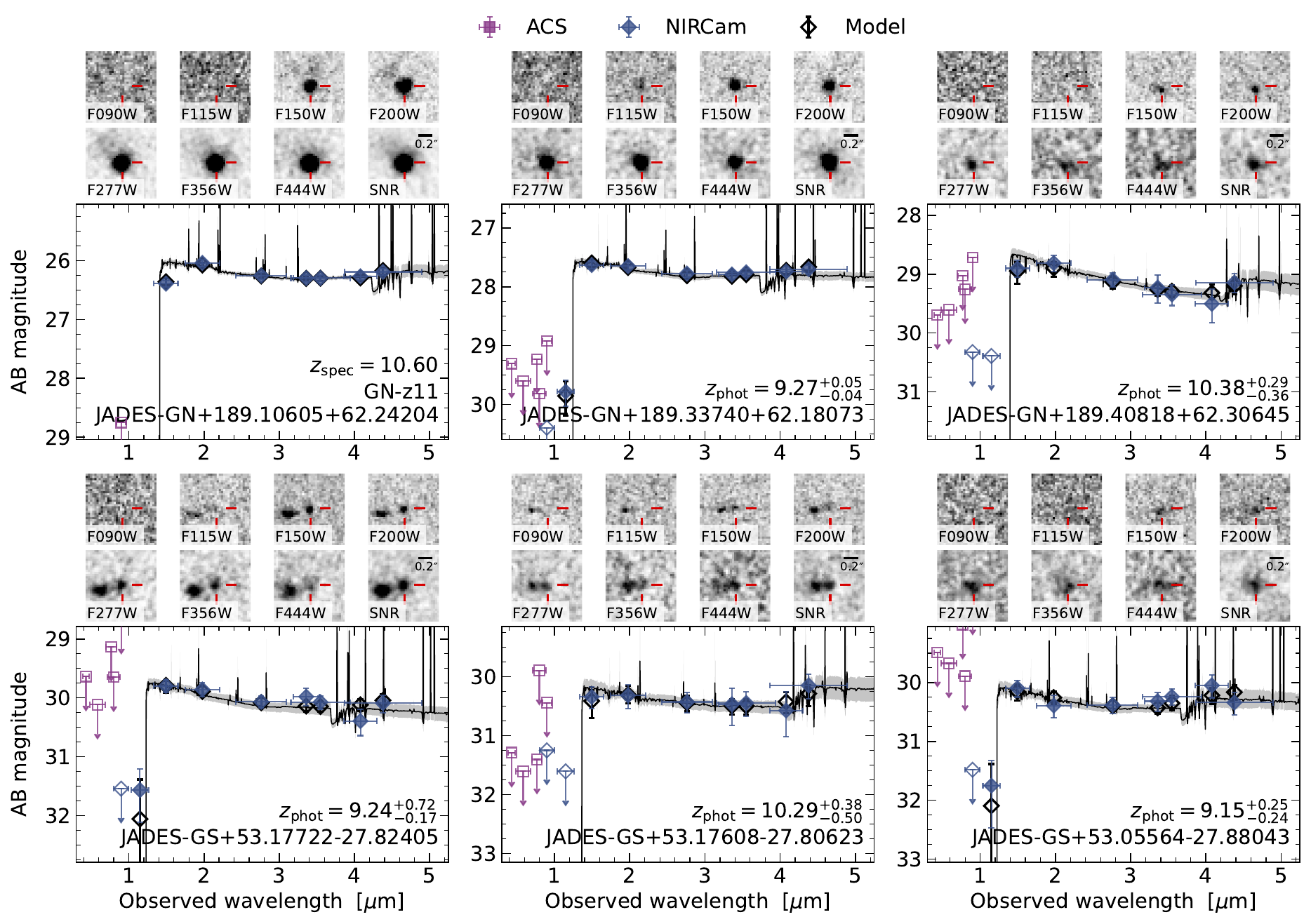}
    \caption{\label{fig:f115w_seds_stamps}Postage stamps and CIRC1 SEDs of a subset of the F115W dropout sample ordered by decreasing UV luminosity from top to bottom and left to right. The postage stamps are 1\arcsec{} on a side and we show a 0.2\arcsec{} scale bar on the SNR image for each object. For the SEDs, we show the observed photometry as the colored points (blue diamonds for NIRCam and purple squares for ACS), the model photometry as the open black diamonds, and the median model spectra with errors as the black lines and grey shaded regions. When the observed photometry has $\text{S/N} < 1$, we show $2\sigma$ upper limits as open symbols. These objects span nearly five magnitudes in UV luminosity, are often observed with extremely red colors spanning the \Lya{} break ($> 2$\,mag), and generally have blue UV continuum slopes. The SEDs and postage stamps for all of our dropout candidates are available online at \url{https://github.com/lwhitler/jades_highz_uvlf}.}
\end{figure*}

\subsection{Spectroscopic Redshifts} \label{subsec:spectroscopy}

The JADES area has been observed extensively with JWST spectroscopy as part of both JADES \citep{eisenstein2023_jades, bunker2024_jades_dr1, deugenio2024} and FRESCO \citep{oesch2023_fresco}, enabling spectroscopic confirmations of galaxies at redshifts as high as $z = 14.2$ \citep{carniani2024}. Of the F115W (F150W) dropout candidates, 22 (four) are spectroscopically confirmed, including GN-z11 \citep[$z_\text{spec} = 10.60$;][]{oesch2016, bunker2023_gnz11} and JADES-GS-z14-0 \citep[$z_\text{spec} = 14.18$;][]{carniani2024, carniani2024_z14_alma, schouws2024}. Throughout the remainder of this work, we use the spectroscopic redshift, $z_\text{spec}$, when available.

\subsection{Measuring UV Luminosities} \label{subsec:Muv_measurement}

We note that we have used a small circular aperture for sample selection to minimize background noise, but this small aperture does not capture the total flux, and eventually total UV luminosity, of large objects or objects with complex morphologies. To measure the total absolute UV magnitude of a given object, we scale the CIRC1-measured UV luminosity up to total UV luminosity using the ratio of the $k = 2.5$ Kron photometry to the CIRC1 photometry for the object under consideration. We choose to rescale the CIRC1 photometry rather than directly use photometry measured in Kron apertures in order to preserve the colors of the CIRC1 SEDs we have used for selection, but we emphasize that our rescaling is designed to recover the normalization of the aperture-corrected Kron photometry. In detail, we first re-fit the CIRC1 SEDs with \beagle{} but restrict the models to high redshifts, then integrate the resulting model spectra over rest-frame wavelengths of $\lambda_\text{rest} = 1450 - 1550$\,\AA{} to obtain a CIRC1-based UV luminosity. If the object has a spectroscopic redshift, we fix the SED model to that redshift. Otherwise, we place a uniform prior on redshift ($z = 8 - 25$ for the F115W dropouts and $z = 10 - 25$ for the F150W dropouts) to ensure that the absolute UV magnitudes we measure reflect the luminosities of these objects at high redshift. Then, we calculate the ratio of the observed Kron fluxes to the CIRC1 fluxes in all of the wide filters at longer wavelengths than the dropout filter for the object under consideration. Finally, we multiply the CIRC1 UV luminosity by the median Kron-to-CIRC1 ratio to obtain the final total UV luminosity, which results in a median change of $\sim 0.3$\ mag, or a factor of $\sim 1.3$. Errors are propagated numerically from the uncertainty on the CIRC1 UV luminosity and the standard deviation of the per-filter Kron-to-CIRC1 ratios.

\section{Sample overview} \label{sec:sample_overview}

In this section, we give a brief summary of the observed properties of our high-redshift candidates. We provide a full description of these properties in an online table (\url{https://github.com/lwhitler/jades_highz_uvlf}) and provide a description of the table columns in Table\ \ref{tab:properties}. We provide a brief description of our F115W and F150W dropout candidates in Sections\ \ref{subsec:f115w_dropouts} and \ref{subsec:f150w_dropouts}. In Sections\ \ref{subsec:hainline2024_comparison} and \ref{subsec:robertson2024_comparison}, we compare our samples with the samples identified by \citet{hainline2024_highz} and \citet{robertson2024_jof}, respectively, who previously searched for high-redshift galaxy candidates in subsets of the data that we use in this work.

\begin{figure*}
    \includegraphics[width=\textwidth]{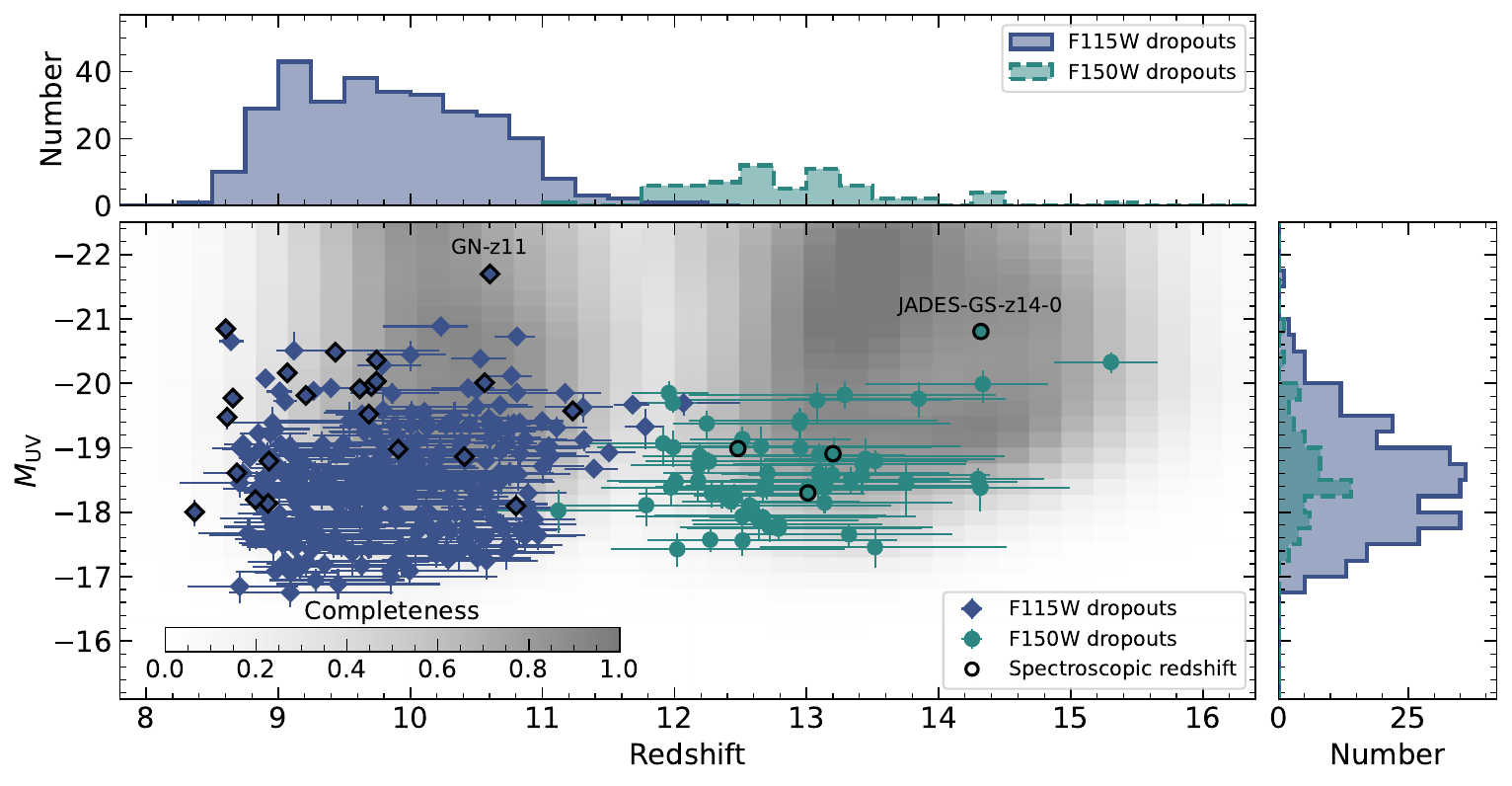}
    \caption{\label{fig:redshift_Muv}The distribution of redshifts (spectroscopic redshifts if available, otherwise photometric redshifts from \beagle) and absolute UV magnitudes for our three dropout samples. We show the F115W dropouts as blue diamonds and the F150W dropouts as dark green circles with corresponding colors for the histograms. Objects with spectroscopic redshifts are outlined in black. We also show the 2D detection and selection completeness measured as described in Section\ \ref{subsec:completeness} as a function of F277W apparent magnitude and redshift in the background. Darker grey indicates a higher completeness. The F115W dropouts span absolute UV magnitudes of $-16.8 \leq \Muv \leq -21.7$ (median $\Muv = -18.5$) and redshifts of $z = 8.4 - 12.1$ (median $z = 9.8$), consistent with expectations from our selection function. The F150W dropouts have absolute UV magnitudes of $-17.4 \leq \Muv \leq -20.8$ (median $\Muv = -18.5$) and redshifts of $z = 11.1 - 15.3$. We note that we attribute the apparent weighting of the observed samples towards the lower redshifts and fainter luminosities of the selection functions to the larger numbers of these objects (i.e. at fixed completeness, we will identify more faint, lower-redshift candidates compared to brighter, higher redshift systems). We note that GN-z11 \citep{oesch2016, bunker2023_gnz11} is the brightest object in the F115W dropout sample and JADES-GS-z14-0 \citep{carniani2024} is the brightest object in the F150W dropout sample.}
\end{figure*}

\begin{figure*}
    \centering
    \includegraphics[width=\textwidth]{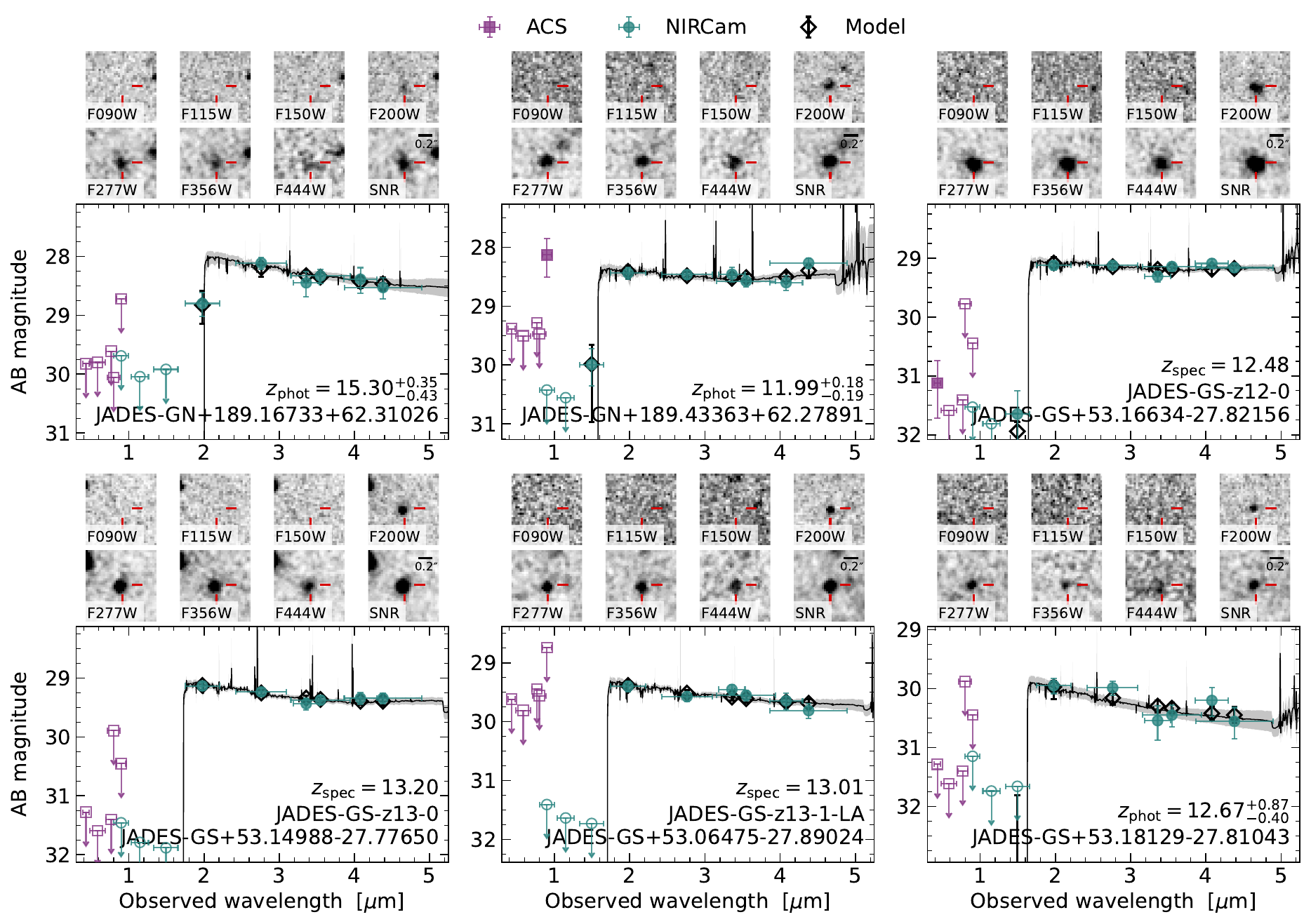}
    \caption{\label{fig:f150w_seds_stamps}The same as Figure\ \ref{fig:f115w_seds_stamps} for a subset of the F150W dropouts. The NIRCam photometry is shown as green circles and the ACS photometry is shown as purple squares. The F150W dropouts span a range of 3.5 magnitudes in UV luminosity and generally have blue UV continuum slopes.}
\end{figure*}

\subsection{F115W dropouts} \label{subsec:f115w_dropouts}

Our 309 F115W dropout candidates span nearly a factor of 100 in flux (almost five magnitudes) with F277W apparent magnitudes ranging from $26.1 \leq m_\text{F277W} \leq 30.9$ (median $m_\text{F277W} = 29.2$; top panel of Figure\ \ref{fig:property_histograms}). As is typical for $z \gtrsim 7$ galaxies \citep[e.g.][]{wilkins2011, finkelstein2012_colors, bouwens2014, cullen2023, austin2024, topping2024, saxena2024}, the rest-frame UV continuum slopes ($\beta$) of the sample, measured by fitting power laws of the form $f_\lambda \propto \lambda^\beta$ to the observed F200W, F277W, and F356W photometry, are generally blue (median $\beta = -2.2$; bottom panel of Figure\ \ref{fig:property_histograms}). To minimize the likelihood of artificially reddening our inferred UV slopes, we do not use F150W (the filter immediately adjacent to the F115W dropout filter) to calculate $\beta$, as this filter may be partially impacted by a \Lya{} break or damped \Lya{} absorption. We show the distributions of F277W apparent magnitudes and rest-UV slopes of the sample in Figure\ \ref{fig:property_histograms} and examples of the images and SEDs of individual objects in Figure\ \ref{fig:f115w_seds_stamps}.

From the SED models, we infer redshifts from \beagle{} ranging from $z_\text{phot} = 8.4 - 12.1$ (median $z_\text{phot} = 9.8$) and absolute UV magnitudes of $-21.7 \leq \Muv \leq -16.8$ (median $\Muv = -18.5$). As shown in Figure\ \ref{fig:redshift_Muv}, these inferred properties are consistent with expectations for the sample based on the selection function calculated as described in Section\ \ref{subsec:completeness}. We also highlight that most objects in the sample have large probabilities of being at $z \geq 8$. All candidates are guaranteed to have integrated probabilities of $\mathcal{P}(z \geq 8) > 0.5$ by the selection (Section\ \ref{sec:selection}) but most easily exceed this requirement and 83\,percent of the candidates without a known spectroscopic redshift have $\mathcal{P}(z \geq 8) > 0.9$. We note that this fraction is primarily driven by objects with luminosities fainter than $\Muv \sim -20$; 100\,percent of objects at $\Muv \lesssim -20$ and $\sim 82$\,percent of fainter objects satisfy $\mathcal{P}(z \geq 8) > 0.9$, with no dependence on luminosity at $\Muv \lesssim -20$.

\subsection{F150W dropouts} \label{subsec:f150w_dropouts}

Our 63 F150W dropout candidates also represent a large range of fluxes (a factor of $\sim 25$, or $\sim 3.5$ magnitudes), and generally have blue rest-UV slopes (measured by fitting a power law to the observed F277W, F356W, and F444W photometry). We show examples of individual objects in Figure\ \ref{fig:f150w_seds_stamps}. As seen in Figure\ \ref{fig:property_histograms}, the F150W dropout sample is observed with a median rest-UV slope of $\beta = -2.1$ and F277W magnitudes of $27.0 \leq m_\text{F277W} \leq 30.5$ (median $m_\text{F277W} = 29.4$). Correspondingly, we measure absolute UV magnitudes of $-20.8 \leq \Muv \leq -17.4$ (median $\Muv = -18.5$) at redshifts of $z_\text{phot} = 11.1 - 15.3$ (median $z = 12.7$); see Figure\ \ref{fig:redshift_Muv}. We again note that many of the F150W dropouts have high-redshift probabilities significantly larger than is required by the selection. Of the objects that are not spectroscopically confirmed, 55\,percent have integrated high-redshift probabilities of $\mathcal{P}(z \geq 10) > 0.9$. Similar to the F115W dropouts, this fraction is dominated by objects with $\Muv \gtrsim -20$, of which $\sim 54$\,percent have $\mathcal{P}(z \geq 10) > 0.9$ (and of these fainter systems, most of the objects with integrated probabilities $\leq 0.9$ have UV luminosities of $-20 \lesssim \Muv \lesssim -19$).

\subsection{Comparison with \citet{hainline2024_highz}} \label{subsec:hainline2024_comparison}

In this work, we have adopted color selections to identify high-redshift galaxy candidates with which to measure the UV luminosity function. We now briefly compare these color-selected samples to the sample of $z > 8$ candidates that were identified in JADES imaging by \citet{hainline2024_highz} using photometric redshifts from \texttt{EAZY}. We note that \citet{hainline2024_highz} searched for candidates at $z > 8$ while our color selections are sensitive to candidates starting at a slightly higher redshift. We also note that the data presented in this work includes JADES imaging observed after the publication of \citet{hainline2024_highz}. Thus, to ensure consistency for this comparison, we only consider the subsets of the two samples that lie in the shared area at $z \geq 9$ (spectroscopic redshifts if available, otherwise best-fit photometric redshifts from \texttt{EAZY} for the \citealt{hainline2024_highz} sample and median photometric redshifts from \beagle{} for the objects in this work). After applying these requirements, we find that the color selections identify a moderately smaller number of $z \geq 9$ galaxy candidates. Out of the parent sample of 372 color-selected candidates we have identified in this work, we find that 271 satisfy these criteria. In comparison, 327 of the full Primary Sample of 717 objects identified by \citet{hainline2024_highz} satisfy the criteria. Of these candidates, 134 are shared between both selection methods.

We attribute this difference in part to the relatively low efficiency of color selections in specific ranges of redshift caused by the progressive redshifting of the \Lya{} break through the dropout filters (see \citealt{hainline2024_highz} for a quantitative discussion of this effect). At the redshifts where color selections are inefficient, the \Lya{} break falls within a filter such that objects do not appear as complete dropouts in that filter. However, if that filter is used to calculate a rest-frame UV color while assuming the object is dropping out in the immediately adjacent, shorter wavelength filter, it appears to be red. Thus, such objects are not identified by either of the adjacent color selections but can be selected by their photometric redshifts. Additionally, we have required the photometry in at least three filters (four for the F115W dropouts) to be measured at $\text{S/N} > 5$, a slightly stronger $\text{S/N}$ requirement than the one adopted by \citet{hainline2024_highz}. 

For the objects that are identified by the color selection and are not included in the photometric redshift-selected sample, almost all have double-peaked redshift probability distributions, with peaks at both at $z < 8$ and $z > 8$. Additionally, we find that approximately half of these sources also have best-fit \texttt{EAZY} redshifts at $z > 8$, but were removed from the \citet{hainline2024_highz} sample by another selection criterion.

\subsection{Comparison with \citet{robertson2024_jof}} \label{subsec:robertson2024_comparison}

The JADES imaging we use in this work encompasses the area of the JOF \citep{eisenstein2023_jof}, in which \citet{robertson2024_jof} conducted a search for high-redshift galaxy candidates using a photometric redshift selection. We identify approximately the same number of $z \gtrsim 11.5$ objects in the JOF area using our F150W dropout color selection, but due to the differences in selection techniques, the objects included in the samples differ.

Of the eleven objects in the \citet{robertson2024_jof} Primary and Auxiliary samples at $z > 11.5$, our color selection shares six. The remaining five are all legitimate dropout candidates but three are removed by the bright neighbor flag that we apply in this work. The last two fall in between our F115W and F150W dropout selections with dropout colors that are slightly too blue to be identified by our color selection; this is the same effect as is seen when comparing our color selected sample to the photometric redshift selected sample of \citet{hainline2024_highz}. We also identify six F150W dropouts that are not included in the sample of \citet{robertson2024_jof} (including the Contributing and Auxiliary Samples), which were removed from the JOF luminosity function sample due to having a best fit photometric redshift from \texttt{EAZY} \citep{brammer2008} of $z < 11.5$ in their catalog or an insufficiently large difference in the goodness-of-fit between the $z > 11$ and $z < 7$ photometric redshift solutions, but all of which also have secondary redshift solutions from \texttt{EAZY} at $z > 12$.

Overall we emphasize that while our sample differs in the specific objects (though not the total number of candidates) from the samples of \citet{hainline2024_highz} of \citet{robertson2024_jof}, the objects identified by all selection criteria are all plausible high-redshift galaxy candidates.

\section{The rest-frame UV luminosity function} \label{sec:uvlf}

With our high-redshift candidate samples in hand, we now measure the UV luminosity function at $z \sim 8.5 - 12$ and $z \sim 12 - 16$ based on the F115W and F150W dropout samples and derive an upper limit at $z \sim 16 - 22.5$ corresponding to our lack of F200W dropouts. We also estimate the binned $z \geq 14$ luminosity function from a subset of the F150W dropout sample. In this section, we describe our methods for quantifying the completeness of our selection, then calculate binned and parametric forms of the UV luminosity function observed in the JADES fields.

\subsection{Completeness} \label{subsec:completeness}

In order to accurately measure the UV luminosity function, we must account for the incompleteness of our selection. To this end, we perform source injection and recovery simulations designed to reproduce the real process of high-redshift galaxy selection as closely as possible. That is, we inject mock sources with a range of SEDs and morphological properties into the real JADES mosaics, then perform source detection, photometric measurements, and sample selection using the same methods as we use for our real data.

When creating artificial sources for injection, we sample properties from distributions designed to represent our current understanding of high redshift galaxy properties and to comfortably encompass the expected parameter space of our selections. Absolute UV magnitudes (\Muv) are sampled from a uniform distribution in the range $-24 \leq \Muv \leq -15$ and redshifts ($z$) are sampled from uniform distributions with lower and upper bounds appropriate for each dropout selection ($7 \leq z \leq 12.5$, $10 \leq z \leq 17$, and $14 \leq z \leq 25$ for the F115W, F150W, and F200W dropout selections, respectively). Then, given $(\Muv, z)$ for each mock source, we construct power law SEDs of the form $f_\lambda \propto \lambda^{\beta}$, where $\beta$ is the rest-frame UV continuum slope. At the redshifts we consider in this work ($z \gtrsim 9$), we expect to be primarily probing the rest-UV with our observational filter set and do not expect significant deviations from a power law shape of the SED due to strong emission lines, as the strongest rest-frame optical nebular emission lines (H$\alpha$, H$\beta$, and the [O\,\textsc{iii}]$\lambda\lambda$4959,5007 doublet) have redshifted out of the filters we consider in this work and the strongest rest-frame UV emission line, \Lya, is observed to be relatively uncommon at these redshifts \citep[e.g.][]{nakane2024, jones2024a, jones2024b, tang2024_lya, napolitano2024_lya}. The rest-UV slope is then determined from the mock source's \Muv{} using the $\beta - \Muv$ relation found by \citet{topping2024}, which follows a similar trend as other JWST measurements of the $\beta - \Muv$ relation \citep[e.g.][]{cullen2023, cullen2024, austin2024}. We note that \citet{topping2024} derived $\beta - \Muv$ relations for F115W dropouts and F150W dropouts but did not fit a $\beta - \Muv$ relation for F200W dropouts, so we adopt the F150W dropout relation for both our F150W and F200W dropout selections. The SED is then normalized to \Muv{} at $\lambda_\text{rest} = 1500$\,\AA{} and redshifted to the observed frame at redshift $z$. Finally, the IGM attenuation model of \citet{inoue2014} is applied and mock source fluxes in the same ACS and NIRCam filters as are in the real photometric catalog are calculated by integrating the SED over the appropriate filter bandpasses. We note that, though the \citet{inoue2014} IGM attenuation model does not explicitly consider the impact of damped \Lya{} absorption, we do not expect strong damped \Lya{} absorption to significantly impact our inferred UV luminosity functions. For the objects in our samples with spectroscopic redshift measurements, we observe only a mild systematic offset between the photometric and spectroscopic redshifts \citep[median $\Delta z \sim 0.2$, consistent with expectations from larger samples][]{heintz2024}, which is expected to have a negligible impact on the inferred UV luminosity ($\Delta \Muv \lesssim 0.05$\,mag).

\begin{figure*}
    \includegraphics[width=\textwidth]{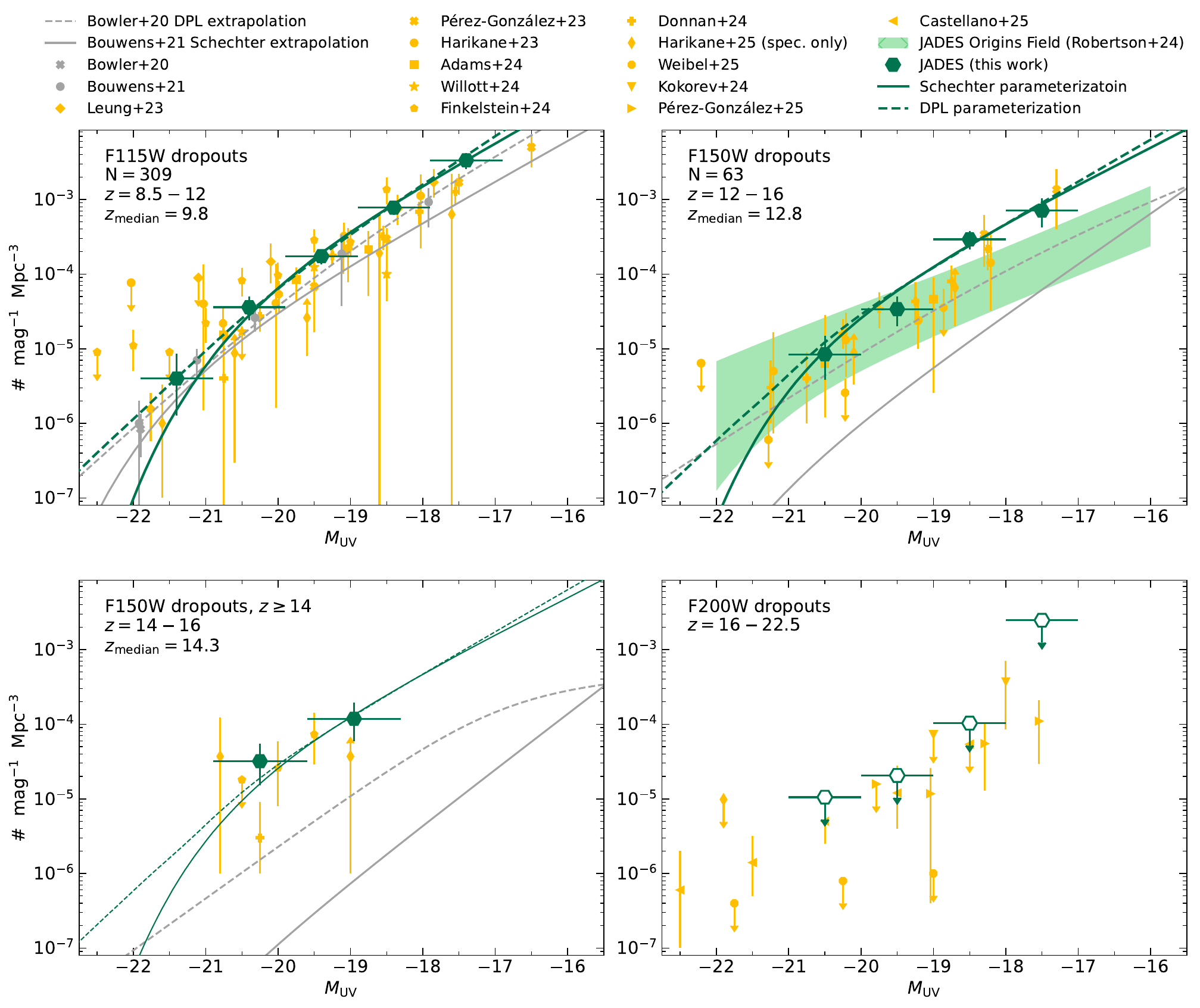}
    \caption{\label{fig:uvlfs}Our measurements of the rest-UV luminosity functions based on JADES data. We show F115W dropout luminosity function ($z_\text{median} = 9.8$, $N = 309$) in the upper left panel, the F150W dropouts ($z_\text{median} = 12.8$, $N = 63$) in the upper right panel, the $z \geq 14$ subset of the F150W dropouts ($z_\text{median} \sim 14.3$) in the lower left panel, and the upper limits based on the absence of F200W dropouts ($z \sim 16 - 22.5$) in the lower right panel. Due to the method we use to calculate the $z \geq 14$ F150W dropout luminosity function, described in Section\ \ref{subsec:binned_lf}, the number of objects is not well-defined. We show our binned measurements as green hexagons, the Schechter fits as solid green lines, and the DPL fits as dashed green lines. For comparison, we show the luminosity functions at the central redshifts of the dropout samples predicted by the fits for the redshift evolution of the Schechter \citep{bouwens2021} and DPL \citep{bowler2020} parameters, which were determined from HST and ground-based data. We also show measurements from \citet{bowler2020, bouwens2021, leung2023, perez-gonzalez2023, harikane2023, harikane2024b, adams2024, willott2024, finkelstein2024, donnan2024, robertson2024_jof, kokorev2024, perez-gonzalez2025, castellano2025, weibel2025} at the redshifts closest to the central redshift of the F115W and F150W dropout samples (typically $z = 9$ or $10$ for the F115W dropouts, $z = 12$ or $12.5$ for the F150W dropouts, $z = 14$ or $14.5$ for the $z \geq 14$ F150W dropouts, and $z = 16, 17$ and $17.5$ for the F200W dropout upper limit). In general, we find broad agreement with existing measurements of the luminosity function with smaller uncertainties.}
\end{figure*}

To simulate source shapes, we assume \citet{sersic1963} profiles with S\'ersic indices ($n$) drawn from a one-sided truncated normal distribution with mean of $\mu_n = 1$, $\sigma_n = 1$, and a minimum of $a_n = 0.5$. We draw the axis ratio (the ratio of the minor axis to the major axis, $q \equiv b / a$) from a truncated normal distribution with mean of $\mu_q = 0.8$, standard deviation of $\sigma_q = 0.4$, minimum of $a_q = 0$, and maximum of $b_q = 1$. Finally, we draw position angles from a uniform distribution ranging from $-90^{\circ}$ to $90^{\circ}$. To set galaxy sizes, we use the rest-UV size-luminosity relation found by \citet{shibuya2015} \citep[which has the same slope and very similar normalizations as $z \gtrsim 8$ size-luminosity relationships derived using JWST;][]{yang2022, ono2023, ono2024, morishita2024} to calculate the half-light radius ($r_\text{half}$) for a mock galaxy with given UV luminosity and redshift. Simulated images of these mock sources are generated using \texttt{GalSim} \citep{rowe2015} and placed into the real JADES mosaics. We then repeat our process of detection, photometric measurements, and selection (Sections\ \ref{subsec:photometry} and \ref{sec:selection}) on the mosaics containing the mock sources. We also measure the absolute UV magnitudes of the recovered mock sources following the same methods we have used for the real objects (Section\ \ref{sec:sample_overview}).

We calculate the combined completeness of our detection and selection methods, $C(\Muv, z) = C_\text{detect}(\Muv, z) \times C_\text{select}(\Muv, z)$, as a function of absolute UV magnitude and redshift marginalized over all other source properties. However, we highlight that the UV luminosity function is a function of true \Muv, $\phi(\Muv^\text{true})$, while we can only observe a recovered absolute UV magnitude, $\Muv^\text{recov}$, and $\Muv^\text{true}$ is not guaranteed to be the same as $\Muv^\text{recov}$. To account for this difference, we quantify our selection completeness as a function of both true and recovered absolute UV magnitude as well as true redshift \citep[similar to methods adopted by, e.g.][]{bouwens2021, leethochawalit2023_hst_uvlf}. Thus, completeness is defined as
\begin{equation} \label{eqn:completeness}
C(\Muv^\text{recov}, \Muv^\text{true}, z) = \frac{N(\Muv^\text{recov}, \Muv^\text{true}, z)}{N(\Muv^\text{true}, z)},
\end{equation}
where $N(\Muv^\text{true}, z)$ is the number of objects that have been injected in a given bin of true absolute UV magnitude and redshift, ($\Muv^\text{true}, z$), and $N(\Muv^\text{recov}, \Muv^\text{true}, z)$ is the number of sources that were originally injected into that same $(\Muv^\text{true}, z)$ bin and were recovered with an absolute UV magnitude of $\Muv^\text{recov}$.

We find that our detection and selection methods recover our injected sources with reasonably high completeness for the luminosities we use to calculate the luminosity function. In the brightest bins and central redshifts of our selection functions where we expect to be most complete, we find a maximum completeness of $\sim 70 - 73$\ percent due to objects overlapping with other sources and photometric scatter. At its central redshifts ($z \sim 9 - 11.5$) and in the deepest subregion of the imaging ($> 40\,000$\,s), our F115W dropout selection is $\sim 30$\ percent complete in the faintest bin we use to calculate the UV luminosity function (centered at $\Muv = -17.4$). For the F150W dropouts, which is maximally complete at $z \sim 12.5 - 15.5$, we find completeness of $\sim 10$\ percent in the faintest bin (centered at $\Muv = -17.5$) and deepest subregion.

Using this definition of the completeness, the effective volume of the survey is also a function of both true and recovered \Muv:
\begin{equation}
V_\text{eff}(\Muv^\text{recov}, \Muv^\text{true}) = \int \frac{dV_c}{dz} C(\Muv^\text{recov}, \Muv^\text{true}, z) \text{d}z,
\end{equation}
where $dV_c / dz$ is the differential comoving volume element in the survey area at redshift $z$. Then, given a UV luminosity function, $\phi(\Muv^\text{true})$, the number of objects expected as a function of $\Muv^\text{recov}$ can be calculated.

Starting from a given bin of true \Muv, $M_{\textsc{uv}, j}^\text{true}$, the number of objects expected to be measured with $M_{\textsc{uv}, i}^\text{recov}$ is $n_{\text{exp}, i, j} = \phi_j V_{\text{eff}, i, j} \Delta M_{\textsc{uv}, i}^\text{recov}$. Then, the total number of objects in the $M_{\textsc{uv}, i}^\text{recov}$ bin is simply the sum of the contributions from all $\Muv^\text{true}$ bins:
\begin{equation} \label{eqn:n_expected}
    n_{\text{exp}, i} =
    \sum_{j} n_{\text{exp}, i, j} = \Big[ \sum_{j} \phi_j V_{\text{eff}, i, j} \Big] \Delta M_{\textsc{uv}, i}^\text{recov}.
\end{equation}
Alternatively, this quantity can be thought of as the $i^\text{th}$ element of the dot product of the binned UV luminosity function, $\bm{\phi}(\bm{\Muv}^\text{true})$, and the matrix of effective volumes, $\bm{V}_\text{eff}(\bm{\Muv}^\text{recov}, \bm{\Muv}^\text{true})$, which produces a vector of expected counts as a function of $\Muv^\text{recov}$:
\begin{align}
    \bm{n_\text{exp}}&(\bm{\Muv}^\text{recov}) = \nonumber \\
    & \big[ \bm{V}_\text{eff}(\bm{\Muv}^\text{recov}, \bm{\Muv}^\text{true}) \bm{\cdot} \bm{\phi}(\bm{\Muv}^\text{true})\big] \odot \bm{\Delta M_{\textsc{uv}}^\text{recov}},
\end{align}
where $\odot$ denotes the element-wise (Hadamard) product of matrices.

Thus, we can now relate the luminosity function as a function of true \Muv{} (an unobservable quantity) to the number of objects expected in the sample as a function of recovered \Muv{} (the observable quantity), and fit the UV luminosity function.

\subsection{The binned UV luminosity function} \label{subsec:binned_lf}

\begin{deluxetable}{cc}
    \renewcommand{\arraystretch}{1.25}
    \tablecaption{\label{tab:binned_lf}The binned number densities measured from the F115W and F150W dropout samples, the $z \geq 14$ subset of the F150W dropouts, and the binned $2\sigma$ upper limits on the F200W dropout luminosity function.}
    \tablehead{\colhead{$M_\textsc{uv}$ [mag]} & $\phi$ [$10^{-5}$\,mag$^{-1}$\,Mpc$^{-3}$]}
\startdata
\multicolumn{2}{c}{F115W dropouts ($z_\mathrm{median} = 9.8$)} \\ \hline
$-21.4 \pm 0.5$ & $0.40_{-0.27}^{+0.45}$ \\
$-20.4 \pm 0.5$ & $3.6_{-1.2}^{+1.4}$ \\
$-19.4 \pm 0.5$ & $17_{-4}^{+4}$ \\
$-18.4 \pm 0.5$ & $78_{-16}^{+16}$ \\
$-17.4 \pm 0.5$ & $330_{-80}^{+80}$ \\ \hline\hline
\multicolumn{2}{c}{F150W dropouts ($z_\mathrm{median} = 12.8$)} \\ \hline
$-20.5 \pm 0.5$ & $0.84_{-0.46}^{+0.65}$ \\
$-19.5 \pm 0.5$ & $3.4_{-1.4}^{+1.6}$ \\
$-18.5 \pm 0.5$ & $29_{-8}^{+9}$ \\
$-17.5 \pm 0.5$ & $71_{-28}^{+33}$ \\ \hline\hline
\multicolumn{2}{c}{F150W dropouts, $z \geq 14$ ($z_\mathrm{median} = 14.3$)} \\ \hline
$-20.2 \pm 0.65$ & $3.2_{-1.7}^{+2.3}$ \\
$-18.9 \pm 0.65$ & $12_{-6}^{+8}$ \\ \hline\hline
\multicolumn{2}{c}{F200W dropouts, $z \sim 16 - 22.5$} \\ \hline
$-20.5 \pm 0.5$ & $< 1.1$ \\
$-19.5 \pm 0.5$ & $< 2.1$ \\
$-18.5 \pm 0.5$ & $< 10$ \\
$-17.5 \pm 0.5$ & $< 250$ \\
\enddata
\end{deluxetable}

We calculate the luminosity function for each of the three dropout samples we consider in this work rather than in bins of redshift to avoid mixing objects between samples. Photometric redshifts can be highly uncertain, so calculating the UV luminosity function in bins of redshift requires joint fitting of the luminosity function at all redshifts under consideration to account for objects scattering between redshift bins.\footnote{For example, if we were to calculate luminosity functions from $z = 9 - 10$ and $z = 10 - 11$, an object with a photometric redshift of $z = 10 \pm 0.5$ may reasonably be expected to contribute to either one.} However, observed candidates do not scatter between the dropout samples, so we can separately calculate the binned luminosity function for each dropout sample (which correspond to redshift ranges of $z \sim 8.5 - 12$, $z \sim 12 - 16$, and $z \sim 16 - 22.5$ for the F115W, F150W, and F200W dropouts, respectively). To investigate the evolution of the luminosity function at $z \geq 14$, we also use a subset of the F150W dropout sample and the corresponding estimate of the completeness of the F150W dropout selection at $z \geq 14$ to estimate a $z \geq 14$ luminosity function.

\begin{figure}
    \centering
    \includegraphics[width=\columnwidth]{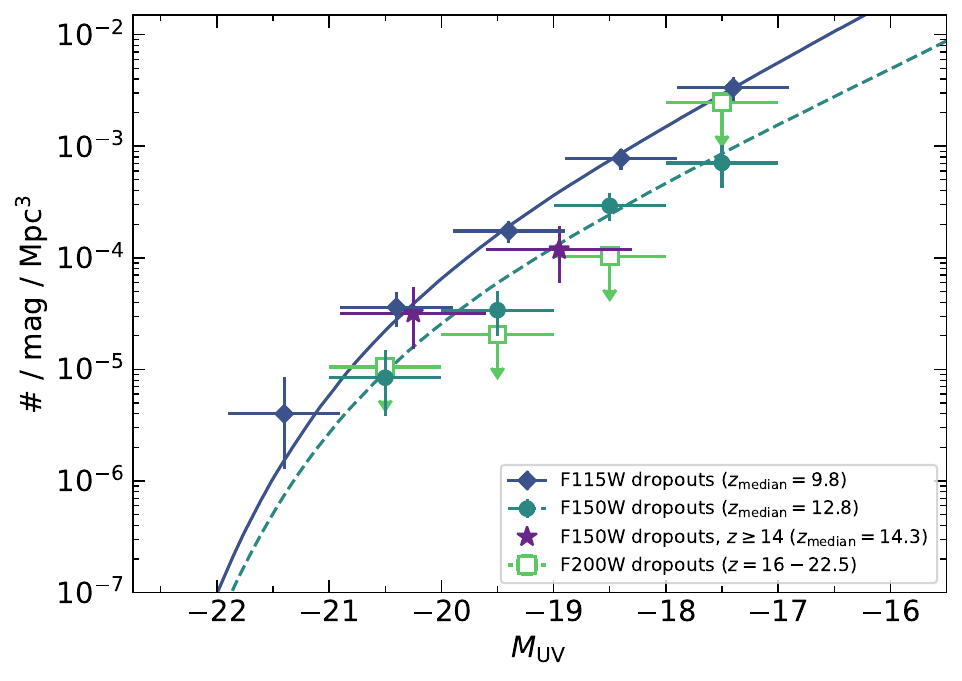}
    \caption{\label{fig:all_uvlfs}All of the UV luminosity functions we have measured in this work. We show the binned values for each of the dropout samples and the Schechter function fits for the F115W and F150W dropout samples. We show the F115W dropout luminosity function using blue diamonds and a solid line, the F150W dropouts as dark green circles and dashed line, and the F200W dropout upper limits as the open, light green squares. We also show the binned $z \geq 14$ luminosity function, calculated as a subset of the F150W dropouts, as purple stars. We find mild evolution between the F115W and F150W dropout luminosity functions, stronger at the faint end than at the bright end, and little to no evolution between the full F150W dropout luminosity function and subset restricted to $z \geq 14$.}
\end{figure}

For the binned F115W dropout, F150W dropout, and $z \geq 14$ F150W dropout UV luminosity functions, we use a Markov chain Monte Carlo (MCMC) algorithm to sample the posterior probability distributions for the number densities in each bin of \Muv. The set of binned number densities, $\{\phi_1, \phi_2, ..., \phi_{n_\text{bins}}\}$, comprises the parameters for which we are fitting. To calculate the upper limit on the F200W dropout UV luminosity function, we adopt the same \Muv{} bins as for the F150W dropout luminosity function, then simply calculate $N_{\text{up}, 2\sigma} /  V_\text{eff} / \Delta\Muv$, where $N_{\text{up}, 2\sigma} = 3.8$ is the single-sided $2\sigma$ upper limit for $n = 0$ assuming Poisson statistics \citep{gehrels1986}.

We adopt a Poisson likelihood function for the number of objects in each bin of \Muv, as is appropriate for the relatively small counts in some of the \Muv{} bins at higher redshifts. Thus, the log-likelihood function for the $i^\text{th}$ \Muv{} bin is
\begin{equation} \label{eqn:binned_likelihood}
    \ln\left(\mathcal{L}_i\right) = n_{\text{obs}, i}\ln\left(n_{\text{exp}, i}\right) - n_{\text{obs}, i} -\ln\left(n_{\text{obs}, i}!\right),
\end{equation}
where $\ln(x)$ is the natural logarithm of $x$ and $x!$ is the factorial of $x$.
$n_{\text{obs}, i}$ is the number of objects that are observed to have absolute UV magnitudes that fall within the $i^\text{th}$ \Muv{} bin and $n_{\text{exp}, i}$ is the expected number of galaxies in the same \Muv{} bin. The total log-likelihood for the luminosity function is then the joint log-likelihood of all of the \Muv{} bins (the sum of Equation\ \eqref{eqn:binned_likelihood} over all \Muv{} bins):
\begin{equation} \label{eqn:total_likelihood}
    \ln\left(\mathcal{L}\right) = \sum_{i} \big[n_{\text{obs}, i} \ln\left(n_{\text{exp}, i}\right) - n_{\text{obs}, i} -\ln\left(n_{\text{obs}, i}!\right)\big].
\end{equation}
We note that, as described in Section\ \ref{subsec:completeness}, the luminosity function is a function of true \Muv, but we will evaluate this likelihood in bins of recovered \Muv{} with $n_{\text{exp}, i}$ calculated from $\bm{\phi}(\bm{\Muv}^\text{true})$ using Equation\ \eqref{eqn:n_expected}.

\begin{deluxetable*}{cccccc}
    \renewcommand{\arraystretch}{1.25}
    \tablecaption{\label{tab:parametric_lf}The parameters for the functional forms of the UV luminosity functions and corresponding UV luminosity densities integrated to a faint limit of $\Muv = -17$ for each of our dropout samples.}
    \tablehead{\colhead{\multirow{2}{*}{Parameterization}} & \colhead{$\phi^*$} & \colhead{$\Muv^*$} & \colhead{\multirow{2}{*}{$\alpha$}} & \colhead{\multirow{2}{*}{$\beta$}} & \colhead{$\rho_\textsc{uv}(\Muv \leq -17)$} \\
& [$10^{-5}$\,mag$^{-1}$\,Mpc$^{-3}$] & [mag] & & & [$10^{25}$erg s$^{-1}$ Hz$^{-1}$ Mpc$^{-3}$]}
\startdata
\multicolumn{6}{c}{F115W dropouts} \\ \hline
Schechter & $10.070_{-6.220}^{+14.304}$ & $-20.32_{-0.50}^{+0.45}$ & $-2.36_{-0.18}^{+0.20}$ & --- & $2.82_{-0.33}^{+0.34}$ \\
DPL & $4.849_{-2.147}^{+5.324}$ & $-20.54_{-0.30}^{+0.39}$ & $-2.60_{-0.19}^{+0.17}$ & $-3.49_{-1.38}^{+0.75}$ & $2.98_{-0.34}^{+0.35}$ \\ \hline
\multicolumn{6}{c}{F150W dropouts} \\ \hline
Schechter & $4.083_{-1.611}^{+2.563}$ & $-20.32^\dagger$ & $-2.23_{-0.30}^{+0.32}$ & --- & $0.90_{-0.17}^{+0.23}$ \\
DPL & $2.263_{-0.941}^{+1.485}$ & $-20.54^\dagger$ & $-2.42_{-0.31}^{+0.32}$ & $-3.50_{-1.56}^{+1.00}$ & $0.98_{-0.20}^{+0.25}$ \\
\enddata
    \tablenotetext{\dagger}{Fixed to the corresponding value of the F115W dropout luminosity function.}
\end{deluxetable*}

We also note that the observed absolute UV magnitude of a real object may be uncertain due to both photometric scatter in the observed SED and uncertainties in the photometric redshift. Thus, we do not assign each observed candidate one value of observed \Muv. Rather, we account for uncertainties in observed \Muv{} by drawing one sample from every observed object's \Muv{} posterior, calculated as described in Section\ \ref{sec:sample_overview}, at each step in the MCMC. This produces a set of observed absolute UV magnitudes (one \Muv{} for each object in the sample), which we use to calculate $n_{\text{obs}, i}$ for the step. For the $z \geq 14$ subset of the F150W dropouts, we perform a similar sampling of both redshift and \Muv{} from each of the F150W dropouts, but only use objects with a sampled redshift of $z \geq 14$ at each step.

We sample the binned luminosity functions using a Metropolis-Hastings algorithm \citep{metropolis1953, hastings1970}. We adopt Gaussian proposal distributions with standard deviations tuned to obtain acceptance rates of $\sim 0.2 - 0.27$, and run the MCMC for $2.5 \times 10^{6}$ steps using 20 walkers. To initialize the samplers, we calculate an initial estimate for the binned number densities by calculating $n_\text{obs} / V_\text{eff} / \Delta\Muv$ (where $n_\text{obs}$ is counted using the median \Muv{} for each object) in each \Muv{} bin, then make small perturbations around this initial guess.

To estimate the final posterior distributions, we concatenate the chains from all 20 walkers, leading to a total of $5 \times 10^{7}$ equally weighted samples. We take the median of the marginalized posterior distribution in each \Muv{} bin as the number density in that bin with uncertainties corresponding to the $16^\text{th}$ and $84^\text{th}$ percentiles. We show the resulting binned UV luminosity functions in Figure\ \ref{fig:uvlfs} and report the values in Table\ \ref{tab:binned_lf} for the F115W, F150W, and F200W dropout samples. For comparison, we also show in Figure\ \ref{fig:uvlfs} the luminosity functions implied by the fits for the redshift evolution of the Schechter \citep{bouwens2021} and double power law \citep[DPL;][]{bowler2020} parameters (based on HST and ground-based data) along with measurements of the UV luminosity function from the literature. In Figure\ \ref{fig:all_uvlfs}, we show a comparison of all three dropout samples along with the $z \geq 14$ subset of the F150W dropouts. For each dropout selection, we show measurements closest to the median redshift of the selection (typically $z = 9$ or $10$ and $z = 12$ or $12.5$ for the F115W and F150W dropout samples, respectively).

Our measurements are generally consistent with most JWST constraints on the UV luminosity function \citep[e.g.][]{harikane2023, leung2023, perez-gonzalez2023, finkelstein2024, willott2024, adams2024, donnan2024}, especially for the F115W dropouts, with smaller uncertainties. While there are fewer measurements at the redshifts of the F150W dropouts and at $z \geq 14$, we also find consistency with the binned data points of \citet{perez-gonzalez2023} at all luminosities where the datasets overlap, and broad agreement with other studies at luminosities brighter than $\Muv \sim -19$. At fainter luminosities, the F150W dropout luminosity function is slightly higher than some studies. Notably, our F150W dropout luminosity function is higher than was measured in the JOF \citep{robertson2024_jof} by a factor of $\sim 3-4$. We attribute this difference to several factors, discussed in more detail in Section\ \ref{subsec:jof_lf_compare}. Finally, our $2\sigma$ upper limit on the F200W dropout luminosity function is approximately consistent with the upper limit measured by \citet{harikane2024b} at $\Muv = -21.9$ and the upper limit measured by \citet{kokorev2024} at $\Muv = -19$. However, we note that the measurement by \citet{kokorev2024} at $\Muv = -18$ based on five $z \gtrsim 16$ galaxy candidates implies the presence of $N = 1 - 7$ $z \gtrsim 16$ galaxies in the JADES fields at $\Muv -18$ (where we found none). This could be due to effects such as cosmic variance or simply Poisson noise, but we note that our upper limit combined with the $\Muv = -18$ measurement by \citet{kokorev2024} requires an extremely rapid increase in the abundance of $z \gtrsim 16$ galaxies between $\Muv \sim -18.5$ and $\Muv \sim -18$. If confirmed to be a physical effect, this is a significant change from the shape of the luminosity function we find at $z \sim 14$ and implies a rapid evolution in star formation processes between $z \sim 14$ and $z \gtrsim 16$, and we emphasize the need for spectroscopic observations for confirmation.

Ultimately, we measure the luminosity function over a large range of luminosity (five magnitudes or a factor of 100 in luminosity for the F115W dropouts, and four magnitudes or a factor of $\sim 40$ in luminosity for the F150W dropouts), enabling us to self-consistently examine the luminosity function and evolution thereof for both bright and faint galaxies. Overall, we measure number densities at $z \sim 10$ that are slightly higher than many pre-JWST measurements of the UV luminosity function at luminosities fainter than $\Muv \sim -21$, a difference that grows larger at increasingly high redshifts \citep[though we note that our findings are broadly consistent with some pre-JWST measurements at $z \sim 10$; e.g.][]{mcleod2016}. We observe essentially no evolution between the full F150W dropout sample and the $z \geq 14$ subset.

\subsection{The parametric UV luminosity function} \label{subsec:parametric_lf}

We also fit parametric forms of the UV luminosity function. We assume a \citet{schechter1976} function\footnote{$
\phi(\Muv)_\text{Schechter} = 0.4\ln(10)\phi^{*} 10^{-0.4(\Muv - \Muv^*)(\alpha + 1)} \times$ \\
\hspace*{3cm} $\exp\left[-10^{-0.4(\Muv - \Muv^*)}\right]$} as our fiducial parameterization, which has long been standard for the high-redshift UV luminosity function. However, at high redshifts ($z \gtrsim 7$), there is evidence that bright ($\Muv \lesssim -22$) galaxies are in excess of the exponential decline of the Schechter function and the high-redshift UV luminosity function may be better represented by a double power law (DPL) parameterization\footnote{$\phi(\Muv)_\text{DPL} = \dfrac{\phi^*}{10^{-0.4(\Muv^* - \Muv)(\alpha + 1)} + 10^{-0.4(\Muv^* - \Muv)(\beta + 1)}}$} \citep[e.g.][]{bowler2020, donnan2024} with characteristic luminosities around $-21 \lesssim \Muv^* \lesssim -20$. While our data do not reach extremely bright luminosities ($\Muv \sim -22$) due to the limited area of the survey, we observe objects at $\Muv \sim -21$, enabling us to also fit a DPL to facilitate comparisons with other studies.

To fit the parametric luminosity function, we again use an MCMC algorithm to sample the probability distributions of the parameters for the functional form under consideration. For the Schechter function, we are fitting for the overall normalization ($\phi^*$), the characteristic UV luminosity ($\Muv^*$), and the faint end slope ($\alpha$). For the DPL, we are fitting for the same parameters plus the bright end slope ($\beta$). We place uniform priors on all parameters with the following ranges: $10^{-8} \leq \phi / (\text{mag}^{-1}\;\text{Mpc}^{-3}) \leq 1$, $-24 \leq \Muv^* \leq -18$, $-3 \leq \alpha \leq -1$, and for the DPL, $-6 \leq \beta \leq -2$. We then use the same Metropolis-Hastings algorithm as described in Section\ \ref{subsec:binned_lf} to sample the function parameters, again with $2.5 \times 10^{6}$ steps and 20 walkers with proposal distributions tuned to obtain acceptance rates of $\sim 0.2 - 0.27$. To use Equation\ \eqref{eqn:total_likelihood} as the likelihood function, we convert the continuous $\phi(\Muv)$ function to a binned luminosity function with bins of width $\Delta\Muv = 0.25$, from which we can calculate expected counts.

For the F115W dropouts, we fit for all of the free parameters of the luminosity function. For the F150W dropouts, we fix the characteristic luminosity ($\Muv^*$) to the value that we find for the F115W dropouts due to the presence of JADES-GS-z14-0 in our sample and the more restricted luminosity range of the F150W dropouts. For the $z \geq 14$ subset of the F150W dropout luminosity function and the F200W dropouts, we fix all parameters except the normalization ($\phi^*$) to the values of the F150W dropout luminosity function. We report the final constraints on the Schechter and DPL parameters in Table\ \ref{tab:parametric_lf} and show the median of the fits marginalized over all of the function parameters in Figure\ \ref{fig:uvlfs}.

For the F115W dropout Schechter function, we find a characteristic UV magnitude of $\Muv^* = -20.32$ (similar to or slightly fainter than other $z \sim 10$ measurements), a normalization of $\phi^* = 10.070 \times 10^{-5}$\,mag$^{-1}$\,Mpc$^{-3}$ (higher than expected from most pre-JWST measurements and consistent with JWST constraints), and a steep faint end slope of $\alpha = -2.36$ (generally consistent with pre-JWST extrapolations but steeper than most JWST measurements). Compared to the Schechter function, when assuming a DPL, we measure a slightly brighter characteristic luminosity ($\Muv^* = -20.54$), correspondingly lower normalization ($\phi^* = 4.849 \times 10^{-5}$\,mag$^{-1}$\,Mpc$^{-3}$), and moderately steeper faint end slope ($\alpha = -2.60$). The normalization of the F150W dropout luminosity function declines by a factor of $\sim 2.1 - 2.3$ from the F115W dropout luminosity function to $\phi^*_\text{Schechter} = 2.2 \times 10^{-5}$\,mag$^{-1}$\,Mpc$^{-3}$ and $\phi^*_\text{DPL} = 2.5 \times 10^{-5}$\,mag$^{-1}$\,Mpc$^{-3}$ for the Schechter function and DPL, respectively. We continue to find steep faint end slopes of $\alpha_\text{Schechter} = -2.29$ and $\alpha_\text{DPL} = -2.41$ at $z \sim 13$. In general, we find similarly high normalizations of the UV luminosity function as other JWST constraints, but slightly steeper faint end slope (see Section \ref{subsec:reionization}), and combined, this leads to a relatively slow decline in the cosmic UV luminosity density over time as discussed below.

\bigskip

We note that in this work, we have fit the UV luminosity function over the entire range of UV luminosities for which we have data. However, as completeness decreases, the measurement of the faint end of the UV luminosity function becomes increasingly sensitive to the details and assumptions of the source injection simulations with which we quantify the completeness of our sample (Section\ \ref{subsec:completeness}). To mitigate this impact, luminosity functions are often measured using luminosity bins with $> 50$\,percent completeness \citep{perez-gonzalez2023, adams2024, donnan2024}. While we do not adopt the same limit as these works, we acknowledge the possible influence of the completeness results on our measured luminosity functions, especially the inferred faint end slopes. To test the impact of our use of luminosities with $< 50$\,percent completeness in fitting the luminosity function, we repeat the measurements described in Sections\ \ref{subsec:binned_lf} and \ref{subsec:parametric_lf} for the F115W and F150W dropout luminosity functions while restricting the range of luminosities to bins with $> 50$\,percent completeness. In general, we find that our results for the faint end slope do not depend strongly on the inclusion of luminosity bins with $< 50$\,percent completeness; for the F115W dropouts, we find faint end slopes of $\alpha_\text{Schechter} = -2.51_{-0.18}^{+0.18}$ and $\alpha_\textsc{dpl} = -2.56_{-0.28}^{+0.30}$ when only considering luminosities with $> 50$\,percent completeness, compared to $\alpha_\text{Schechter} = -2.38_{-0.18}^{+0.20}$ and $\alpha_\textsc{dpl} = -2.60_{-0.19}^{+0.17}$ over the full dynamic range of UV luminosity. Similarly, for the F150W dropouts, we find $\alpha_\text{Schechter} = -2.25_{-0.39}^{+0.43}$ (originally $\alpha_\text{Schechter} = -2.23_{-0.30}^{+0.31}$) and $\alpha_\textsc{dpl} = -2.41_{-0.38}^{+0.49}$ (originally $\alpha_\text{Schechter} = -2.42_{-0.31}^{+0.32}$). Thus, we consider the UV luminosity function results measured using the full dynamic range of UV luminosity as fiducial.

\subsection{The cosmic UV luminosity density} \label{subsec:rho_uv}

\begin{figure}
    \centering
    \includegraphics[width=\columnwidth]{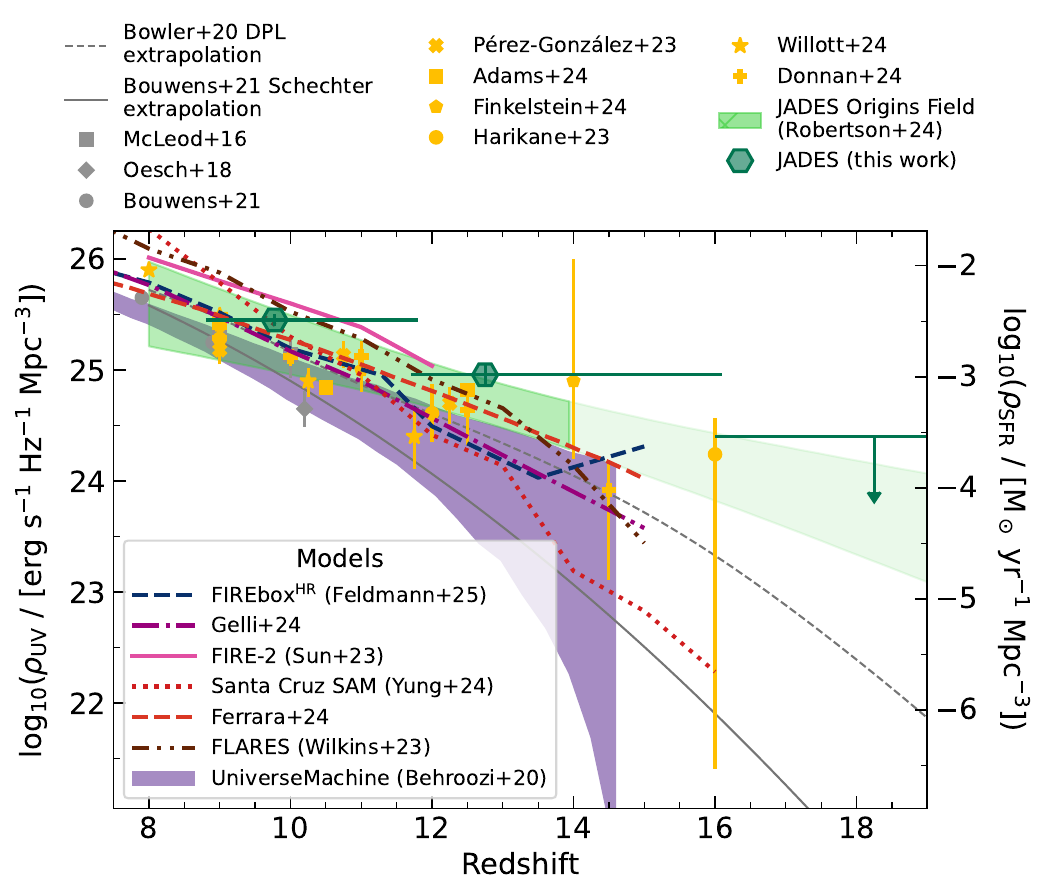}
    \caption{\label{fig:rho_uv}The cosmic UV luminosity density integrated to $\Muv = -17$ (left $y$-axis) and converted to $\rho_\textsc{sfr}$ (right $y$-axis) with the factor $\mathcal{K} = 1.15 \times 10^{-28}$\,(M$_\odot$\,yr$^{-1}$)\,/\,(erg s$^{-1}$ Hz$^{-1}$) \citep{madau2014}. For the F115W and F150W dropouts, we show the results of integrating the Schechter function fits. For the F200W dropout upper limit, we integrate the binned upper limits of the luminosity function between $\Muv = -21$ and $\Muv = -17$. For comparison, we show observational constraints based on HST and ground-based data \citep{mcleod2016, oesch2018, bouwens2021} and JWST data \citep{perez-gonzalez2023, adams2024, finkelstein2024, harikane2024a, willott2024, donnan2024, robertson2024_jof}. We also show theoretical predictions from a variety of models \citep{behroozi2020, wilkins2023, sun2023, ferrara2024, yung2024, gelli2024, feldmann2025}. Consistent with other JWST measurements, we observe an excess in the cosmic UV luminosity density that is in increasingly strong tension with models at $z \sim 10$.}
\end{figure}

We calculate the cosmic UV luminosity density, $\rho_\textsc{uv}$, as the luminosity-weighted integral of the function parameterizations of the F115W and F150W dropout luminosity functions. In order to quantify the uncertainty on $\rho_\textsc{uv}$, we calculate $\rho_\textsc{uv}$ for each of the $5 \times 10^7$ sets of Schechter or DPL parameter samples for the F115W and F150W dropout luminosity functions, then take the median, $16^\text{th}$, and $84^\text{th}$ percentiles as the values and uncertainties for $\rho_\textsc{uv}$. For the F200W dropout upper limit, we take the luminosity-weighted integral of the binned values of the upper limit on the luminosity function between $\Muv = -21$ and $\Muv = -17$. We show the resulting values of $\rho_\textsc{uv}$ calculated from the Schechter functions in Figure\ \ref{fig:rho_uv}, and report the results of both the Schechter and DPL forms of the luminosity function in Table\ \ref{tab:parametric_lf}. The DPL parameterization systematically implies moderately higher values of $\rho_\textsc{uv}$ than the Schechter functions, but both parameterizations are consistent within errors. We also convert $\rho_\textsc{uv}$ to the cosmic star formation rate density ($\rho_\textsc{sfr}$) using the commonly adopted conversion factor $\mathcal{K} = 1.15 \times 10^{-28}$\,(M$_\odot$\,yr$^{-1}$)\,/\,(erg s$^{-1}$ Hz$^{-1}$) \citep{madau2014}, which assumes a \citet{salpeter1955} stellar initial mass function. We report the values of $\rho_\textsc{uv}$ obtained when integrating to the frequently assumed faint limit of $\Muv = -17$ and emphasize that the JADES luminosity function measurement extends down to $\Muv = -17$; thus, calculating $\rho_\textsc{uv}(\Muv \leq -17)$ does not require any extrapolation.

From the F115W dropouts ($z \sim 10$), we find a cosmic UV luminosity density of $\rho_\textsc{uv} = 2.82 \times 10^{25}$\,erg\,s$^{-1}$\,Hz$^{-1}$\,Mpc$^{-3}$, which declines by a factor of $\sim \text{three}$ between $z \sim 10$ and $z \sim 13$ (probed by the F150W dropouts) to $\rho_\textsc{uv} = 0.93 \times 10^{25}$\,erg\,s$^{-1}$\,Hz$^{-1}$\,Mpc$^{-3}$, then further declines by another factor of $\gtrsim 4$ to be $\rho_\textsc{uv} < 2.51 \times 10^{24}$\,erg\,s$^{-1}$\,Hz$^{-1}$\,Mpc$^{-3}$ for the F200W dropout upper limit. As expected from the high normalization and steep faint end slopes, these UV luminosity densities are slightly higher than measurements based on HST and ground-based data at $z \sim 10$, with increasing tension at $z \gtrsim 12$.

\begin{figure*}
    \includegraphics[width=\textwidth]{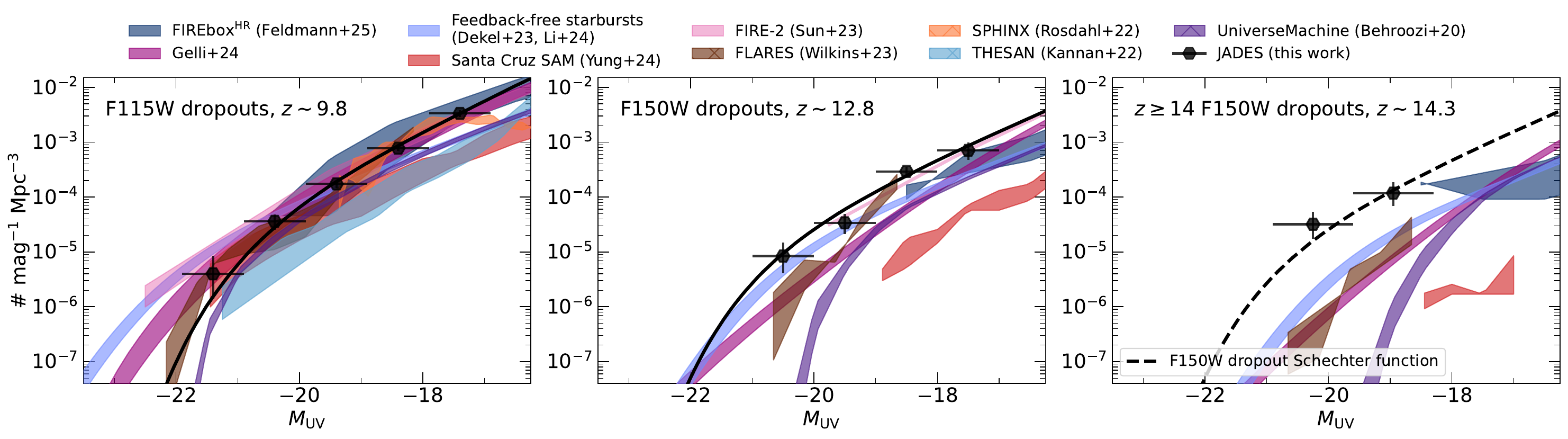}
    \caption{\label{fig:uvlf_theory}A comparison of our measured F115W dropout, F150W dropout, and $z \geq 14$ F150W dropout luminosity functions with various theoretical models as colored regions (with hatches for models from before JWST, \citealt{behroozi2020, kannan2022, rosdahl2022, wilkins2023}; solid colors for models published after JWST, \citealt{dekel2023, sun2023, yung2024, li2024, gelli2024, feldmann2025}). We show the binned data as black hexagons and the Schechter fits as black lines. We do not fit a Schechter function to the $z \geq 14$ luminosity function but show the F150W dropout luminosity function as a dashed line for comparison. The upper and lower bounds of the model regions correspond to the predicted luminosity functions for the redshifts that bracket the central redshift of our observed luminosity function (e.g. we often show $z = 9$ and $z = 10$ for the F115W dropouts). Our observed luminosity functions are consistent with some models at $z \sim 10$. However, at $z \gtrsim 12$, the observations lie increasingly above model predictions at all luminosities we have examined in this work.}
\end{figure*}

\subsection{Comparison with the JOF UV luminosity function} \label{subsec:jof_lf_compare}

We now briefly compare our luminosity function to that measured by \citet{robertson2024_jof} in the JOF. As introduced in Section\ \ref{subsec:binned_lf}, we find higher number densities at $\Muv \gtrsim -19$ in this work compared to \citet{robertson2024_jof}, which may be due to several factors.

First, while both the JOF sample and our sample include the very bright, $z \sim 14$ galaxy JADES-GS-z14-0, the JADES area we consider in this work is nearly 20 times larger than the JOF area (i.e. the surface density implied by JADES is lower than that implied by JOF). This then raises the median of the bright end of the JOF luminosity function above what we measure over the full JADES area. Then, due to the parameterization of the JOF luminosity function as an evolving Schechter function, which explicitly ties the behavior of the bright and faint regimes of the luminosity function together, this influences the luminosity function measured in the JOF area to trend towards a shallower faint end slope and therefore lower number densities at $\Muv \gtrsim -19$ in the absence of large numbers of faint objects. In other words, when assuming a parametric form of the luminosity function, the presence of an unusually bright object (JADES-GS-z14-0) driving the bright end to higher number densities induces a difference between the JOF luminosity function and this work at the faint end. This difference may be further influenced by the fitting method that \citet{robertson2024_jof} used that quantitatively accounts for how low luminosity objects with uncertain colors or photometric redshifts may scatter into higher redshift samples, which our analysis does not explicitly model.

Second, the method used by \citet{robertson2024_jof} to calculate the luminosity function uses the full photometric redshift likelihoods of the observed objects. In comparison, we use the entire photometric redshift probability distributions for selection and remove objects that we find have an integrated probability $\geq 50$\,percent of being at low redshift (based on \beagle{} models; Section\ \ref{sec:selection}). We then consider only high-redshift solutions when calculating our luminosity function (Section\ \ref{subsec:binned_lf}), which assumes that our selection identifies high-redshift candidates with a low contamination rate (noting that we find a relatively small $\sim 10$\,percent spectroscopic contamination rate for the objects in our sample that have spectroscopic redshifts). Incorporating the photometric redshift distributions while fitting the luminosity function accounts for redshift uncertainties probabilistically, which may lead to differences, particularly at the faint end where photometric measurements and therefore inferred quantities are more uncertain. For instance, there are five objects in our sample in the JOF area with $M_\textsc{uv} > -18.5$ that do not satisfy the selection criteria used by \citet{robertson2024_jof} and are not used in the calculation of the JOF luminosity function. Of those objects, \citet{robertson2024_jof} found that two are best fit at low redshifts of $z < 4$ (based on \texttt{EAZY}) and the other three are best fit at high redshift but have low redshift peaks in their redshift probability distributions such that the difference in the goodness-of-fit between the best-fit low- and high-redshift solutions is $\Delta \chi^2 < 4$. In this work, though they are less well characterized than brighter systems, these objects satisfy our photometric redshift requirement (having an integrated low-redshift probability of less than 50\,percent from \beagle{} models), and ultimately, illustrate the need for spectroscopy to confirm the redshifts of faint objects.

Third, we calculate both our detection and selection completeness correction with end-to-end source injection and recovery simulations. In comparison, \citet{robertson2024_jof} used source injection simulations to measure detection completeness and then used Monte Carlo simulations of SED models to compute their selection completeness. Both methods may face different systematic uncertainties.

\section{Discussion} \label{sec:discussion}

In this work, we have used deep JWST imaging from JADES to identify galaxies to faint UV magnitudes of $\Muv \sim -17$ and constrain the $z \gtrsim 9$ UV luminosity function down to luminosities up to 40 times fainter than the characteristic luminosity at these redshifts. Using this powerful dataset, we can now begin to examine implications for early galaxy evolution and the reionization process.

\subsection{Comparison with theoretical predictions} \label{subsec:theory_comparison}

\begin{figure}
    \includegraphics[width=\columnwidth]{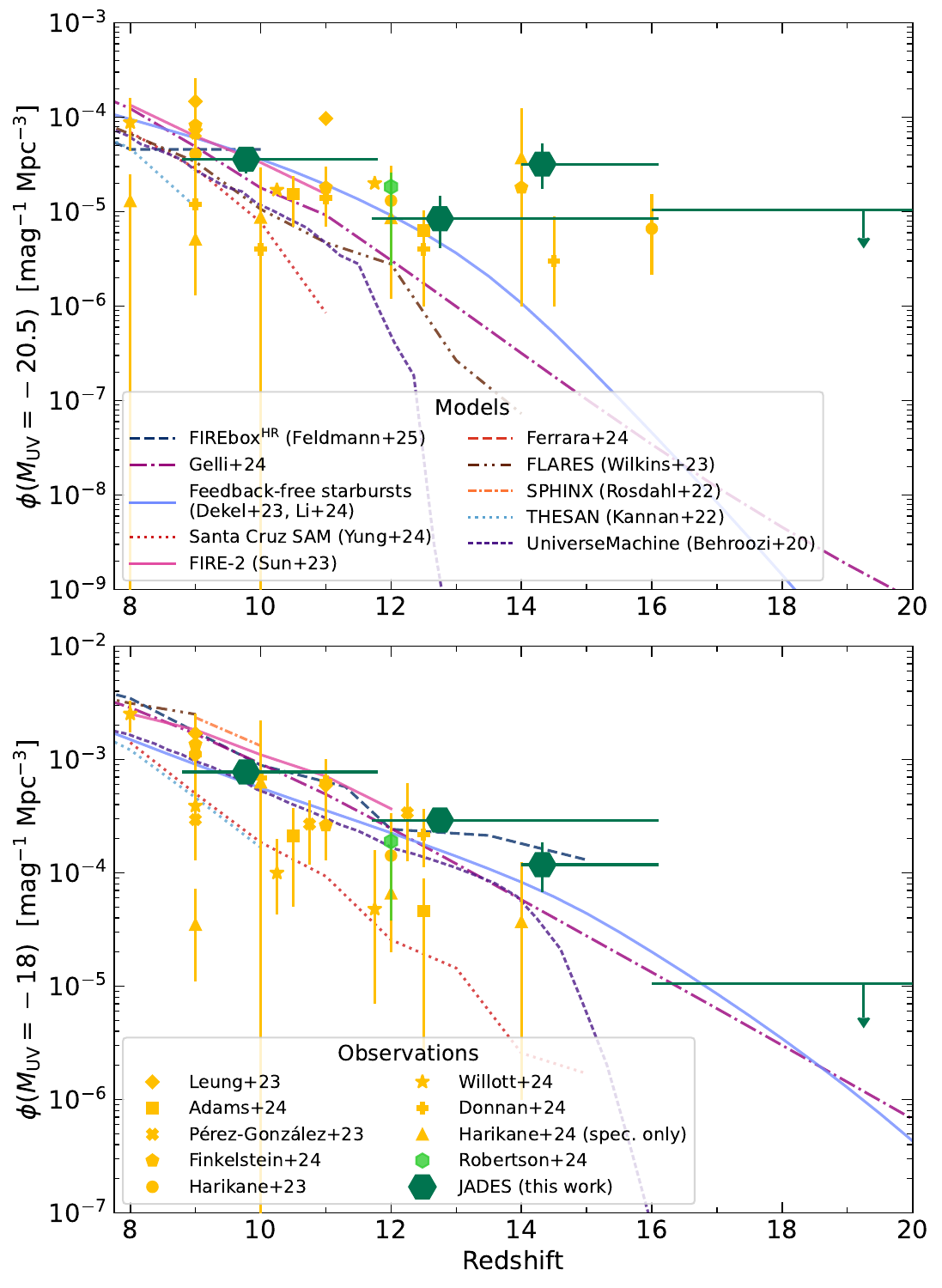}
    \caption{\label{fig:uvlf_z_evol}The redshift evolution of the number densities of bright ($\Muv = -20.5$, top) and faint ($\Muv = -18$, bottom) galaxies. We show our measurements from the binned luminosity functions as dark green hexagons (we note that the bright end of the $z \geq 14$ F150W dropout luminosity function, with a median redshift of $z \sim 14.3$, is dominated by JADES-GS-z14-0 and may not be broadly representative of the full galaxy population). We also show the measurement from the JADES Origins Field \citep{robertson2024_jof} as a light green hexagon, a compilation of measurements from the literature as yellow points (symbols as in Figure\ \ref{fig:uvlfs}), and a comparison to the same models as in Figure\ \ref{fig:uvlf_theory}. Consistent with many other observations of the high-redshift UV luminosity function from JWST, we find an excess of bright galaxies over all models at $z \gtrsim 12$. We also find a similar, though smaller, excess over most models at the faint end of the luminosity function.}
\end{figure}

Within just a few months of starting scientific operations, JWST enabled the discovery of several $\Muv \lesssim -20$ galaxies at $z \gtrsim 10$ in relatively small survey areas \citep[e.g.][]{naidu2022, castellano2022, donnan2023, finkelstein2022_maisie}. Though comprising only a few objects with large uncertainties from both Poisson errors and cosmic variance, this tentatively suggested the existence of a surprisingly large number of bright galaxies. Subsequently, statistical measurements of the UV luminosity function have confirmed the presence of a large bright galaxy population at $z \gtrsim 10$ \citep[e.g.][]{adams2023, harikane2023, harikane2024a, harikane2024b, finkelstein2023_ceers, mcleod2024, donnan2024}. However, while there is now a general consensus that JWST has observed an excess of bright ($\Muv \lesssim -20$) galaxies at $z \gtrsim 10$ relative to expectations from pre-JWST theoretical models, it is less clear if that excess is also observed in the $\Muv \gtrsim -18$ galaxy population.

In Figure\ \ref{fig:uvlf_theory}, we compare our measurement of the UV luminosity function to a variety of theoretical models. We emphasize that our measurement of the luminosity function simultaneously reaches both relatively bright luminosities ($\Muv \sim -22$) and the faintest luminosities accessible to date in blank fields ($\Muv \sim -17$), allowing us to probe the full shape of the high-redshift luminosity function. We refer back to Figure\ \ref{fig:rho_uv} for comparisons to model predictions of the cosmic UV luminosity density.

As has been found by other JWST studies of the luminosity function, we find a significant excess of $\Muv \lesssim -20$ galaxies over expectations from many models at $z \gtrsim 12$. Notably, this excess is observed not only for models calibrated to data before JWST (FLARES, SPHINX, THESAN, and UniverseMachine, shown with hatches in Figure\ \ref{fig:uvlf_theory}), but is also present for some of those that successfully reproduce JWST observations at $z \sim 10$ (top panel of Figure\ \ref{fig:uvlf_z_evol}). Moreover, we highlight that we also observe high number densities at faint UV luminosities ($-18 \lesssim \Muv \lesssim -17$), especially at higher redshifts (bottom panel of Figure\ \ref{fig:uvlf_z_evol}), though the excess is smaller at the faint end than the bright end. That is, the galaxy population at any luminosity within the regime we have probed within this work ($\Muv \lesssim -17$) is at least slightly larger than most theoretical predictions and the abundance of $\Muv \lesssim -20$ galaxies is just one realization of a systematic trend that extends to fainter luminosities. The exact degree of this excess, as well as its luminosity and redshift dependence, can help constrain the physical mechanisms that regulate star formation in the early Universe.

Many of the processes proposed to explain the abundance of $\Muv \lesssim -20$ galaxies at $z \gtrsim 10$, such as bursty star formation \citep[e.g.][]{mason2023, mirocha2023, shen2023, sun2023, kravtsov2024, gelli2024} and/or a higher star formation efficiency than expected at high redshift \citep[e.g.][]{dekel2023, li2024, harikane2023, ceverino2024, feldmann2025}, either of which may lead to an increase in the average light-to-stellar mass ratios with redshift \cite[e.g.][]{donnan2024}, predict different behavior for the faint end of the luminosity function. For example, as seen in Figure\ \ref{fig:uvlf_theory}, the faint end slope of the UV luminosity function predicted by the feedback-free starburst model \citep{dekel2023, li2024} is shallower than many other models that can match the $z \sim 10$ UV luminosity function at $\Muv \sim -21$. Thus, though the predicted bright end of the higher redshift, $z \sim 13$ luminosity function approximately matches our measurement, it becomes inconsistent at $\Muv \sim -18$. In contrast, the steep slope of the predicted $z \sim 13$ luminosity function from FLARES \citep{lovell2021, vijayan2021, wilkins2023} is in strong tension with our observations at $\Muv \sim -21$ but agreement at $\Muv \sim -18$. Moreover, the FIREbox$^\textsc{hr}$ model \citep{feldmann2025} explicitly predicts an increasingly steep faint end slope at higher redshifts. In the future, full constraints on the abundance of both bright and faint galaxies, and ultimately the shape of the luminosity function and evolution thereof, will be crucial for understanding star formation processes in early galaxies.

\subsection{Implications for reionization} \label{subsec:reionization}

In addition to providing insights into early galaxy evolution, the UV luminosity function has important implications for cosmic reionization, as it contributes to setting the total number of ionizing photons produced by galaxies that are available to reionize the Universe. Notably, in this work, we have measured not only a high normalization of the $z \gtrsim 9$ UV luminosity function consistent with other measurements from JWST, but also a steeper faint end slope than has been found by many other studies. Together, this implies the existence of a large population of UV-faint galaxies and a commensurately large non-ionizing UV luminosity density at very early times. If this large population of far-UV-continuum--faint galaxies also produces ionizing photons that can escape into the IGM at any significant rate, such systems could be appreciably reionizing the Universe as early as $z \sim 14$.

\begin{figure*}
    \includegraphics[width=\textwidth]{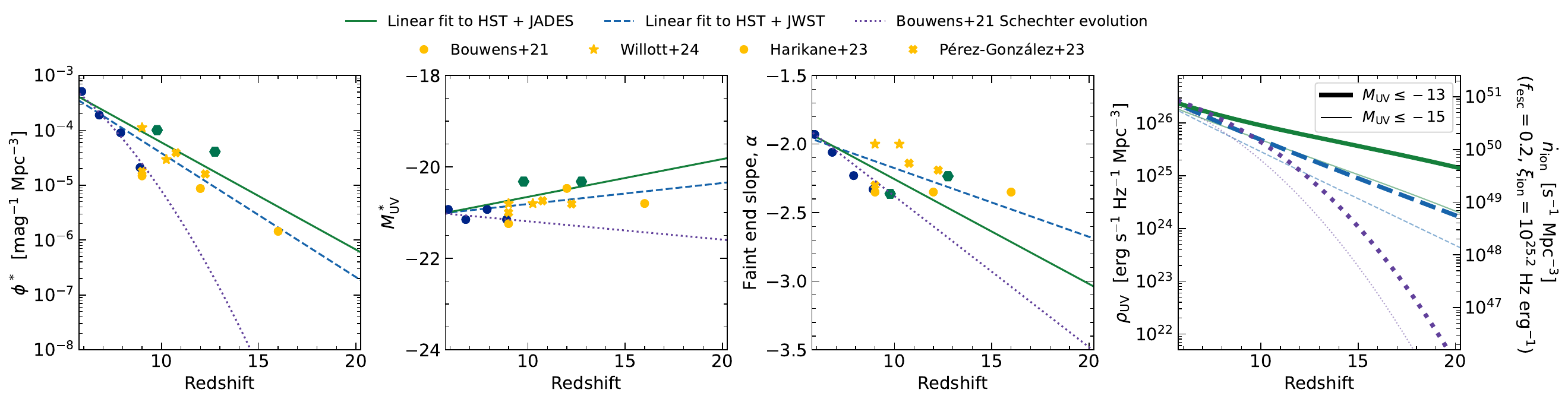}
    \caption{\label{fig:schechter_evol}The redshift evolution of the Schechter parameters and the cosmic UV luminosity density. From left to right, we show the normalization, the characteristic luminosity, the faint end slope, and the cosmic UV luminosity density. On the right $y$-axis of the rightmost panel, we also show the ionizing emissivity, $\dot{n}_\text{ion}$, that corresponds to the cosmic UV luminosity assuming a constant ionizing photon production efficiency of $\xi_\text{ion} = 10^{25.2}$\,Hz\,erg$^{-1}$ and escape fraction of $f_\text{esc}$, which were shown to be sufficient to reionize the Universe given luminosity function constraints from HST \citep{robertson2013}. We show three different evolutionary fits: HST only \citep[][dotted purple line and dark blue circles]{bouwens2021}, HST and a compilation of JWST data including this work (dashed blue line, yellow symbols), and HST and JADES only (solid green line, green hexagons). The two fits that include JWST data have higher normalizations than the HST-only fit at $z > 10$ but slightly shallower faint end slopes. However, the HST and JADES fit predicts steeper faint end slopes than when including other JWST data. Together, the higher normalization and steep faint end slope suggests the presence of a large population of faint galaxies producing copious amounts of non-ionizing UV photons at early times (as seen in the cosmic UV luminosity density), which may play a significant role in reionization if they also produce ionizing UV radiation that can escape into the IGM.}
\end{figure*}

To investigate the implications of the faint galaxy population for reionization, we calculate the reionization history as the time evolution of the volume-averaged fraction of neutral hydrogen in the IGM, $\overline{x}_\textsc{hi}$. In practice, we quantify the neutral fraction in terms of the volume-averaged ionized hydrogen fraction, $\overline{x}_\textsc{hii}$, as $\overline{x}_\textsc{hi} = 1 - \overline{x}_\textsc{hii}$. At its simplest, $\overline{x}_\textsc{hii}$ is governed by the following differential equation \citep{madau1999}, where dots denote time derivatives:
\begin{equation} \label{eqn:dxHII_dt}
    \dot{\overline{x}}_\textsc{hii} = \frac{\dot{n}_\text{ion}(t)}{\langle n_\textsc{h} \rangle} - \frac{\overline{x}_\textsc{hii}(t)}{t_\text{rec}(t)},
\end{equation}
which describes the competing effects of photoionizations of intergalactic neutral hydrogen and recombinations of free electrons with protons to (re)form neutral hydrogen. The average comoving number density of hydrogen, $\langle n_\textsc{h} \rangle = X_p \Omega_\text{b}\rho_\text{c} / m_\textsc{h}$, depends on the primordial mass fraction of hydrogen ($X_p = 0.75$), the fractional baryon density parameter ($\Omega_b$), the critical density ($\rho_\text{c}$), and the mass of hydrogen ($m_\textsc{h}$). The timescale of recombinations in the IGM at a given time, $t_\text{rec}(t)$, can be calculated as
\begin{equation} \label{eqn:t_rec}
    t_\text{rec}(t) = [C_\textsc{hii} \alpha_\textsc{b}(T) n_e(1 + z(t))^3 ]^{-1}.
\end{equation}
$C_\textsc{hii} \equiv \langle n_\textsc{h}^2 \rangle / \langle n_\textsc{h} \rangle^2$ is a ``clumping factor'' that accounts for inhomogeneities in the IGM and we choose to fix $C_\textsc{hii} = 3$ \citep[consistent with theoretical expectations from simulations; e.g.][]{shull2012, finlator2012, kaurov2015, gorce2018}. $\alpha_\textsc{b}(T)$ is the case B recombination coefficient for hydrogen and we assume a temperature of $T = 10^4$\,K, giving $\alpha_\textsc{b}(10^4\,\text{K}) = 2.59 \times 10^{-13}$\,cm$^{3}$\,s$^{-1}$ \citep{draine2011}. $n_e = (1 + Y_p  / 4X_p) \langle n_\textsc{h} \rangle$ is the comoving free electron number density assuming single ionized helium and a primordial helium mass fraction of $Y_p = 1 - X_p = 0.25$, and $z(t)$ is the redshift at time $t$.

Finally, $\dot{n}_\text{ion}(t)$ is the hydrogen ionizing photon production rate per unit comoving volume at a given time (i.e. the ionizing photon emissivity) that depends on the non-ionizing UV luminosity of the galaxy population, the ionizing photon production rate per unit non-ionizing UV luminosity (i.e. the ionizing photon production efficiency; $\xi_\text{ion}$), and the fraction of all hydrogen ionizing photons produced by the ionizing sources, here assumed to be galaxies, that escape into the IGM (i.e. the escape fraction, $f_\text{esc}$). When both $\xi_\text{ion}$ and $f_\text{esc}$ are independent of UV luminosity, the non-ionizing UV luminosity term is the cosmic UV luminosity density, $\rho_\textsc{uv} = \int_{L_\textsc{uv}^\text{faint}}^{\infty} \phi(L_\textsc{uv}) L_\textsc{uv} \text{d}L_\textsc{uv}$, giving
\begin{equation} \label{eqn:ndot_ion_rho_uv}
    \dot{n}_\text{ion}(t) = \rho_\textsc{uv}(t)\xi_\text{ion}f_\text{esc}.
\end{equation}
In this work, we also allow $\xi_\text{ion}$ to depend on UV luminosity (and can also allow $f_\text{esc}$ to be luminosity dependent), so Equation\ \eqref{eqn:ndot_ion_rho_uv} becomes
\begin{equation} \label{eqn:ndot_ion_phi}
    \dot{n}_\text{ion}(t) = \int_{L_\textsc{uv}^\text{faint}}^{\infty} \phi(L_\textsc{uv}; t)\,L_\textsc{uv}\,\xi_\text{ion}(L_\textsc{uv})f_\text{esc}\,\text{d}L_\textsc{uv},
\end{equation}
where $\phi(L_\textsc{uv}; t)$ is the UV luminosity function at time $t$. To isolate the effects of the UV luminosity function on the reionization timeline, we choose to assume a constant escape fraction with $f_\text{esc} = 0.1$ as our fiducial value \citep[approximately similar to Lyman continuum escape fractions observed in samples with similar UV luminosities and UV continuum slopes as we study in this work; e.g.][]{izotov2021, chisholm2022, mascia2024, rinaldi2024} and adopt the $\xi_\text{ion} - \Muv$ relation at $z \sim 6$ found by \citet{endsley2024_jades}. This relation has the primary features of a decrease in the average and an increase in the scatter of $\xi_\text{ion}$ with decreasing UV luminosity (where the average declines from $\xi_\text{ion} = 10^{25.8}$\,Hz\,erg$^{-1}$ to $\xi_\text{ion} = 10^{25.3}$\,Hz\,erg$^{-1}$ in the luminosity range that we probe in this work). We note that even at the faintest UV luminosities studied, the median $\xi_\text{ion}$ is moderately higher than the often used ``canonical'' value of $\xi_\text{ion} = 10^{25.2}$\,Hz\,erg$^{-1}$ informed by \citet{bruzual2003} stellar population synthesis models matched to observed UV colors (\citealt{robertson2013}, though also see \citealt{bouwens2016, debarros2019, lam2019, rinaldi2024} for discussions of higher values of $\xi_\text{ion}$ for high-redshift star forming galaxies).

\begin{figure*}
    \includegraphics[width=\textwidth]{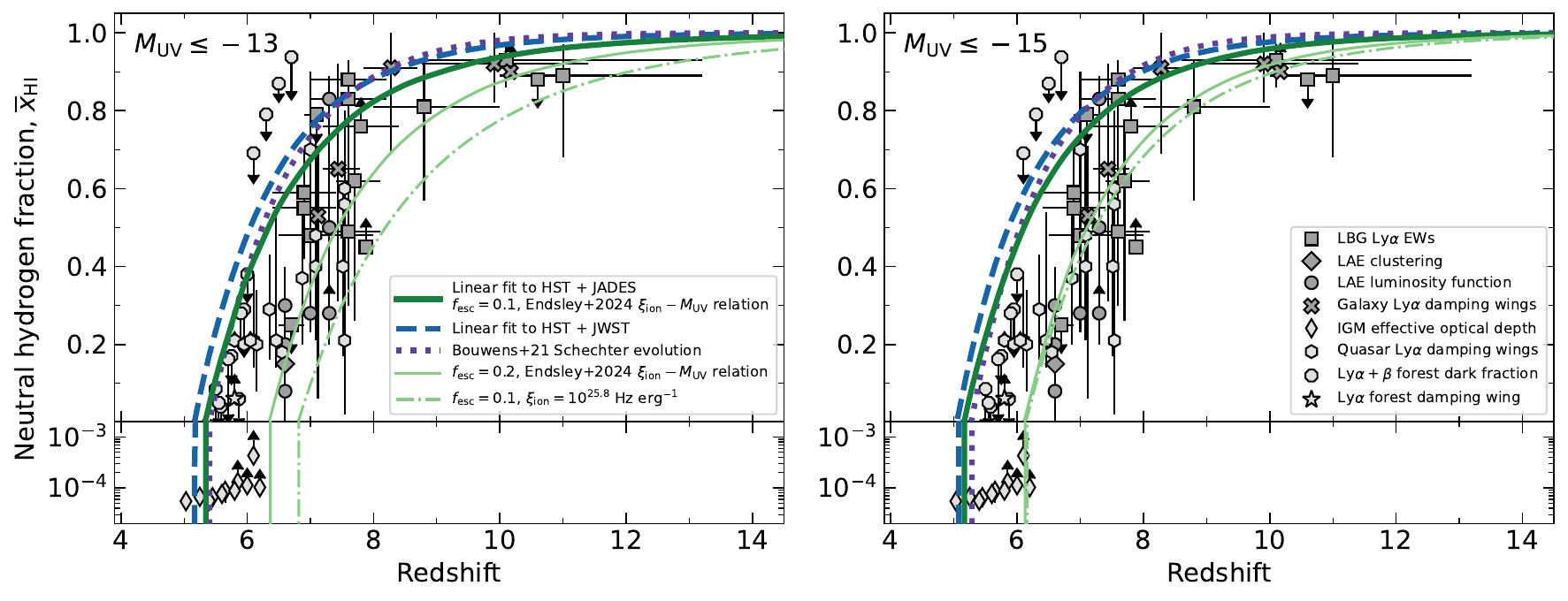}
    \caption{\label{fig:reionization_history}The timeline of reionization implied by the redshift evolution of the UV luminosity function. We show the results of all three luminosity function evolutionary fits (HST only as the dotted purple line, HST+JWST as the dashed blue line, and HST+JADES as the solid green line) for our fiducial assumptions of $f_\text{esc} = 0.1$ and the $\xi_\text{ion} - \Muv$ relation found by \citealt{endsley2024_jades}. To show the impact of the escape fraction and ionizing photon production production efficiency, we also show the reionization timeline assuming the HST+JADES UV luminosity function evolution for $f_\text{esc} = 0.2$ (thin, solid, light green line) and constant $\xi_\text{ion} = 10^{25.8}$\,Hz\,erg$^{-1}$ (thin, dot-dashed, light green line). For comparison, we show observational constraints on the neutral fraction from galaxies \citep{ouchi2010, ouchi2018, konno2014, inoue2018, mason2018a, mason2019_kmos, hoag2019, whitler2020, jung2020, morales2021, goto2021, bolan2022, ning2022, morishita2023, hsiao2023, bruton2023, nakane2024, umeda2024, tang2024_lya} and quasars \citep{fan2006, mcgreer2015, greig2017, greig2019, greig2024, banados2018, davies2018, yang2020_damping_wing, yang2020_tau_eff, wang2020, zhu2022, zhu2024, jin2023, durovcikova24}. When assuming an escape fraction of $f_\text{esc} = 0.2$ or constant $\xi_\text{ion} = 10^{25.8}$\,Hz\,erg$^{-1}$, the large faint galaxy population produces more ionizing photons than is necessary to drive reionization, ending reionization earlier than is allowed by observational constraints \citep[e.g.][]{atek2024, munoz2024}. In contrast, reionization ends around $z \sim 5.3$ for all models with our fiducial assumptions, approximately consistent with precise constraints on the end of reionization from quasars. However, though the end of reionization is largely insensitive to the UV luminosity function evolution, the HST+JADES fit starts reionization slightly earlier, leading to a neutral fraction of $\overline{x}_\textsc{hi} \sim 0.9 - 0.95$ at $z = 10$.}
\end{figure*}

To obtain $\phi(L_\textsc{uv}; t)$, we assume a Schechter function and fit the redshift evolution of the Schechter parameters, assuming that $\log_{10}(\phi^*)$, $\alpha$, and $\Muv^*$ all evolve linearly with redshift. As we wish to evaluate the impact of the steep faint end slope we have found in this work, we perform two fits for the redshift evolution of the Schechter parameters, which will determine $\phi(L_\textsc{uv}; t)$: one fit to a compilation of measurements of the Schechter parameters from the literature at $z \sim 3 - 15$ from both HST and JWST and a second fit using only the JADES data combined with HST. We also compare to the Schechter parameter evolution found by \citet{bouwens2021}, who fit $\log_{10}(\phi^*)$ as a quadratic function with redshift and $\alpha$ and $\Muv^*$ as linear functions.

We show the resulting evolutionary fits for the Schechter parameters along with the implied cosmic UV luminosity density integrated to $\Muv = -13$ in Figure\ \ref{fig:schechter_evol}. For illustrative purposes, we also show the ionizing photon emissivity corresponding to the cosmic UV luminosity density for fixed $f_\text{esc} = 0.2$ and $\xi_\text{ion} = 10^{25.2}$\,Hz\,erg$^{-1}$ as the right axis of the rightmost panel. Though we do not assume these values when calculating the reionization history, it has been shown that these assumptions are sufficient to reionize the Universe when integrating UV luminosity functions from HST down to $\Muv = -13$ \citep{robertson2013, robertson2015}. When fitting the evolution of the Schechter function parameters with a large compilation of JWST data combined with HST, we find a slower redshift evolution of the faint end slope than when we use only JADES data combined with HST measurements; this is due to the slightly shallower faint end slopes found by many JWST studies compared to the steep faint end slopes we find from JADES at $z \gtrsim 9$. At $z = 14$, the HST and JADES fit predicts a faint end slope of $\alpha \sim -2.7$ while the fit including HST and other JWST data gives $\alpha \sim -2.4$. Combined with the high overall normalization of the luminosity function at $z > 12$, this implies that the cosmic UV luminosity density can remain large at early times, potentially allowing the escape fractions and/or ionizing photon production efficiencies necessary to drive reionization to be lower than were thought necessary prior to JWST.

With all components of Equation\ \eqref{eqn:dxHII_dt} thus calculated, we can integrate forward in time to obtain the reionization history. We start the integration at $z = 30$ and show the resulting reionization histories in Figure\ \ref{fig:reionization_history} alongside observational constraints on the neutral fraction from galaxies and quasars. We note that the reionization history does not significantly depend on the starting redshift due to the normalization of the luminosity function decreasing to be vanishingly small at $z \gtrsim 20$. We show the results of integrating Equation\ \eqref{eqn:ndot_ion_phi} (i.e. all of the reionization histories) assuming each of the luminosity function evolution fits assuming an escape fraction of $f_\text{esc} = 0.1$. We also show the reionization history assuming $f_\text{esc} = 0.2$ and the HST and JADES luminosity function evolution fit to demonstrate the impact of changing the escape fraction.

We first highlight that, despite the steep faint end slope of the luminosity function and the implied large population of faint galaxies, this simple estimate of the reionization timeline is naturally in broad agreement with observational constraints on the end of reionization \citep[see][and references therein]{fan2023}. When considering galaxies down to luminosities of $\Muv = -13$ and an escape fraction $f_\text{esc} = 0.1$, all three luminosity function fits complete reionization at $z \sim 5.2 - 5.4$. When only considering contributions from $\Muv \leq -15$ galaxies, reionization ends at $z \sim 5 - 5.3$.

We acknowledge that the ionizing photon emissivity depends on $\xi_\text{ion}$ and $f_\text{esc}$ as well as $\rho_\textsc{uv}$. We have adopted specific parameterizations for both of these quantities and choosing other physically plausible values can impact the reionization history. For example, at fixed escape fraction, adopting the \citet{simmonds2024} relation for $\xi_\text{ion}(\Muv, z)$ ends reionization slightly later than our fiducial model based on \citet{endsley2024_jades}. Alternatively, increasing the escape fraction to $f_\text{esc} = 0.2$ or fixing $\xi_\text{ion}$ to a constant value of $\xi_\text{ion} = 10^{25.8}$\,Hz\,erg$^{-1}$ (consistent with the findings of \citealt{atek2024} for a sample of faint galaxies) increases $\dot{n}_\text{ion}$ to the extent of driving reionization faster than is allowed by strong constraints on the timing of reionization placed by the cosmic microwave background \citep[see discussion by][]{munoz2024}. However, we note that recent studies of $\xi_\text{ion}$ with large samples from JWST \citep[e.g.][]{simmonds2024, begley2025, pahl2025} generally find lower values of $\xi_\text{ion}$ ($\sim 10^{25.3}$\,Hz\,erg$^{-1}$) than the early results available to \citet{munoz2024}. The high values of $\xi_\text{ion}$ found by early studies are often attributed to selection and/or measurement bias towards star-forming galaxies with detectable emission lines, which likely have higher ionizing photon production efficiencies on average than the full population \citep[as noted by e.g.][]{simmonds2024, pahl2025}. Thus, overall, we conclude that current observations of the ionizing properties of high-redshift galaxies are broadly consistent with other constraints on the reionization timeline such that galaxies can drive the reionization process without overproducing ionizing photons, even if there is a large population of faint galaxies.

While all of the luminosity function fits predict similar evolutions of the neutral fraction near the end of reionization ($z \sim 6$), the early stages of reionization proceed differently. If $f_\text{esc} = 0.1$, both the HST-based luminosity function evolution \citep{bouwens2021} and our evolutionary fit to HST and many JWST luminosity functions lead to a nearly entirely neutral IGM at $z = 10$ ($\overline{x}_\textsc{hi} = 0.97 - 0.98$ for both $\Muv \leq -15$ and $\Muv \leq -13$). However, when considering HST combined only with JADES, the combination of a steep faint end slope and high normalization leads to reionization starting earlier, and correspondingly, progressing slightly more slowly in order to end at $z \sim 5.3$. Under these circumstances, though the IGM is very neutral at $z \geq 10$, it is not quite \textit{fully} neutral ($\overline{x}_\textsc{hi} = 0.91$ and $\overline{x}_\textsc{hi} = 0.95$ at $z = 10$ when integrating to $\Muv = -13$ and $\Muv = -15$, respectively) -- a scenario that would also be consistent with the remarkable discoveries of two galaxies emitting \Lya{} at $z > 10$ \citep{bunker2023_gnz11, witstok2024} and the implied neutral fraction of $\overline{x}_\textsc{hi} < 1$ \citep[e.g.][]{bruton2023, nakane2024, tang2024_lya, umeda2024}.

Above, we have explored the role of the UV luminosity function in determining the reionization timeline. Given empirically motivated assumptions for the production and escape of ionizing photons from galaxies and the evolution of the UV luminosity function implied by our findings in this work, we have found that it is feasible for the IGM to be appreciably ionized only $\sim 400$\,Myr after the Big Bang while still being compatible with observations at the conclusion of reionization. However, fully determining the reionization timeline will require even better constraints on the abundances and properties of early galaxies. In the future, combining deep imaging with JWST in lensing fields, which can reach extremely faint luminosities to constrain the faint end turnover and increase the dynamic range of luminosities available to measure the faint end slope, with imaging in blank fields over larger areas, which provides insights into the brighter galaxies. Together, these observations will provide robust constraints on the shape of the entire luminosity function and enable us to directly characterize the ionizing properties of early galaxies.

\section{Summary} \label{sec:summary}

In this work, we have used deep NIRCam imaging from the JADES program to measure the high-redshift UV luminosity function down to faint luminosities of $\Muv \sim -17$. We search for high-redshift galaxy candidates at $z \sim 8.5 - 22.5$ (and identify candidates at $z \sim 8.5 - 16$), measure their properties, infer both binned and parametric forms of the UV luminosity function, and discuss the implications of our results for early galaxy evolution and cosmic reionization. We summarize our key findings below.

\begin{enumerate}[(i)]
    \item We conduct a search for $z \sim 8.5 - 22.5$ galaxy candidates using three dropout filters: F115W (probing $z \sim 8.5 - 12$), F150W ($z \sim 12 - 16$), and F200W ($z \sim 16 - 22.5$). We color select 309 F115W dropouts, 63 F150W dropouts, and zero F200W dropouts for a total of 372 candidates over 163\,arcmin$^{2}$ of JADES NIRCam imaging. The objects we select span nearly five magnitudes and generally have the blue rest-UV continuum slopes of $\beta \lesssim -2$ that are expected for high-redshift galaxies.
    \item We use four bins of redshift to calculate the UV luminosity function: three corresponding to the dropout filters used for selection and one isolating the subset of $z \sim 14 - 16$ from the F150W dropouts. At $z \sim 8.5 - 12$ (F115W dropouts), we measure the luminosity function at $-22 \lesssim \Muv \lesssim -17$. At $z \sim 12 - 16$ (all F150W dropouts) and $z \sim 14 - 16$ (the $z \geq 14$ subset of the F150W dropouts), we constrain the luminosity function at $-21 \lesssim \Muv \lesssim -17$ and $-21 \lesssim \Muv \lesssim -18$, respectively. We also place an upper limit on the $z \sim 16 - 22.5$ UV luminosity function at $-21 \lesssim \Muv \lesssim -17$ based on the absence of F200W dropouts. Using this large dynamic range of UV luminosities enabled by the deep JADES imaging over a moderately large area, we can simultaneously constrain the bright and faint end of the luminosity function. We find number densities in general agreement with other measurements of the high-redshift UV luminosity function from JWST but measure moderately steeper faint end slopes ($\alpha \sim -2.3$ at $z \sim 13$).
    \item Our measurement of the UV luminosity function is broadly consistent with models at $z \sim 10$, but lies above theoretical expectations for the UV luminosity function at $z \gtrsim 12$ at all luminosities. Thus, for a given parent population, the average UV luminosity is higher than expected, and the abundant bright galaxy population is a realization of a systematic trend that extends to fainter luminosities.
    \item The combination of a high normalization and steep faint end slope implies that there may be a large population of faint galaxies contributing ionizing photons towards ionizing the Universe. We therefore estimate the reionization history with our measurement of the UV luminosity function and physically motivated assumptions for ionizing photon production efficiencies and escape fractions. Given these simple assumptions, the reionization process can end at $z \sim 5.3$ (consistent with observations at the end of reionization), even with a large faint galaxy population. However, the reionization process can start earlier, leading to volume-averaged neutral hydrogen fractions of $\overline{x}_\textsc{hi} < 1$ even at $z \sim 10$, which may be consistent with observations of individual Ly$\alpha$ emitters at these redshifts.
    \item We highlight the importance of future deep observations to fully characterize the faint galaxy population and their contribution to reionization. Observations in lensing fields that are sensitive to significantly fainter luminosities than we have probed in this work, combined with additional deep imaging in blank fields covering larger areas, will be crucial to fully and robustly constrain the faint end slope and the turnover of the UV luminosity function.
\end{enumerate}


\section*{Acknowledgments}

The authors thank the SPHINX collaboration, the THESAN collaboration, Peter Behroozi, Robert Feldmann, Viola Gelli, Zhaozhou Li, Jason Sun, Steve Wilkins, and L.Y. Aaron Yung for providing their theoretical predictions for the luminosity function. LW also thanks Nick Schragal for helpful conversations throughout this work (and his dice collection for providing random number generator seeds).

LW, BR, MR, KNH, EE, DJE, JMH, ZJ, BDJ, and CNAW acknowledge support from the JWST/NIRCam contract to the University of Arizona, NAS5-02015. LW also acknowledges support from the National Science Foundation Graduate Research Fellowship under Grant No. DGE-2137419. BR acknowledges support from JWST Program 3215. WMB acknowledges support by a research grant (VIL54489) from VILLUM FONDEN. AJB and JC acknowledge funding from the ``FirstGalaxies'' Advanced Grant from the European Research Council (ERC) under the European Union's Horizon 2020 research and innovation program (Grant agreement No. 789056). S. Carniani acknowledges support by European Union's HE ERC Starting Grant No. 101040227 - WINGS. ECL acknowledges support of an STFC Webb Fellowship (ST/W001438/1). DJE is supported as a Simons Investigator. ZJ acknowledges funding from program JWST-GO-1963 that is provided by NASA through a grant from the Space Telescope Science Institute, which is operated by the Association of Universities for Research in Astronomy, Inc., under NASA contract NAS 5-03127. PGP-G acknowledges support from grant PID2022-139567NB-I00 funded by Spanish Ministerio de Ciencia, Innovaci\'on y Universidades MICIU/AEI/10.13039/501100011033, and the European Union FEDER program \textit{Una manera de hacer Europa}. ST acknowledges support by the Royal Society Research Grant G125142. The research of CCW is supported by NOIRLab, which is managed by the Association of Universities for Research in Astronomy (AURA) under a cooperative agreement with the National Science Foundation. JW gratefully acknowledges support from the Cosmic Dawn Center through the DAWN Fellowship. The Cosmic Dawn Center (DAWN) is funded by the Danish National Research Foundation under grant No. 140.

This work is based on observations made with the NASA/ESA/CSA JWST. The data were obtained from the Mikulski Archive for Space Telescopes at the Space Telescope Science Institute, which is operated by the Association of Universities for Research in Astronomy, Inc., under NASA contract NAS 5-03127 for JWST. These observations are associated with programs 1180, 1181, 1210, 1286, 1287, 1895, 1963, and 3215. The authors acknowledge the FRESCO team led by PI Pascal Oesch for developing their observing program with a zero-exclusive-access period. We also acknowledge the Hubble Legacy Fields team for the reduction and release of the HST ACS imaging we use in this work.

The JADES Collaboration acknowledges support from JWST/NIRCam contract to the University of Arizona, NAS5-02015. This project made use of the \textit{lux} supercomputer at UC Santa Cruz, funded by NSF MRI grant AST-1828315, as well as High Performance Computing (HPC) resources supported by the University of Arizona TRIF, UITS, and Research, Innovation, and Impact (RII) and maintained by the UArizona Research Technologies department.

We respectfully acknowledge the University of Arizona is on the land and territories of Indigenous peoples. Today, Arizona is home to 22 federally recognized tribes, with Tucson being home to the O’odham and the Yaqui. Committed to diversity and inclusion, the University strives to build sustainable relationships with sovereign Native Nations and Indigenous communities through education offerings, partnerships, and community service.

\vspace{5mm}
\facilities{HST (ACS), JWST (NIRCam, NIRSpec)}
The JWST JADES data used in this work can be found in MAST: \dataset[10.17909/8tdj-8n28]{https://dx.doi.org/10.17909/8tdj-8n28}.
\software{\texttt{numpy} \citep{harris2020}, \texttt{matplotlib} \citep{hunter2007}, \texttt{astropy} \citep{astropy2013, astropy2018, astropy2022}, \texttt{BEAGLE} \citep{chevallard2016}, \texttt{multinest} \citep{feroz2008, feroz2009, feroz2019}}


\bibliography{refs}{}

\begin{thebibliography}{}
\expandafter\ifx\csname natexlab\endcsname\relax\def\natexlab#1{#1}\fi
\providecommand{\url}[1]{\href{#1}{#1}}
\providecommand{\dodoi}[1]{doi:~\href{http://doi.org/#1}{\nolinkurl{#1}}}
\providecommand{\doeprint}[1]{\href{http://ascl.net/#1}{\nolinkurl{http://ascl.net/#1}}}
\providecommand{\doarXiv}[1]{\href{https://arxiv.org/abs/#1}{\nolinkurl{https://arxiv.org/abs/#1}}}

\bibitem[{{Adams} {et~al.}(2023){Adams}, {Conselice}, {Ferreira}, {Austin}, {Trussler}, {Juod{\v{z}}balis}, {Wilkins}, {Caruana}, {Dayal}, {Verma}, \& {Vijayan}}]{adams2023}
{Adams}, N.~J., {Conselice}, C.~J., {Ferreira}, L., {et~al.} 2023, \mnras, 518, 4755, \dodoi{10.1093/mnras/stac3347}

\bibitem[{{Adams} {et~al.}(2024){Adams}, {Conselice}, {Austin}, {Harvey}, {Ferreira}, {Trussler}, {Juod{\v{z}}balis}, {Li}, {Windhorst}, {Cohen}, {Jansen}, {Summers}, {Tompkins}, {Driver}, {Robotham}, {D'Silva}, {Yan}, {Coe}, {Frye}, {Grogin}, {Koekemoer}, {Marshall}, {Pirzkal}, {Ryan}, {Maksym}, {Rutkowski}, {Willmer}, {Hammel}, {Nonino}, {Bhatawdekar}, {Wilkins}, {Bradley}, {Broadhurst}, {Cheng}, {Dole}, {Hathi}, \& {Zitrin}}]{adams2024}
{Adams}, N.~J., {Conselice}, C.~J., {Austin}, D., {et~al.} 2024, \apj, 965, 169, \dodoi{10.3847/1538-4357/ad2a7b}

\bibitem[{{Akhlaghi} \& {Ichikawa}(2015)}]{akhlaghi2015}
{Akhlaghi}, M., \& {Ichikawa}, T. 2015, \apjs, 220, 1, \dodoi{10.1088/0067-0049/220/1/1}

\bibitem[{{Arrabal Haro} {et~al.}(2023){Arrabal Haro}, {Dickinson}, {Finkelstein}, {Kartaltepe}, {Donnan}, {Burgarella}, {Carnall}, {Cullen}, {Dunlop}, {Fern{\'a}ndez}, {Fujimoto}, {Jung}, {Krips}, {Larson}, {Papovich}, {P{\'e}rez-Gonz{\'a}lez}, {Amor{\'\i}n}, {Bagley}, {Buat}, {Casey}, {Chworowsky}, {Cohen}, {Ferguson}, {Giavalisco}, {Huertas-Company}, {Hutchison}, {Kocevski}, {Koekemoer}, {Lucas}, {McLeod}, {McLure}, {Pirzkal}, {Seill{\'e}}, {Trump}, {Weiner}, {Wilkins}, \& {Zavala}}]{arrabalharo2023_ddt}
{Arrabal Haro}, P., {Dickinson}, M., {Finkelstein}, S.~L., {et~al.} 2023, \nat, 622, 707, \dodoi{10.1038/s41586-023-06521-7}

\bibitem[{{Astropy Collaboration} {et~al.}(2013){Astropy Collaboration}, {Robitaille}, {Tollerud}, {Greenfield}, {Droettboom}, {Bray}, {Aldcroft}, {Davis}, {Ginsburg}, {Price-Whelan}, {Kerzendorf}, {Conley}, {Crighton}, {Barbary}, {Muna}, {Ferguson}, {Grollier}, {Parikh}, {Nair}, {Unther}, {Deil}, {Woillez}, {Conseil}, {Kramer}, {Turner}, {Singer}, {Fox}, {Weaver}, {Zabalza}, {Edwards}, {Azalee Bostroem}, {Burke}, {Casey}, {Crawford}, {Dencheva}, {Ely}, {Jenness}, {Labrie}, {Lim}, {Pierfederici}, {Pontzen}, {Ptak}, {Refsdal}, {Servillat}, \& {Streicher}}]{astropy2013}
{Astropy Collaboration}, {Robitaille}, T.~P., {Tollerud}, E.~J., {et~al.} 2013, \aap, 558, A33, \dodoi{10.1051/0004-6361/201322068}

\bibitem[{{Astropy Collaboration} {et~al.}(2018){Astropy Collaboration}, {Price-Whelan}, {Sip{\H{o}}cz}, {G{\"u}nther}, {Lim}, {Crawford}, {Conseil}, {Shupe}, {Craig}, {Dencheva}, {Ginsburg}, {VanderPlas}, {Bradley}, {P{\'e}rez-Su{\'a}rez}, {de Val-Borro}, {Aldcroft}, {Cruz}, {Robitaille}, {Tollerud}, {Ardelean}, {Babej}, {Bach}, {Bachetti}, {Bakanov}, {Bamford}, {Barentsen}, {Barmby}, {Baumbach}, {Berry}, {Biscani}, {Boquien}, {Bostroem}, {Bouma}, {Brammer}, {Bray}, {Breytenbach}, {Buddelmeijer}, {Burke}, {Calderone}, {Cano Rodr{\'\i}guez}, {Cara}, {Cardoso}, {Cheedella}, {Copin}, {Corrales}, {Crichton}, {D'Avella}, {Deil}, {Depagne}, {Dietrich}, {Donath}, {Droettboom}, {Earl}, {Erben}, {Fabbro}, {Ferreira}, {Finethy}, {Fox}, {Garrison}, {Gibbons}, {Goldstein}, {Gommers}, {Greco}, {Greenfield}, {Groener}, {Grollier}, {Hagen}, {Hirst}, {Homeier}, {Horton}, {Hosseinzadeh}, {Hu}, {Hunkeler}, {Ivezi{\'c}}, {Jain}, {Jenness}, {Kanarek}, {Kendrew}, {Kern}, {Kerzendorf}, {Khvalko}, {King}, {Kirkby}, {Kulkarni},
  {Kumar}, {Lee}, {Lenz}, {Littlefair}, {Ma}, {Macleod}, {Mastropietro}, {McCully}, {Montagnac}, {Morris}, {Mueller}, {Mumford}, {Muna}, {Murphy}, {Nelson}, {Nguyen}, {Ninan}, {N{\"o}the}, {Ogaz}, {Oh}, {Parejko}, {Parley}, {Pascual}, {Patil}, {Patil}, {Plunkett}, {Prochaska}, {Rastogi}, {Reddy Janga}, {Sabater}, {Sakurikar}, {Seifert}, {Sherbert}, {Sherwood-Taylor}, {Shih}, {Sick}, {Silbiger}, {Singanamalla}, {Singer}, {Sladen}, {Sooley}, {Sornarajah}, {Streicher}, {Teuben}, {Thomas}, {Tremblay}, {Turner}, {Terr{\'o}n}, {van Kerkwijk}, {de la Vega}, {Watkins}, {Weaver}, {Whitmore}, {Woillez}, {Zabalza}, \& {Astropy Contributors}}]{astropy2018}
{Astropy Collaboration}, {Price-Whelan}, A.~M., {Sip{\H{o}}cz}, B.~M., {et~al.} 2018, \aj, 156, 123, \dodoi{10.3847/1538-3881/aabc4f}

\bibitem[{{Astropy Collaboration} {et~al.}(2022){Astropy Collaboration}, {Price-Whelan}, {Lim}, {Earl}, {Starkman}, {Bradley}, {Shupe}, {Patil}, {Corrales}, {Brasseur}, {N{\"o}the}, {Donath}, {Tollerud}, {Morris}, {Ginsburg}, {Vaher}, {Weaver}, {Tocknell}, {Jamieson}, {van Kerkwijk}, {Robitaille}, {Merry}, {Bachetti}, {G{\"u}nther}, {Aldcroft}, {Alvarado-Montes}, {Archibald}, {B{\'o}di}, {Bapat}, {Barentsen}, {Baz{\'a}n}, {Biswas}, {Boquien}, {Burke}, {Cara}, {Cara}, {Conroy}, {Conseil}, {Craig}, {Cross}, {Cruz}, {D'Eugenio}, {Dencheva}, {Devillepoix}, {Dietrich}, {Eigenbrot}, {Erben}, {Ferreira}, {Foreman-Mackey}, {Fox}, {Freij}, {Garg}, {Geda}, {Glattly}, {Gondhalekar}, {Gordon}, {Grant}, {Greenfield}, {Groener}, {Guest}, {Gurovich}, {Handberg}, {Hart}, {Hatfield-Dodds}, {Homeier}, {Hosseinzadeh}, {Jenness}, {Jones}, {Joseph}, {Kalmbach}, {Karamehmetoglu}, {Ka{\l}uszy{\'n}ski}, {Kelley}, {Kern}, {Kerzendorf}, {Koch}, {Kulumani}, {Lee}, {Ly}, {Ma}, {MacBride}, {Maljaars}, {Muna}, {Murphy}, {Norman},
  {O'Steen}, {Oman}, {Pacifici}, {Pascual}, {Pascual-Granado}, {Patil}, {Perren}, {Pickering}, {Rastogi}, {Roulston}, {Ryan}, {Rykoff}, {Sabater}, {Sakurikar}, {Salgado}, {Sanghi}, {Saunders}, {Savchenko}, {Schwardt}, {Seifert-Eckert}, {Shih}, {Jain}, {Shukla}, {Sick}, {Simpson}, {Singanamalla}, {Singer}, {Singhal}, {Sinha}, {Sip{\H{o}}cz}, {Spitler}, {Stansby}, {Streicher}, {{\v{S}}umak}, {Swinbank}, {Taranu}, {Tewary}, {Tremblay}, {de Val-Borro}, {Van Kooten}, {Vasovi{\'c}}, {Verma}, {de Miranda Cardoso}, {Williams}, {Wilson}, {Winkel}, {Wood-Vasey}, {Xue}, {Yoachim}, {Zhang}, {Zonca}, \& {Astropy Project Contributors}}]{astropy2022}
{Astropy Collaboration}, {Price-Whelan}, A.~M., {Lim}, P.~L., {et~al.} 2022, \apj, 935, 167, \dodoi{10.3847/1538-4357/ac7c74}

\bibitem[{{Atek} {et~al.}(2024){Atek}, {Labb{\'e}}, {Furtak}, {Chemerynska}, {Fujimoto}, {Setton}, {Miller}, {Oesch}, {Bezanson}, {Price}, {Dayal}, {Zitrin}, {Kokorev}, {Weaver}, {Brammer}, {Dokkum}, {Williams}, {Cutler}, {Feldmann}, {Fudamoto}, {Greene}, {Leja}, {Maseda}, {Muzzin}, {Pan}, {Papovich}, {Nelson}, {Nanayakkara}, {Stark}, {Stefanon}, {Suess}, {Wang}, \& {Whitaker}}]{atek2024}
{Atek}, H., {Labb{\'e}}, I., {Furtak}, L.~J., {et~al.} 2024, \nat, 626, 975, \dodoi{10.1038/s41586-024-07043-6}

\bibitem[{{Austin} {et~al.}(2024){Austin}, {Conselice}, {Adams}, {Harvey}, {Duan}, {Trussler}, {Li}, {Juodzbalis}, {Ormerod}, {Ferreira}, {Westcott}, {Harris}, {Wilkins}, {Bhatawdekar}, {Caruana}, {Coe}, {Cohen}, {Driver}, {D'Silva}, {Frye}, {Furtak}, {Grogin}, {Hathi}, {Holwerda}, {Jansen}, {Koekemoer}, {Marshall}, {Nonino}, {Ortiz}, {Pirzkal}, {Robotham}, {Ryan}, {Summers}, {Willmer}, {Windhorst}, {Yan}, \& {Zackrisson}}]{austin2024}
{Austin}, D., {Conselice}, C.~J., {Adams}, N.~J., {et~al.} 2024, arXiv e-prints, arXiv:2404.10751, \dodoi{10.48550/arXiv.2404.10751}

\bibitem[{{Ba{\~n}ados} {et~al.}(2018){Ba{\~n}ados}, {Venemans}, {Mazzucchelli}, {Farina}, {Walter}, {Wang}, {Decarli}, {Stern}, {Fan}, {Davies}, {Hennawi}, {Simcoe}, {Turner}, {Rix}, {Yang}, {Kelson}, {Rudie}, \& {Winters}}]{banados2018}
{Ba{\~n}ados}, E., {Venemans}, B.~P., {Mazzucchelli}, C., {et~al.} 2018, \nat, 553, 473, \dodoi{10.1038/nature25180}

\bibitem[{{Begley} {et~al.}(2025){Begley}, {McLure}, {Cullen}, {McLeod}, {Dunlop}, {Carnall}, {Stanton}, {Shapley}, {Cochrane}, {Donnan}, {Ellis}, {Fontana}, {Grogin}, \& {Koekemoer}}]{begley2025}
{Begley}, R., {McLure}, R.~J., {Cullen}, F., {et~al.} 2025, \mnras, 537, 3245, \dodoi{10.1093/mnras/staf211}

\bibitem[{{Behroozi} {et~al.}(2020){Behroozi}, {Conroy}, {Wechsler}, {Hearin}, {Williams}, {Moster}, {Yung}, {Somerville}, {Gottl{\"o}ber}, {Yepes}, \& {Endsley}}]{behroozi2020}
{Behroozi}, P., {Conroy}, C., {Wechsler}, R.~H., {et~al.} 2020, \mnras, 499, 5702, \dodoi{10.1093/mnras/staa3164}

\bibitem[{{Bertin} \& {Arnouts}(1996)}]{bertin1996}
{Bertin}, E., \& {Arnouts}, S. 1996, \aaps, 117, 393, \dodoi{10.1051/aas:1996164}

\bibitem[{{Bolan} {et~al.}(2022){Bolan}, {Lemaux}, {Mason}, {Brada{\v{c}}}, {Treu}, {Strait}, {Pelliccia}, {Pentericci}, \& {Malkan}}]{bolan2022}
{Bolan}, P., {Lemaux}, B.~C., {Mason}, C., {et~al.} 2022, \mnras, 517, 3263, \dodoi{10.1093/mnras/stac1963}

\bibitem[{{Bouwens} {et~al.}(2016){Bouwens}, {Smit}, {Labb{\'e}}, {Franx}, {Caruana}, {Oesch}, {Stefanon}, \& {Rasappu}}]{bouwens2016}
{Bouwens}, R.~J., {Smit}, R., {Labb{\'e}}, I., {et~al.} 2016, \apj, 831, 176, \dodoi{10.3847/0004-637X/831/2/176}

\bibitem[{{Bouwens} {et~al.}(2011){Bouwens}, {Illingworth}, {Oesch}, {Labb{\'e}}, {Trenti}, {van Dokkum}, {Franx}, {Stiavelli}, {Carollo}, {Magee}, \& {Gonzalez}}]{bouwens2011}
{Bouwens}, R.~J., {Illingworth}, G.~D., {Oesch}, P.~A., {et~al.} 2011, \apj, 737, 90, \dodoi{10.1088/0004-637X/737/2/90}

\bibitem[{{Bouwens} {et~al.}(2014){Bouwens}, {Illingworth}, {Oesch}, {Labb{\'e}}, {van Dokkum}, {Trenti}, {Franx}, {Smit}, {Gonzalez}, \& {Magee}}]{bouwens2014}
---. 2014, \apj, 793, 115, \dodoi{10.1088/0004-637X/793/2/115}

\bibitem[{{Bouwens} {et~al.}(2015){Bouwens}, {Illingworth}, {Oesch}, {Trenti}, {Labb{\'e}}, {Bradley}, {Carollo}, {van Dokkum}, {Gonzalez}, {Holwerda}, {Franx}, {Spitler}, {Smit}, \& {Magee}}]{bouwens2015_uvlf}
---. 2015, \apj, 803, 34, \dodoi{10.1088/0004-637X/803/1/34}

\bibitem[{{Bouwens} {et~al.}(2021){Bouwens}, {Oesch}, {Stefanon}, {Illingworth}, {Labb{\'e}}, {Reddy}, {Atek}, {Montes}, {Naidu}, {Nanayakkara}, {Nelson}, \& {Wilkins}}]{bouwens2021}
{Bouwens}, R.~J., {Oesch}, P.~A., {Stefanon}, M., {et~al.} 2021, \aj, 162, 47, \dodoi{10.3847/1538-3881/abf83e}

\bibitem[{{Bouwens} {et~al.}(2023){Bouwens}, {Stefanon}, {Brammer}, {Oesch}, {Herard-Demanche}, {Illingworth}, {Matthee}, {Naidu}, {van Dokkum}, \& {van Leeuwen}}]{bouwens2023}
{Bouwens}, R.~J., {Stefanon}, M., {Brammer}, G., {et~al.} 2023, \mnras, 523, 1036, \dodoi{10.1093/mnras/stad1145}

\bibitem[{{Bowler} {et~al.}(2020){Bowler}, {Jarvis}, {Dunlop}, {McLure}, {McLeod}, {Adams}, {Milvang-Jensen}, \& {McCracken}}]{bowler2020}
{Bowler}, R.~A.~A., {Jarvis}, M.~J., {Dunlop}, J.~S., {et~al.} 2020, \mnras, 493, 2059, \dodoi{10.1093/mnras/staa313}

\bibitem[{{Bowler} {et~al.}(2014){Bowler}, {Dunlop}, {McLure}, {Rogers}, {McCracken}, {Milvang-Jensen}, {Furusawa}, {Fynbo}, {Taniguchi}, {Afonso}, {Bremer}, \& {Le F{\`e}vre}}]{bowler2014}
{Bowler}, R.~A.~A., {Dunlop}, J.~S., {McLure}, R.~J., {et~al.} 2014, \mnras, 440, 2810, \dodoi{10.1093/mnras/stu449}

\bibitem[{{Bowler} {et~al.}(2015){Bowler}, {Dunlop}, {McLure}, {McCracken}, {Milvang-Jensen}, {Furusawa}, {Taniguchi}, {Le F{\`e}vre}, {Fynbo}, {Jarvis}, \& {H{\"a}u{\ss}ler}}]{bowler2015}
---. 2015, \mnras, 452, 1817, \dodoi{10.1093/mnras/stv1403}

\bibitem[{{Boyett} {et~al.}(2024){Boyett}, {Bunker}, {Curtis-Lake}, {Chevallard}, {Cameron}, {Jones}, {Saxena}, {Charlot}, {Curti}, {Wallace}, {Arribas}, {Carniani}, {Willott}, {Alberts}, {Eisenstein}, {Hainline}, {Hausen}, {Johnson}, {Rieke}, {Robertson}, {Stark}, {Tacchella}, {Williams}, {Chen}, {Egami}, {Endsley}, {Kumari}, {Laseter}, {Looser}, {Maseda}, {Scholtz}, {Shivaei}, {Simmonds}, {Smit}, {{\"U}bler}, \& {Witstok}}]{boyett2024}
{Boyett}, K., {Bunker}, A.~J., {Curtis-Lake}, E., {et~al.} 2024, \mnras, 535, 1796, \dodoi{10.1093/mnras/stae2430}

\bibitem[{Bradley {et~al.}(2023)Bradley, Sipőcz, Robitaille, Tollerud, Vinícius, Deil, Barbary, Wilson, Busko, Donath, Günther, Cara, Lim, Meßlinger, Conseil, Bostroem, Droettboom, Bray, Bratholm, Jamieson, Ginsburg, Barentsen, Craig, Morris, Perrin, Rathi, Pascual, Perren, Georgiev, \& Kerzendorf}]{bradley2023_photutils}
Bradley, L., Sipőcz, B., Robitaille, T., {et~al.} 2023, astropy/photutils: 1.9.0, 1.9.0,  Zenodo, \dodoi{10.5281/zenodo.8248020}

\bibitem[{{Brammer} {et~al.}(2008){Brammer}, {van Dokkum}, \& {Coppi}}]{brammer2008}
{Brammer}, G.~B., {van Dokkum}, P.~G., \& {Coppi}, P. 2008, \apj, 686, 1503, \dodoi{10.1086/591786}

\bibitem[{{Bruton} {et~al.}(2023){Bruton}, {Lin}, {Scarlata}, \& {Hayes}}]{bruton2023}
{Bruton}, S., {Lin}, Y.-H., {Scarlata}, C., \& {Hayes}, M.~J. 2023, \apjl, 949, L40, \dodoi{10.3847/2041-8213/acd5d0}

\bibitem[{{Bruzual} \& {Charlot}(2003)}]{bruzual2003}
{Bruzual}, G., \& {Charlot}, S. 2003, \mnras, 344, 1000, \dodoi{10.1046/j.1365-8711.2003.06897.x}

\bibitem[{{Bunker} {et~al.}(2004){Bunker}, {Stanway}, {Ellis}, \& {McMahon}}]{bunker2004}
{Bunker}, A.~J., {Stanway}, E.~R., {Ellis}, R.~S., \& {McMahon}, R.~G. 2004, \mnras, 355, 374, \dodoi{10.1111/j.1365-2966.2004.08326.x}

\bibitem[{{Bunker} {et~al.}(2010){Bunker}, {Wilkins}, {Ellis}, {Stark}, {Lorenzoni}, {Chiu}, {Lacy}, {Jarvis}, \& {Hickey}}]{bunker2010}
{Bunker}, A.~J., {Wilkins}, S., {Ellis}, R.~S., {et~al.} 2010, \mnras, 409, 855, \dodoi{10.1111/j.1365-2966.2010.17350.x}

\bibitem[{{Bunker} {et~al.}(2023){Bunker}, {Saxena}, {Cameron}, {Willott}, {Curtis-Lake}, {Jakobsen}, {Carniani}, {Smit}, {Maiolino}, {Witstok}, {Curti}, {D'Eugenio}, {Jones}, {Ferruit}, {Arribas}, {Charlot}, {Chevallard}, {Giardino}, {de Graaff}, {Looser}, {L{\"u}tzgendorf}, {Maseda}, {Rawle}, {Rix}, {Del Pino}, {Alberts}, {Egami}, {Eisenstein}, {Endsley}, {Hainline}, {Hausen}, {Johnson}, {Rieke}, {Rieke}, {Robertson}, {Shivaei}, {Stark}, {Sun}, {Tacchella}, {Tang}, {Williams}, {Willmer}, {Baker}, {Baum}, {Bhatawdekar}, {Bowler}, {Boyett}, {Chen}, {Circosta}, {Helton}, {Ji}, {Kumari}, {Lyu}, {Nelson}, {Parlanti}, {Perna}, {Sandles}, {Scholtz}, {Suess}, {Topping}, {{\"U}bler}, {Wallace}, \& {Whitler}}]{bunker2023_gnz11}
{Bunker}, A.~J., {Saxena}, A., {Cameron}, A.~J., {et~al.} 2023, \aap, 677, A88, \dodoi{10.1051/0004-6361/202346159}

\bibitem[{{Bunker} {et~al.}(2024){Bunker}, {Cameron}, {Curtis-Lake}, {Jakobsen}, {Carniani}, {Curti}, {Witstok}, {Maiolino}, {D'Eugenio}, {Looser}, {Willott}, {Bonaventura}, {Hainline}, {{\"U}bler}, {Willmer}, {Saxena}, {Smit}, {Alberts}, {Arribas}, {Baker}, {Baum}, {Bhatawdekar}, {Bowler}, {Boyett}, {Charlot}, {Chen}, {Chevallard}, {Circosta}, {DeCoursey}, {de Graaff}, {Egami}, {Eisenstein}, {Endsley}, {Ferruit}, {Giardino}, {Hausen}, {Helton}, {Hviding}, {Ji}, {Johnson}, {Jones}, {Kumari}, {Laseter}, {L{\"u}tzgendorf}, {Maseda}, {Nelson}, {Parlanti}, {Perna}, {Rauscher}, {Rawle}, {Rix}, {Rieke}, {Robertson}, {Rodr{\'\i}guez Del Pino}, {Sandles}, {Scholtz}, {Sharpe}, {Skarbinski}, {Stark}, {Sun}, {Tacchella}, {Topping}, {Villanueva}, {Wallace}, {Williams}, \& {Woodrum}}]{bunker2024_jades_dr1}
{Bunker}, A.~J., {Cameron}, A.~J., {Curtis-Lake}, E., {et~al.} 2024, \aap, 690, A288, \dodoi{10.1051/0004-6361/202347094}

\bibitem[{{Carniani} {et~al.}(2024{\natexlab{a}}){Carniani}, {Hainline}, {D'Eugenio}, {Eisenstein}, {Jakobsen}, {Witstok}, {Johnson}, {Chevallard}, {Maiolino}, {Helton}, {Willott}, {Robertson}, {Alberts}, {Arribas}, {Baker}, {Bhatawdekar}, {Boyett}, {Bunker}, {Cameron}, {Cargile}, {Charlot}, {Curti}, {Curtis-Lake}, {Egami}, {Giardino}, {Isaak}, {Ji}, {Jones}, {Maseda}, {Parlanti}, {Rawle}, {Rieke}, {Rieke}, {Rodr{\'\i}guez Del Pino}, {Saxena}, {Scholtz}, {Smit}, {Sun}, {Tacchella}, {{\"U}bler}, {Venturi}, {Williams}, \& {Willmer}}]{carniani2024}
{Carniani}, S., {Hainline}, K., {D'Eugenio}, F., {et~al.} 2024{\natexlab{a}}, arXiv e-prints, arXiv:2405.18485, \dodoi{10.48550/arXiv.2405.18485}

\bibitem[{{Carniani} {et~al.}(2024{\natexlab{b}}){Carniani}, {D'Eugenio}, {Ji}, {Parlanti}, {Scholtz}, {Sun}, {Venturi}, {Bakx}, {Curti}, {Maiolino}, {Tacchella}, {Zavala}, {Hainline}, {Witstok}, {Johnson}, {Alberts}, {Bunker}, {Charlot}, {Eisenstein}, {Helton}, {Jakobsen}, {Kumari}, {Robertson}, {Saxena}, {{\"U}bler}, {Williams}, {Willmer}, \& {Willott}}]{carniani2024_z14_alma}
{Carniani}, S., {D'Eugenio}, F., {Ji}, X., {et~al.} 2024{\natexlab{b}}, arXiv e-prints, arXiv:2409.20533, \dodoi{10.48550/arXiv.2409.20533}

\bibitem[{{Castellano} {et~al.}(2022){Castellano}, {Fontana}, {Treu}, {Santini}, {Merlin}, {Leethochawalit}, {Trenti}, {Vanzella}, {Mestric}, {Bonchi}, {Belfiori}, {Nonino}, {Paris}, {Polenta}, {Roberts-Borsani}, {Boyett}, {Brada{\v{c}}}, {Calabr{\`o}}, {Glazebrook}, {Grillo}, {Mascia}, {Mason}, {Mercurio}, {Morishita}, {Nanayakkara}, {Pentericci}, {Rosati}, {Vulcani}, {Wang}, \& {Yang}}]{castellano2022}
{Castellano}, M., {Fontana}, A., {Treu}, T., {et~al.} 2022, \apjl, 938, L15, \dodoi{10.3847/2041-8213/ac94d0}

\bibitem[{{Castellano} {et~al.}(2023){Castellano}, {Fontana}, {Treu}, {Merlin}, {Santini}, {Bergamini}, {Grillo}, {Rosati}, {Acebron}, {Leethochawalit}, {Paris}, {Bonchi}, {Belfiori}, {Calabr{\`o}}, {Correnti}, {Nonino}, {Polenta}, {Trenti}, {Boyett}, {Brammer}, {Broadhurst}, {Caminha}, {Chen}, {Filippenko}, {Fortuni}, {Glazebrook}, {Mascia}, {Mason}, {Menci}, {Meneghetti}, {Mercurio}, {Metha}, {Morishita}, {Nanayakkara}, {Pentericci}, {Roberts-Borsani}, {Roy}, {Vanzella}, {Vulcani}, {Yang}, \& {Wang}}]{castellano2023_lf}
---. 2023, \apjl, 948, L14, \dodoi{10.3847/2041-8213/accea5}

\bibitem[{{Castellano} {et~al.}(2024){Castellano}, {Napolitano}, {Fontana}, {Roberts-Borsani}, {Treu}, {Vanzella}, {Zavala}, {Arrabal Haro}, {Calabr{\`o}}, {Llerena}, {Mascia}, {Merlin}, {Paris}, {Pentericci}, {Santini}, {Bakx}, {Bergamini}, {Cupani}, {Dickinson}, {Filippenko}, {Glazebrook}, {Grillo}, {Kelly}, {Malkan}, {Mason}, {Morishita}, {Nanayakkara}, {Rosati}, {Sani}, {Wang}, \& {Yoon}}]{castellano2024_ghz2}
{Castellano}, M., {Napolitano}, L., {Fontana}, A., {et~al.} 2024, \apj, 972, 143, \dodoi{10.3847/1538-4357/ad5f88}

\bibitem[{{Castellano} {et~al.}(2025){Castellano}, {Fontana}, {Merlin}, {Santini}, {Napolitano}, {Menci}, {Calabr{\`o}}, {Paris}, {Pentericci}, {Zavala}, {Dickinson}, {Finkelstein}, {Treu}, {Amorin}, {Arrabal Haro}, {Bergamini}, {Bisigello}, {Daddi}, {Dayal}, {Dekel}, {Ferrara}, {Fortuni}, {Gandolfi}, {Giavalisco}, {Grillo}, {Guida}, {Hathi}, {Holwerda}, {Koekemoer}, {Kokorev}, {Li}, {Llerena}, {Lucas}, {Mascia}, {Metha}, {Morishita}, {Nanayakkara}, {Pacucci}, {P{\'e}rez-Gonz{\'a}lez}, {Roberts-Borsani}, {Rodighiero}, {Rosati}, {Salazar}, {Schneider}, {Somerville}, {Taylor}, {Trenti}, {Trinca}, {Wang}, {Watson}, {Yang}, \& {Yung}}]{castellano2025}
{Castellano}, M., {Fontana}, A., {Merlin}, E., {et~al.} 2025, arXiv e-prints, arXiv:2504.05893, \dodoi{10.48550/arXiv.2504.05893}

\bibitem[{{Ceverino} {et~al.}(2024){Ceverino}, {Nakazato}, {Yoshida}, {Klessen}, \& {Glover}}]{ceverino2024}
{Ceverino}, D., {Nakazato}, Y., {Yoshida}, N., {Klessen}, R.~S., \& {Glover}, S.~C.~O. 2024, \aap, 689, A244, \dodoi{10.1051/0004-6361/202450224}

\bibitem[{{Chabrier}(2003)}]{chabrier2003}
{Chabrier}, G. 2003, \pasp, 115, 763, \dodoi{10.1086/376392}

\bibitem[{{Chevallard} \& {Charlot}(2016)}]{chevallard2016}
{Chevallard}, J., \& {Charlot}, S. 2016, \mnras, 462, 1415, \dodoi{10.1093/mnras/stw1756}

\bibitem[{{Chisholm} {et~al.}(2022){Chisholm}, {Saldana-Lopez}, {Flury}, {Schaerer}, {Jaskot}, {Amor{\'\i}n}, {Atek}, {Finkelstein}, {Fleming}, {Ferguson}, {Fern{\'a}ndez}, {Giavalisco}, {Hayes}, {Heckman}, {Henry}, {Ji}, {Marques-Chaves}, {Mauerhofer}, {McCandliss}, {Oey}, {{\"O}stlin}, {Rutkowski}, {Scarlata}, {Thuan}, {Trebitsch}, {Wang}, {Worseck}, \& {Xu}}]{chisholm2022}
{Chisholm}, J., {Saldana-Lopez}, A., {Flury}, S., {et~al.} 2022, \mnras, 517, 5104, \dodoi{10.1093/mnras/stac2874}

\bibitem[{{Cueto} {et~al.}(2024){Cueto}, {Hutter}, {Dayal}, {Gottl{\"o}ber}, {Heintz}, {Mason}, {Trebitsch}, \& {Yepes}}]{cueto2024}
{Cueto}, E.~R., {Hutter}, A., {Dayal}, P., {et~al.} 2024, \aap, 686, A138, \dodoi{10.1051/0004-6361/202349017}

\bibitem[{{Cullen} {et~al.}(2023){Cullen}, {McLure}, {McLeod}, {Dunlop}, {Donnan}, {Carnall}, {Bowler}, {Begley}, {Hamadouche}, \& {Stanton}}]{cullen2023}
{Cullen}, F., {McLure}, R.~J., {McLeod}, D.~J., {et~al.} 2023, \mnras, 520, 14, \dodoi{10.1093/mnras/stad073}

\bibitem[{{Cullen} {et~al.}(2024){Cullen}, {McLeod}, {McLure}, {Dunlop}, {Donnan}, {Carnall}, {Keating}, {Magee}, {Arellano-Cordova}, {Bowler}, {Begley}, {Flury}, {Hamadouche}, \& {Stanton}}]{cullen2024}
{Cullen}, F., {McLeod}, D.~J., {McLure}, R.~J., {et~al.} 2024, \mnras, 531, 997, \dodoi{10.1093/mnras/stae1211}

\bibitem[{{Curtis-Lake} {et~al.}(2023){Curtis-Lake}, {Carniani}, {Cameron}, {Charlot}, {Jakobsen}, {Maiolino}, {Bunker}, {Witstok}, {Smit}, {Chevallard}, {Willott}, {Ferruit}, {Arribas}, {Bonaventura}, {Curti}, {D'Eugenio}, {Franx}, {Giardino}, {Looser}, {L{\"u}tzgendorf}, {Maseda}, {Rawle}, {Rix}, {Rodr{\'\i}guez del Pino}, {{\"U}bler}, {Sirianni}, {Dressler}, {Egami}, {Eisenstein}, {Endsley}, {Hainline}, {Hausen}, {Johnson}, {Rieke}, {Robertson}, {Shivaei}, {Stark}, {Tacchella}, {Williams}, {Willmer}, {Bhatawdekar}, {Bowler}, {Boyett}, {Chen}, {de Graaff}, {Helton}, {Hviding}, {Jones}, {Kumari}, {Lyu}, {Nelson}, {Perna}, {Sandles}, {Saxena}, {Suess}, {Sun}, {Topping}, {Wallace}, \& {Whitler}}]{curtis-lake2023}
{Curtis-Lake}, E., {Carniani}, S., {Cameron}, A., {et~al.} 2023, Nature Astronomy, 7, 622, \dodoi{10.1038/s41550-023-01918-w}

\bibitem[{{Davies} {et~al.}(2018){Davies}, {Hennawi}, {Ba{\~n}ados}, {Simcoe}, {Decarli}, {Fan}, {Farina}, {Mazzucchelli}, {Rix}, {Venemans}, {Walter}, {Wang}, \& {Yang}}]{davies2018}
{Davies}, F.~B., {Hennawi}, J.~F., {Ba{\~n}ados}, E., {et~al.} 2018, \apj, 864, 143, \dodoi{10.3847/1538-4357/aad7f8}

\bibitem[{{De Barros} {et~al.}(2019){De Barros}, {Oesch}, {Labb{\'e}}, {Stefanon}, {Gonz{\'a}lez}, {Smit}, {Bouwens}, \& {Illingworth}}]{debarros2019}
{De Barros}, S., {Oesch}, P.~A., {Labb{\'e}}, I., {et~al.} 2019, \mnras, 489, 2355, \dodoi{10.1093/mnras/stz940}

\bibitem[{{DeCoursey} {et~al.}(2023{\natexlab{a}}){DeCoursey}, {Egami}, {Rieke}, {DeFour-Remy}, {Khairnar}, {Ma}, {Sun}, {Eisenstein}, {Robertson}, {Johnson}, \& {Tacchella}}]{decoursey2023_gs}
{DeCoursey}, C., {Egami}, E., {Rieke}, M., {et~al.} 2023{\natexlab{a}}, Transient Name Server AstroNote, 16, 1

\bibitem[{{DeCoursey} {et~al.}(2023{\natexlab{b}}){DeCoursey}, {Egami}, {Rieke}, {DeFour-Remy}, {Khairnar}, {Ma}, {Sun}, {Willmer}, {Rest}, {Pierel}, {Engesser}, {Hainline}, {Helton}, {Eisenstein}, {Robertson}, {Johnson}, {Tacchella}, {Hausen}, \& {Williams}}]{decoursey2023_gn}
---. 2023{\natexlab{b}}, Transient Name Server AstroNote, 164, 1

\bibitem[{{DeCoursey} {et~al.}(2023{\natexlab{c}}){DeCoursey}, {Sun}, {Egami}, {Rieke}, {Hainline}, {DeFour-Remy}, {Rest}, {Pierel}, {Engesser}, {Eisenstein}, {Robertson}, {Johnson}, {Tacchella}, {Willmer}, {Willot}, {Hausen}, {Helton}, {Williams}, {Fox}, {Strolger}, \& {Siebert}}]{decoursey2023_3215}
{DeCoursey}, C., {Sun}, F., {Egami}, E., {et~al.} 2023{\natexlab{c}}, Transient Name Server AstroNote, 275, 1

\bibitem[{{Dekel} {et~al.}(2023){Dekel}, {Sarkar}, {Birnboim}, {Mandelker}, \& {Li}}]{dekel2023}
{Dekel}, A., {Sarkar}, K.~C., {Birnboim}, Y., {Mandelker}, N., \& {Li}, Z. 2023, \mnras, 523, 3201, \dodoi{10.1093/mnras/stad1557}

\bibitem[{{D'Eugenio} {et~al.}(2024{\natexlab{a}}){D'Eugenio}, {Maiolino}, {Carniani}, {Chevallard}, {Curtis-Lake}, {Witstok}, {Charlot}, {Baker}, {Arribas}, {Boyett}, {Bunker}, {Curti}, {Eisenstein}, {Hainline}, {Ji}, {Johnson}, {Kumari}, {Looser}, {Nakajima}, {Nelson}, {Rieke}, {Robertson}, {Scholtz}, {Smit}, {Sun}, {Venturi}, {Tacchella}, {{\"U}bler}, {Willmer}, \& {Willott}}]{deugenio2024_carbon}
{D'Eugenio}, F., {Maiolino}, R., {Carniani}, S., {et~al.} 2024{\natexlab{a}}, \aap, 689, A152, \dodoi{10.1051/0004-6361/202348636}

\bibitem[{{D'Eugenio} {et~al.}(2024{\natexlab{b}}){D'Eugenio}, {Cameron}, {Scholtz}, {Carniani}, {Willott}, {Curtis-Lake}, {Bunker}, {Parlanti}, {Maiolino}, {Willmer}, {Jakobsen}, {Robertson}, {Johnson}, {Tacchella}, {Cargile}, {Rawle}, {Arribas}, {Chevallard}, {Curti}, {Egami}, {Eisenstein}, {Kumari}, {Looser}, {Rieke}, {Rodr{\'\i}guez Del Pino}, {Saxena}, {{\"U}bler}, {Venturi}, {Witstok}, {Baker}, {Bhatawdekar}, {Bonaventura}, {Boyett}, {Charlot}, {Danhaive}, {Hainline}, {Hausen}, {Helton}, {Ji}, {Ji}, {Jones}, {Joud{\v{z}}balis}, {Maseda}, {P{\'e}rez-Gonz{\'a}lez}, {Perna}, {Pusk{\'a}s}, {Shivaei}, {Silcock}, {Simmonds}, {Smit}, {Sun}, {Villanueva}, {Williams}, \& {Zhu}}]{deugenio2024}
{D'Eugenio}, F., {Cameron}, A.~J., {Scholtz}, J., {et~al.} 2024{\natexlab{b}}, arXiv e-prints, arXiv:2404.06531, \dodoi{10.48550/arXiv.2404.06531}

\bibitem[{{Donnan} {et~al.}(2023){Donnan}, {McLeod}, {Dunlop}, {McLure}, {Carnall}, {Begley}, {Cullen}, {Hamadouche}, {Bowler}, {Magee}, {McCracken}, {Milvang-Jensen}, {Moneti}, \& {Targett}}]{donnan2023}
{Donnan}, C.~T., {McLeod}, D.~J., {Dunlop}, J.~S., {et~al.} 2023, \mnras, 518, 6011, \dodoi{10.1093/mnras/stac3472}

\bibitem[{{Donnan} {et~al.}(2024){Donnan}, {McLure}, {Dunlop}, {McLeod}, {Magee}, {Arellano-C{\'o}rdova}, {Barrufet}, {Begley}, {Bowler}, {Carnall}, {Cullen}, {Ellis}, {Fontana}, {Illingworth}, {Grogin}, {Hamadouche}, {Koekemoer}, {Liu}, {Mason}, {Santini}, \& {Stanton}}]{donnan2024}
{Donnan}, C.~T., {McLure}, R.~J., {Dunlop}, J.~S., {et~al.} 2024, \mnras, 533, 3222, \dodoi{10.1093/mnras/stae2037}

\bibitem[{{Draine}(2011)}]{draine2011}
{Draine}, B.~T. 2011, {Physics of the Interstellar and Intergalactic Medium}

\bibitem[{{Eisenstein} {et~al.}(2023{\natexlab{a}}){Eisenstein}, {Willott}, {Alberts}, {Arribas}, {Bonaventura}, {Bunker}, {Cameron}, {Carniani}, {Charlot}, {Curtis-Lake}, {D'Eugenio}, {Endsley}, {Ferruit}, {Giardino}, {Hainline}, {Hausen}, {Jakobsen}, {Johnson}, {Maiolino}, {Rieke}, {Rieke}, {Rix}, {Robertson}, {Stark}, {Tacchella}, {Williams}, {Willmer}, {Baker}, {Baum}, {Bhatawdekar}, {Boyett}, {Chen}, {Chevallard}, {Circosta}, {Curti}, {Danhaive}, {DeCoursey}, {de Graaff}, {Dressler}, {Egami}, {Helton}, {Hviding}, {Ji}, {Jones}, {Kumari}, {L{\"u}tzgendorf}, {Laseter}, {Looser}, {Lyu}, {Maseda}, {Nelson}, {Parlanti}, {Perna}, {Pusk{\'a}s}, {Rawle}, {Rodr{\'\i}guez Del Pino}, {Sandles}, {Saxena}, {Scholtz}, {Sharpe}, {Shivaei}, {Silcock}, {Simmonds}, {Skarbinski}, {Smit}, {Stone}, {Suess}, {Sun}, {Tang}, {Topping}, {{\"U}bler}, {Villanueva}, {Wallace}, {Whitler}, {Witstok}, \& {Woodrum}}]{eisenstein2023_jades}
{Eisenstein}, D.~J., {Willott}, C., {Alberts}, S., {et~al.} 2023{\natexlab{a}}, arXiv e-prints, arXiv:2306.02465, \dodoi{10.48550/arXiv.2306.02465}

\bibitem[{{Eisenstein} {et~al.}(2023{\natexlab{b}}){Eisenstein}, {Johnson}, {Robertson}, {Tacchella}, {Hainline}, {Jakobsen}, {Maiolino}, {Bonaventura}, {Bunker}, {Cameron}, {Cargile}, {Curtis-Lake}, {Hausen}, {Pusk{\'a}s}, {Rieke}, {Sun}, {Willmer}, {Willott}, {Alberts}, {Arribas}, {Baker}, {Baum}, {Bhatawdekar}, {Carniani}, {Charlot}, {Chen}, {Chevallard}, {Curti}, {DeCoursey}, {D'Eugenio}, {de Graaff}, {Egami}, {Helton}, {Ji}, {Jones}, {Kumari}, {L{\"u}tzgendorf}, {Laseter}, {Looser}, {Lyu}, {Maseda}, {Nelson}, {Parlanti}, {Rauscher}, {Rawle}, {Rieke}, {Rix}, {Rujopakarn}, {Sandles}, {Saxena}, {Scholtz}, {Sharpe}, {Shivaei}, {Simmonds}, {Smit}, {Topping}, {{\"U}bler}, {Venturi}, {Williams}, {Witstok}, \& {Woodrum}}]{eisenstein2023_jof}
{Eisenstein}, D.~J., {Johnson}, B.~D., {Robertson}, B., {et~al.} 2023{\natexlab{b}}, arXiv e-prints, arXiv:2310.12340, \dodoi{10.48550/arXiv.2310.12340}

\bibitem[{{Ellis} {et~al.}(2013){Ellis}, {McLure}, {Dunlop}, {Robertson}, {Ono}, {Schenker}, {Koekemoer}, {Bowler}, {Ouchi}, {Rogers}, {Curtis-Lake}, {Schneider}, {Charlot}, {Stark}, {Furlanetto}, \& {Cirasuolo}}]{ellis2013}
{Ellis}, R.~S., {McLure}, R.~J., {Dunlop}, J.~S., {et~al.} 2013, \apjl, 763, L7, \dodoi{10.1088/2041-8205/763/1/L7}

\bibitem[{{Endsley} {et~al.}(2023){Endsley}, {Stark}, {Whitler}, {Topping}, {Chen}, {Plat}, {Chisholm}, \& {Charlot}}]{endsley2023}
{Endsley}, R., {Stark}, D.~P., {Whitler}, L., {et~al.} 2023, \mnras, 524, 2312, \dodoi{10.1093/mnras/stad1919}

\bibitem[{{Endsley} {et~al.}(2024){Endsley}, {Stark}, {Whitler}, {Topping}, {Johnson}, {Robertson}, {Tacchella}, {Alberts}, {Baker}, {Bhatawdekar}, {Boyett}, {Bunker}, {Cameron}, {Carniani}, {Charlot}, {Chen}, {Chevallard}, {Curtis-Lake}, {Danhaive}, {Egami}, {Eisenstein}, {Hainline}, {Helton}, {Ji}, {Looser}, {Maiolino}, {Nelson}, {Pusk{\'a}s}, {Rieke}, {Rieke}, {Rix}, {Sandles}, {Saxena}, {Simmonds}, {Smit}, {Sun}, {Williams}, {Willmer}, {Willott}, \& {Witstok}}]{endsley2024_jades}
---. 2024, \mnras, 533, 1111, \dodoi{10.1093/mnras/stae1857}

\bibitem[{{Fan} {et~al.}(2023){Fan}, {Ba{\~n}ados}, \& {Simcoe}}]{fan2023}
{Fan}, X., {Ba{\~n}ados}, E., \& {Simcoe}, R.~A. 2023, \araa, 61, 373, \dodoi{10.1146/annurev-astro-052920-102455}

\bibitem[{{Fan} {et~al.}(2006){Fan}, {Strauss}, {Becker}, {White}, {Gunn}, {Knapp}, {Richards}, {Schneider}, {Brinkmann}, \& {Fukugita}}]{fan2006}
{Fan}, X., {Strauss}, M.~A., {Becker}, R.~H., {et~al.} 2006, \aj, 132, 117, \dodoi{10.1086/504836}

\bibitem[{{Feldmann} {et~al.}(2025){Feldmann}, {Boylan-Kolchin}, {Bullock}, {{\c{C}}atmabacak}, {Faucher-Gigu{\`e}re}, {Hayward}, {Kere{\v{s}}}, {Lazar}, {Liang}, {Moreno}, {Oesch}, {Quataert}, {Shen}, \& {Sun}}]{feldmann2025}
{Feldmann}, R., {Boylan-Kolchin}, M., {Bullock}, J.~S., {et~al.} 2025, \mnras, 536, 988, \dodoi{10.1093/mnras/stae2633}

\bibitem[{{Feroz} \& {Hobson}(2008)}]{feroz2008}
{Feroz}, F., \& {Hobson}, M.~P. 2008, \mnras, 384, 449, \dodoi{10.1111/j.1365-2966.2007.12353.x}

\bibitem[{{Feroz} {et~al.}(2009){Feroz}, {Hobson}, \& {Bridges}}]{feroz2009}
{Feroz}, F., {Hobson}, M.~P., \& {Bridges}, M. 2009, \mnras, 398, 1601, \dodoi{10.1111/j.1365-2966.2009.14548.x}

\bibitem[{{Feroz} {et~al.}(2019){Feroz}, {Hobson}, {Cameron}, \& {Pettitt}}]{feroz2019}
{Feroz}, F., {Hobson}, M.~P., {Cameron}, E., \& {Pettitt}, A.~N. 2019, The Open Journal of Astrophysics, 2, 10, \dodoi{10.21105/astro.1306.2144}

\bibitem[{{Ferrara}(2024)}]{ferrara2024}
{Ferrara}, A. 2024, \aap, 684, A207, \dodoi{10.1051/0004-6361/202348321}

\bibitem[{{Ferrara} {et~al.}(2023){Ferrara}, {Pallottini}, \& {Dayal}}]{ferrara2023}
{Ferrara}, A., {Pallottini}, A., \& {Dayal}, P. 2023, \mnras, 522, 3986, \dodoi{10.1093/mnras/stad1095}

\bibitem[{{Ferruit} {et~al.}(2022){Ferruit}, {Jakobsen}, {Giardino}, {Rawle}, {Alves de Oliveira}, {Arribas}, {Beck}, {Birkmann}, {B{\"o}ker}, {Bunker}, {Charlot}, {de Marchi}, {Franx}, {Henry}, {Karakla}, {Kassin}, {Kumari}, {L{\'o}pez-Caniego}, {L{\"u}tzgendorf}, {Maiolino}, {Manjavacas}, {Marston}, {Moseley}, {Muzerolle}, {Pirzkal}, {Rauscher}, {Rix}, {Sabbi}, {Sirianni}, {te Plate}, {Valenti}, {Willott}, \& {Zeidler}}]{ferruit2022}
{Ferruit}, P., {Jakobsen}, P., {Giardino}, G., {et~al.} 2022, \aap, 661, A81, \dodoi{10.1051/0004-6361/202142673}

\bibitem[{{Finkelstein} {et~al.}(2012){Finkelstein}, {Papovich}, {Salmon}, {Finlator}, {Dickinson}, {Ferguson}, {Giavalisco}, {Koekemoer}, {Reddy}, {Bassett}, {Conselice}, {Dunlop}, {Faber}, {Grogin}, {Hathi}, {Kocevski}, {Lai}, {Lee}, {McLure}, {Mobasher}, \& {Newman}}]{finkelstein2012_colors}
{Finkelstein}, S.~L., {Papovich}, C., {Salmon}, B., {et~al.} 2012, \apj, 756, 164, \dodoi{10.1088/0004-637X/756/2/164}

\bibitem[{{Finkelstein} {et~al.}(2015){Finkelstein}, {Ryan}, {Papovich}, {Dickinson}, {Song}, {Somerville}, {Ferguson}, {Salmon}, {Giavalisco}, {Koekemoer}, {Ashby}, {Behroozi}, {Castellano}, {Dunlop}, {Faber}, {Fazio}, {Fontana}, {Grogin}, {Hathi}, {Jaacks}, {Kocevski}, {Livermore}, {McLure}, {Merlin}, {Mobasher}, {Newman}, {Rafelski}, {Tilvi}, \& {Willner}}]{finkelstein2015}
{Finkelstein}, S.~L., {Ryan}, Russell~E., J., {Papovich}, C., {et~al.} 2015, \apj, 810, 71, \dodoi{10.1088/0004-637X/810/1/71}

\bibitem[{{Finkelstein} {et~al.}(2022{\natexlab{a}}){Finkelstein}, {Bagley}, {Song}, {Larson}, {Papovich}, {Dickinson}, {Finkelstein}, {Koekemoer}, {Pirzkal}, {Somerville}, {Yung}, {Behroozi}, {Ferguson}, {Giavalisco}, {Grogin}, {Hathi}, {Hutchison}, {Jung}, {Kocevski}, {Kawinwanichakij}, {Rojas-Ruiz}, {Ryan}, {Snyder}, \& {Tacchella}}]{finkelstein2022_hst}
{Finkelstein}, S.~L., {Bagley}, M., {Song}, M., {et~al.} 2022{\natexlab{a}}, \apj, 928, 52, \dodoi{10.3847/1538-4357/ac3aed}

\bibitem[{{Finkelstein} {et~al.}(2022{\natexlab{b}}){Finkelstein}, {Bagley}, {Arrabal Haro}, {Dickinson}, {Ferguson}, {Kartaltepe}, {Papovich}, {Burgarella}, {Kocevski}, {Huertas-Company}, {Iyer}, {Koekemoer}, {Larson}, {P{\'e}rez-Gonz{\'a}lez}, {Rose}, {Tacchella}, {Wilkins}, {Chworowsky}, {Medrano}, {Morales}, {Somerville}, {Yung}, {Fontana}, {Giavalisco}, {Grazian}, {Grogin}, {Kewley}, {Kirkpatrick}, {Kurczynski}, {Lotz}, {Pentericci}, {Pirzkal}, {Ravindranath}, {Ryan}, {Trump}, {Yang}, {Almaini}, {Amor{\'\i}n}, {Annunziatella}, {Backhaus}, {Barro}, {Behroozi}, {Bell}, {Bhatawdekar}, {Bisigello}, {Bromm}, {Buat}, {Buitrago}, {Calabr{\`o}}, {Casey}, {Castellano}, {Ch{\'a}vez Ortiz}, {Ciesla}, {Cleri}, {Cohen}, {Cole}, {Cooke}, {Cooper}, {Cooray}, {Costantin}, {Cox}, {Croton}, {Daddi}, {Dav{\'e}}, {de La Vega}, {Dekel}, {Elbaz}, {Estrada-Carpenter}, {Faber}, {Fern{\'a}ndez}, {Finkelstein}, {Freundlich}, {Fujimoto}, {Garc{\'\i}a-Argum{\'a}nez}, {Gardner}, {Gawiser}, {G{\'o}mez-Guijarro}, {Guo}, {Hamblin},
  {Hamilton}, {Hathi}, {Holwerda}, {Hirschmann}, {Hutchison}, {Jaskot}, {Jha}, {Jogee}, {Juneau}, {Jung}, {Kassin}, {Le Bail}, {Leung}, {Lucas}, {Magnelli}, {Mantha}, {Matharu}, {McGrath}, {McIntosh}, {Merlin}, {Mobasher}, {Newman}, {Nicholls}, {Pandya}, {Rafelski}, {Ronayne}, {Santini}, {Seill{\'e}}, {Shah}, {Shen}, {Simons}, {Snyder}, {Stanway}, {Straughn}, {Teplitz}, {Vanderhoof}, {Vega-Ferrero}, {Wang}, {Weiner}, {Willmer}, {Wuyts}, {Zavala}, \& {Ceers Team}}]{finkelstein2022_maisie}
{Finkelstein}, S.~L., {Bagley}, M.~B., {Arrabal Haro}, P., {et~al.} 2022{\natexlab{b}}, \apjl, 940, L55, \dodoi{10.3847/2041-8213/ac966e}

\bibitem[{{Finkelstein} {et~al.}(2023){Finkelstein}, {Bagley}, {Ferguson}, {Wilkins}, {Kartaltepe}, {Papovich}, {Yung}, {Arrabal Haro}, {Behroozi}, {Dickinson}, {Kocevski}, {Koekemoer}, {Larson}, {Le Bail}, {Morales}, {P{\'e}rez-Gonz{\'a}lez}, {Burgarella}, {Dav{\'e}}, {Hirschmann}, {Somerville}, {Wuyts}, {Bromm}, {Casey}, {Fontana}, {Fujimoto}, {Gardner}, {Giavalisco}, {Grazian}, {Grogin}, {Hathi}, {Hutchison}, {Jha}, {Jogee}, {Kewley}, {Kirkpatrick}, {Long}, {Lotz}, {Pentericci}, {Pierel}, {Pirzkal}, {Ravindranath}, {Ryan}, {Trump}, {Yang}, {Bhatawdekar}, {Bisigello}, {Buat}, {Calabr{\`o}}, {Castellano}, {Cleri}, {Cooper}, {Croton}, {Daddi}, {Dekel}, {Elbaz}, {Franco}, {Gawiser}, {Holwerda}, {Huertas-Company}, {Jaskot}, {Leung}, {Lucas}, {Mobasher}, {Pandya}, {Tacchella}, {Weiner}, \& {Zavala}}]{finkelstein2023_ceers}
{Finkelstein}, S.~L., {Bagley}, M.~B., {Ferguson}, H.~C., {et~al.} 2023, \apjl, 946, L13, \dodoi{10.3847/2041-8213/acade4}

\bibitem[{{Finkelstein} {et~al.}(2024){Finkelstein}, {Leung}, {Bagley}, {Dickinson}, {Ferguson}, {Papovich}, {Akins}, {Arrabal Haro}, {Dav{\'e}}, {Dekel}, {Kartaltepe}, {Kocevski}, {Koekemoer}, {Pirzkal}, {Somerville}, {Yung}, {Amor{\'\i}n}, {Backhaus}, {Behroozi}, {Bisigello}, {Bromm}, {Casey}, {Ch{\'a}vez Ortiz}, {Cheng}, {Chworowsky}, {Cleri}, {Cooper}, {Davis}, {de la Vega}, {Elbaz}, {Franco}, {Fontana}, {Fujimoto}, {Giavalisco}, {Grogin}, {Holwerda}, {Huertas-Company}, {Hirschmann}, {Iyer}, {Jogee}, {Jung}, {Larson}, {Lucas}, {Mobasher}, {Morales}, {Morley}, {Mukherjee}, {P{\'e}rez-Gonz{\'a}lez}, {Ravindranath}, {Rodighiero}, {Rowland}, {Tacchella}, {Taylor}, {Trump}, \& {Wilkins}}]{finkelstein2024}
{Finkelstein}, S.~L., {Leung}, G. C.~K., {Bagley}, M.~B., {et~al.} 2024, \apjl, 969, L2, \dodoi{10.3847/2041-8213/ad4495}

\bibitem[{{Finlator} {et~al.}(2012){Finlator}, {Oh}, {{\"O}zel}, \& {Dav{\'e}}}]{finlator2012}
{Finlator}, K., {Oh}, S.~P., {{\"O}zel}, F., \& {Dav{\'e}}, R. 2012, \mnras, 427, 2464, \dodoi{10.1111/j.1365-2966.2012.22114.x}

\bibitem[{{Fiore} {et~al.}(2023){Fiore}, {Ferrara}, {Bischetti}, {Feruglio}, \& {Travascio}}]{fiore2023}
{Fiore}, F., {Ferrara}, A., {Bischetti}, M., {Feruglio}, C., \& {Travascio}, A. 2023, \apjl, 943, L27, \dodoi{10.3847/2041-8213/acb5f2}

\bibitem[{{Fujimoto} {et~al.}(2023){Fujimoto}, {Arrabal Haro}, {Dickinson}, {Finkelstein}, {Kartaltepe}, {Larson}, {Burgarella}, {Bagley}, {Behroozi}, {Chworowsky}, {Hirschmann}, {Trump}, {Wilkins}, {Yung}, {Koekemoer}, {Papovich}, {Pirzkal}, {Ferguson}, {Fontana}, {Grogin}, {Grazian}, {Kewley}, {Kocevski}, {Lotz}, {Pentericci}, {Ravindranath}, {Somerville}, {Amorin}, {Backhaus}, {Calabro}, {Casey}, {Cooper}, {Franco}, {Giavalisco}, {Hathi}, {Harish}, {Hutchison}, {Iyer}, {Jung}, {Lucas}, \& {Zavala}}]{fujimoto2023}
{Fujimoto}, S., {Arrabal Haro}, P., {Dickinson}, M., {et~al.} 2023, arXiv e-prints, arXiv:2301.09482, \dodoi{10.48550/arXiv.2301.09482}

\bibitem[{{Gaia Collaboration} {et~al.}(2018){Gaia Collaboration}, {Brown}, {Vallenari}, {Prusti}, {de Bruijne}, {Babusiaux}, {Bailer-Jones}, {Biermann}, {Evans}, {Eyer}, {Jansen}, {Jordi}, {Klioner}, {Lammers}, {Lindegren}, {Luri}, {Mignard}, {Panem}, {Pourbaix}, {Randich}, {Sartoretti}, {Siddiqui}, {Soubiran}, {van Leeuwen}, {Walton}, {Arenou}, {Bastian}, {Cropper}, {Drimmel}, {Katz}, {Lattanzi}, {Bakker}, {Cacciari}, {Casta{\~n}eda}, {Chaoul}, {Cheek}, {De Angeli}, {Fabricius}, {Guerra}, {Holl}, {Masana}, {Messineo}, {Mowlavi}, {Nienartowicz}, {Panuzzo}, {Portell}, {Riello}, {Seabroke}, {Tanga}, {Th{\'e}venin}, {Gracia-Abril}, {Comoretto}, {Garcia-Reinaldos}, {Teyssier}, {Altmann}, {Andrae}, {Audard}, {Bellas-Velidis}, {Benson}, {Berthier}, {Blomme}, {Burgess}, {Busso}, {Carry}, {Cellino}, {Clementini}, {Clotet}, {Creevey}, {Davidson}, {De Ridder}, {Delchambre}, {Dell'Oro}, {Ducourant}, {Fern{\'a}ndez-Hern{\'a}ndez}, {Fouesneau}, {Fr{\'e}mat}, {Galluccio}, {Garc{\'\i}a-Torres},
  {Gonz{\'a}lez-N{\'u}{\~n}ez}, {Gonz{\'a}lez-Vidal}, {Gosset}, {Guy}, {Halbwachs}, {Hambly}, {Harrison}, {Hern{\'a}ndez}, {Hestroffer}, {Hodgkin}, {Hutton}, {Jasniewicz}, {Jean-Antoine-Piccolo}, {Jordan}, {Korn}, {Krone-Martins}, {Lanzafame}, {Lebzelter}, {L{\"o}ffler}, {Manteiga}, {Marrese}, {Mart{\'\i}n-Fleitas}, {Moitinho}, {Mora}, {Muinonen}, {Osinde}, {Pancino}, {Pauwels}, {Petit}, {Recio-Blanco}, {Richards}, {Rimoldini}, {Robin}, {Sarro}, {Siopis}, {Smith}, {Sozzetti}, {S{\"u}veges}, {Torra}, {van Reeven}, {Abbas}, {Abreu Aramburu}, {Accart}, {Aerts}, {Altavilla}, {{\'A}lvarez}, {Alvarez}, {Alves}, {Anderson}, {Andrei}, {Anglada Varela}, {Antiche}, {Antoja}, {Arcay}, {Astraatmadja}, {Bach}, {Baker}, {Balaguer-N{\'u}{\~n}ez}, {Balm}, {Barache}, {Barata}, {Barbato}, {Barblan}, {Barklem}, {Barrado}, {Barros}, {Barstow}, {Bartholom{\'e} Mu{\~n}oz}, {Bassilana}, {Becciani}, {Bellazzini}, {Berihuete}, {Bertone}, {Bianchi}, {Bienaym{\'e}}, {Blanco-Cuaresma}, {Boch}, {Boeche}, {Bombrun}, {Borrachero},
  {Bossini}, {Bouquillon}, {Bourda}, {Bragaglia}, {Bramante}, {Breddels}, {Bressan}, {Brouillet}, {Br{\"u}semeister}, {Brugaletta}, {Bucciarelli}, {Burlacu}, {Busonero}, {Butkevich}, {Buzzi}, {Caffau}, {Cancelliere}, {Cannizzaro}, {Cantat-Gaudin}, {Carballo}, {Carlucci}, {Carrasco}, {Casamiquela}, {Castellani}, {Castro-Ginard}, {Charlot}, {Chemin}, {Chiavassa}, {Cocozza}, {Costigan}, {Cowell}, {Crifo}, {Crosta}, {Crowley}, {Cuypers}, {Dafonte}, {Damerdji}, {Dapergolas}, {David}, {David}, {de Laverny}, {De Luise}, {De March}, {de Martino}, {de Souza}, {de Torres}, {Debosscher}, {del Pozo}, {Delbo}, {Delgado}, {Delgado}, {Di Matteo}, {Diakite}, {Diener}, {Distefano}, {Dolding}, {Drazinos}, {Dur{\'a}n}, {Edvardsson}, {Enke}, {Eriksson}, {Esquej}, {Eynard Bontemps}, {Fabre}, {Fabrizio}, {Faigler}, {Falc{\~a}o}, {Farr{\`a}s Casas}, {Federici}, {Fedorets}, {Fernique}, {Figueras}, {Filippi}, {Findeisen}, {Fonti}, {Fraile}, {Fraser}, {Fr{\'e}zouls}, {Gai}, {Galleti}, {Garabato}, {Garc{\'\i}a-Sedano}, {Garofalo},
  {Garralda}, {Gavel}, {Gavras}, {Gerssen}, {Geyer}, {Giacobbe}, {Gilmore}, {Girona}, {Giuffrida}, {Glass}, {Gomes}, {Granvik}, {Gueguen}, {Guerrier}, {Guiraud}, {Guti{\'e}rrez-S{\'a}nchez}, {Haigron}, {Hatzidimitriou}, {Hauser}, {Haywood}, {Heiter}, {Helmi}, {Heu}, {Hilger}, {Hobbs}, {Hofmann}, {Holland}, {Huckle}, {Hypki}, {Icardi}, {Jan{\ss}en}, {Jevardat de Fombelle}, {Jonker}, {Juh{\'a}sz}, {Julbe}, {Karampelas}, {Kewley}, {Klar}, {Kochoska}, {Kohley}, {Kolenberg}, {Kontizas}, {Kontizas}, {Koposov}, {Kordopatis}, {Kostrzewa-Rutkowska}, {Koubsky}, {Lambert}, {Lanza}, {Lasne}, {Lavigne}, {Le Fustec}, {Le Poncin-Lafitte}, {Lebreton}, {Leccia}, {Leclerc}, {Lecoeur-Taibi}, {Lenhardt}, {Leroux}, {Liao}, {Licata}, {Lindstr{\o}m}, {Lister}, {Livanou}, {Lobel}, {L{\'o}pez}, {Managau}, {Mann}, {Mantelet}, {Marchal}, {Marchant}, {Marconi}, {Marinoni}, {Marschalk{\'o}}, {Marshall}, {Martino}, {Marton}, {Mary}, {Massari}, {Matijevi{\v{c}}}, {Mazeh}, {McMillan}, {Messina}, {Michalik}, {Millar}, {Molina}, {Molinaro},
  {Moln{\'a}r}, {Montegriffo}, {Mor}, {Morbidelli}, {Morel}, {Morris}, {Mulone}, {Muraveva}, {Musella}, {Nelemans}, {Nicastro}, {Noval}, {O'Mullane}, {Ord{\'e}novic}, {Ord{\'o}{\~n}ez-Blanco}, {Osborne}, {Pagani}, {Pagano}, {Pailler}, {Palacin}, {Palaversa}, {Panahi}, {Pawlak}, {Piersimoni}, {Pineau}, {Plachy}, {Plum}, {Poggio}, {Poujoulet}, {Pr{\v{s}}a}, {Pulone}, {Racero}, {Ragaini}, {Rambaux}, {Ramos-Lerate}, {Regibo}, {Reyl{\'e}}, {Riclet}, {Ripepi}, {Riva}, {Rivard}, {Rixon}, {Roegiers}, {Roelens}, {Romero-G{\'o}mez}, {Rowell}, {Royer}, {Ruiz-Dern}, {Sadowski}, {Sagrist{\`a} Sell{\'e}s}, {Sahlmann}, {Salgado}, {Salguero}, {Sanna}, {Santana-Ros}, {Sarasso}, {Savietto}, {Schultheis}, {Sciacca}, {Segol}, {Segovia}, {S{\'e}gransan}, {Shih}, {Siltala}, {Silva}, {Smart}, {Smith}, {Solano}, {Solitro}, {Sordo}, {Soria Nieto}, {Souchay}, {Spagna}, {Spoto}, {Stampa}, {Steele}, {Steidelm{\"u}ller}, {Stephenson}, {Stoev}, {Suess}, {Surdej}, {Szabados}, {Szegedi-Elek}, {Tapiador}, {Taris}, {Tauran}, {Taylor},
  {Teixeira}, {Terrett}, {Teyssandier}, {Thuillot}, {Titarenko}, {Torra Clotet}, {Turon}, {Ulla}, {Utrilla}, {Uzzi}, {Vaillant}, {Valentini}, {Valette}, {van Elteren}, {Van Hemelryck}, {van Leeuwen}, {Vaschetto}, {Vecchiato}, {Veljanoski}, {Viala}, {Vicente}, {Vogt}, {von Essen}, {Voss}, {Votruba}, {Voutsinas}, {Walmsley}, {Weiler}, {Wertz}, {Wevers}, {Wyrzykowski}, {Yoldas}, {{\v{Z}}erjal}, {Ziaeepour}, {Zorec}, {Zschocke}, {Zucker}, {Zurbach}, \& {Zwitter}}]{gaia-collaboration2018}
{Gaia Collaboration}, {Brown}, A.~G.~A., {Vallenari}, A., {et~al.} 2018, \aap, 616, A1, \dodoi{10.1051/0004-6361/201833051}

\bibitem[{{Gardner} {et~al.}(2006){Gardner}, {Mather}, {Clampin}, {Doyon}, {Greenhouse}, {Hammel}, {Hutchings}, {Jakobsen}, {Lilly}, {Long}, {Lunine}, {McCaughrean}, {Mountain}, {Nella}, {Rieke}, {Rieke}, {Rix}, {Smith}, {Sonneborn}, {Stiavelli}, {Stockman}, {Windhorst}, \& {Wright}}]{gardner2006}
{Gardner}, J.~P., {Mather}, J.~C., {Clampin}, M., {et~al.} 2006, \ssr, 123, 485, \dodoi{10.1007/s11214-006-8315-7}

\bibitem[{{Gardner} {et~al.}(2023){Gardner}, {Mather}, {Abbott}, {Abell}, {Abernathy}, {Abney}, {Abraham}, {Abraham}, {Abul-Huda}, {Acton}, {Adams}, {Adams}, {Adler}, {Adriaensen}, {Aguilar}, {Ahmed}, {Ahmed}, {Ahmed}, {Albat}, {Albert}, {Alberts}, {Aldridge}, {Allen}, {Allen}, {Altenburg}, {Altunc}, {Alvarez}, {{\'A}lvarez-M{\'a}rquez}, {Alves de Oliveira}, {Ambrose}, {Anandakrishnan}, {Andersen}, {Anderson}, {Anderson}, {Anderson}, {Anderson}, {Aprea}, {Archer}, {Arenberg}, {Argyriou}, {Arribas}, {Artigau}, {Arvai}, {Atcheson}, {Atkinson}, {Averbukh}, {Aymergen}, {Bacinski}, {Baggett}, {Bagnasco}, {Baker}, {Balzano}, {Banks}, {Baran}, {Barker}, {Barrett}, {Barringer}, {Barto}, {Bast}, {Baudoz}, {Baum}, {Beatty}, {Beaulieu}, {Bechtold}, {Beck}, {Beddard}, {Beichman}, {Bellagama}, {Bely}, {Berger}, {Bergeron}, {Bernier}, {Bertch}, {Beskow}, {Betz}, {Biagetti}, {Birkmann}, {Bjorklund}, {Blackwood}, {Blazek}, {Blossfeld}, {Bluth}, {Boccaletti}, {Boegner}, {Bohlin}, {Boia}, {B{\"o}ker}, {Bonaventura}, {Bond},
  {Bosley}, {Boucarut}, {Bouchet}, {Bouwman}, {Bower}, {Bowers}, {Bowers}, {Boyce}, {Boyer}, {Boyer}, {Boyer}, {Boyer}, {Bradley}, {Brady}, {Brandl}, {Brannen}, {Breda}, {Bremmer}, {Brennan}, {Bresnahan}, {Bright}, {Broiles}, {Bromenschenkel}, {Brooks}, {Brooks}, {Brown}, {Brown}, {Brown}, {Bruce}, {Bryson}, {Bujanda}, {Bullock}, {Bunker}, {Bureo}, {Burt}, {Bush}, {Bushouse}, {Bussman}, {Cabaud}, {Cale}, {Calhoon}, {Calvani}, {Canipe}, {Caputo}, {Cara}, {Carey}, {Case}, {Cesari}, {Cetorelli}, {Chance}, {Chandler}, {Chaney}, {Chapman}, {Charlot}, {Chayer}, {Cheezum}, {Chen}, {Chen}, {Cherinka}, {Chichester}, {Chilton}, {Chittiraibalan}, {Clampin}, {Clark}, {Clark}, {Clark}, {Claybrooks}, {Cleveland}, {Cohen}, {Cohen}, {Col{\'o}n}, {Coleman}, {Colina}, {Comber}, {Comeau}, {Comer}, {Conde Reis}, {Connolly}, {Conroy}, {Contos}, {Contreras}, {Cook}, {Cooper}, {Cooper}, {Correia}, {Correnti}, {Cossou}, {Costanza}, {Coulais}, {Cox}, {Coyle}, {Cracraft}, {Crew}, {Curtis}, {Cusveller}, {Da Costa Maciel}, {Dailey},
  {Daugeron}, {Davidson}, {Davies}, {Davis}, {Davis}, {Day}, {de Chambure}, {de Jong}, {De Marchi}, {Dean}, {Decker}, {Delisa}, {Dell}, {Dellagatta}, {Dembinska}, {Demosthenes}, {Dencheva}, {Deneu}, {DePriest}, {Deschenes}, {Dethienne}, {Detre}, {Diaz}, {Dicken}, {DiFelice}, {Dillman}, {Disharoon}, {Dixon}, {Doggett}, {Dominguez}, {Donaldson}, {Doria-Warner}, {Santos}, {Doty}, {Douglas}, {Doyon}, {Dressler}, {Driggers}, {Driggers}, {Dunn}, {DuPrie}, {Dupuis}, {Durning}, {Dutta}, {Earl}, {Eccleston}, {Ecobichon}, {Egami}, {Ehrenwinkler}, {Eisenhamer}, {Eisenhower}, {Eisenstein}, {El Hamel}, {Elie}, {Elliott}, {Elliott}, {Engesser}, {Espinoza}, {Etienne}, {Etxaluze}, {Evans}, {Fabreguettes}, {Falcolini}, {Falini}, {Fatig}, {Feeney}, {Feinberg}, {Fels}, {Ferdous}, {Ferguson}, {Ferrarese}, {Ferreira}, {Ferruit}, {Ferry}, {Filippazzo}, {Firre}, {Fix}, {Flagey}, {Flanagan}, {Fleming}, {Florian}, {Flynn}, {Foiadelli}, {Fontaine}, {Fontanella}, {Forshay}, {Fortner}, {Fox}, {Framarini}, {Francisco}, {Franck}, {Franx},
  {Franz}, {Friedman}, {Friend}, {Frost}, {Fu}, {Fullerton}, {Gaillard}, {Galkin}, {Gallagher}, {Galyer}, {Garc{\'\i}a Mar{\'\i}n}, {Gardner}, {Garland}, {Garrett}, {Gasman}, {G{\'a}sp{\'a}r}, {Gastaud}, {Gaudreau}, {Gauthier}, {Geers}, {Geithner}, {Gennaro}, {Gerber}, {Gereau}, {Giampaoli}, {Giardino}, {Gibbons}, {Gilbert}, {Gilman}, {Girard}, {Giuliano}, {Gkountis}, {Glasse}, {Glassmire}, {Glauser}, {Glazer}, {Goldberg}, {Golimowski}, {Gonzaga}, {Gordon}, {Gordon}, {Goudfrooij}, {Gough}, {Graham}, {Grau}, {Green}, {Greene}, {Greene}, {Greenfield}, {Greenhouse}, {Greve}, {Greville}, {Grimaldi}, {Groe}, {Groebner}, {Grumm}, {Grundy}, {G{\"u}del}, {Guillard}, {Guldalian}, {Gunn}, {Gurule}, {Gutman}, {Guy}, {Guyot}, {Hack}, {Haderlein}, {Hagan}, {Hagedorn}, {Hainline}, {Haley}, {Hami}, {Hamilton}, {Hammann}, {Hammel}, {Hanley}, {Hansen}, {Hardy}, {Harnisch}, {Harr}, {Harris}, {Hart}, {Hartig}, {Hasan}, {Hashim}, {Hashimoto}, {Haskins}, {Hawkins}, {Hayden}, {Hayden}, {Healy}, {Hecht}, {Heeg}, {Hejal}, {Helm},
  {Hengemihle}, {Henning}, {Henry}, {Henry}, {Henshaw}, {Hernandez}, {Herrington}, {Heske}, {Hesman}, {Hickey}, {Hilbert}, {Hines}, {Hinz}, {Hirsch}, {Hitcho}, {Hodapp}, {Hodge}, {Hoffman}, {Holfeltz}, {Holler}, {Hoppa}, {Horner}, {Howard}, {Howard}, {Huber}, {Hunkeler}, {Hunter}, {Hunter}, {Hurd}, {Hurst}, {Hutchings}, {Hylan}, {Ignat}, {Illingworth}, {Irish}, {Isaacs}, {Jackson}, {Jaffe}, {Jahic}, {Jahromi}, {Jakobsen}, {James}, {James}, {James}, {Jamieson}, {Jandra}, {Jayawardhana}, {Jedrzejewski}, {Jeffers}, {Jensen}, {Joanne}, {Johns}, {Johnson}, {Johnson}, {Johnson}, {Johnson}, {Johnson}, {Johnson}, {Johnstone}, {Jollet}, {Jones}, {Jones}, {Jones}, {Jones}, {Jones}, {Jordan}, {Jordan}, {Jue}, {Jurkowski}, {Justis}, {Justtanont}, {Kaleida}, {Kalirai}, {Kalmanson}, {Kaltenegger}, {Kammerer}, {Kan}, {Kanarek}, {Kao}, {Karakla}, {Karl}, {Kassin}, {Kauffman}, {Kavanagh}, {Kelley}, {Kelly}, {Kendrew}, {Kennedy}, {Kenny}, {Keski-Kuha}, {Keyes}, {Khan}, {Kidwell}, {Kimble}, {King}, {King}, {Kinzel}, {Kirk},
  {Kirkpatrick}, {Klaassen}, {Klingemann}, {Klintworth}, {Knapp}, {Knight}, {Knollenberg}, {Knutsen}, {Koehler}, {Koekemoer}, {Kofler}, {Kontson}, {Kovacs}, {Kozhurina-Platais}, {Krause}, {Kriss}, {Krist}, {Kristoffersen}, {Krogel}, {Krueger}, {Kulp}, {Kumari}, {Kwan}, {Kyprianou}, {Labador}, {Labiano}, {Lafreni{\`e}re}, {Lagage}, {Laidler}, {Laine}, {Laird}, {Lajoie}, {Lallo}, {Lam}, {LaMassa}, {Lambros}, {Lampenfield}, {Lander}, {Langston}, {Larson}, {Larson}, {LaVerghetta}, {Law}, {Lawrence}, {Lee}, {Lee}, {Lee}, {Leisenring}, {Leveille}, {Levenson}, {Levi}, {Levine}, {Lewis}, {Lewis}, {Lewis}, {Libralato}, {Lidon}, {Liebrecht}, {Lightsey}, {Lilly}, {Lim}, {Lim}, {Ling}, {Link}, {Link}, {Lipinski}, {Liu}, {Lo}, {Lobmeyer}, {Logue}, {Long}, {Long}, {Long}, {Long}, {L{\'o}pez-Caniego}, {Lotz}, {Love-Pruitt}, {Lubskiy}, {Luers}, {Luetgens}, {Luevano}, {Lui}, {Lund}, {Lundquist}, {Lunine}, {L{\"u}tzgendorf}, {Lynch}, {MacDonald}, {MacDonald}, {Macias}, {Macklis}, {Maghami}, {Maharaja}, {Maiolino},
  {Makrygiannis}, {Malla}, {Malumuth}, {Manjavacas}, {Marini}, {Marrione}, {Marston}, {Martel}, {Martin}, {Martin}, {Martinez}, {Maschmann}, {Masci}, {Masetti}, {Maszkiewicz}, {Matthews}, {Matuskey}, {McBrayer}, {McCarthy}, {McCaughrean}, {McClare}, {McClare}, {McCloskey}, {McClurg}, {McCoy}, {McElwain}, {McGregor}, {McGuffey}, {McKay}, {McKenzie}, {McLean}, {McMaster}, {McNeil}, {De Meester}, {Mehalick}, {Meixner}, {Mel{\'e}ndez}, {Menzel}, {Menzel}, {Merz}, {Mesterharm}, {Meyer}, {Meyett}, {Meza}, {Midwinter}, {Milam}, {Miller}, {Miller}, {Miskey}, {Misselt}, {Mitchell}, {Mohan}, {Montoya}, {Moran}, {Morishita}, {Moro-Mart{\'\i}n}, {Morrison}, {Morrison}, {Morse}, {Moschos}, {Moseley}, {Mosier}, {Mosner}, {Mountain}, {Muckenthaler}, {Mueller}, {Mueller}, {Muhiem}, {M{\"u}hlmann}, {Mullally}, {Mullen}, {Munger}, {Murphy}, {Murray}, {Muzerolle}, {Mycroft}, {Myers}, {Myers}, {Myers}, {Myers}, {Myrick}, {Nagle}, {Nayak}, {Naylor}, {Neff}, {Nelan}, {Nella}, {Nguyen}, {Nguyen}, {Nickson}, {Nidhiry}, {Niedner},
  {Nieto-Santisteban}, {Nikolov}, {Nishisaka}, {Noriega-Crespo}, {Nota}, {O'Mara}, {Oboryshko}, {O'Brien}, {Ochs}, {Offenberg}, {Ogle}, {Ohl}, {Olmsted}, {Osborne}, {O'Shaughnessy}, {{\"O}stlin}, {O'Sullivan}, {Otor}, {Ottens}, {Ouellette}, {Outlaw}, {Owens}, {Pacifici}, {Page}, {Paranilam}, {Park}, {Parrish}, {Paschal}, {Patapis}, {Patel}, {Patrick}, {Pattishall}, {Paul}, {Paul}, {Pauly}, {Pavlovsky}, {Pe{\~n}a-Guerrero}, {Pedder}, {Peek}, {Pelham}, {Penanen}, {Perriello}, {Perrin}, {Perrine}, {Perrygo}, {Peslier}, {Petach}, {Peterson}, {Pfarr}, {Pierson}, {Pietraszkiewicz}, {Pilchen}, {Pipher}, {Pirzkal}, {Pitman}, {Player}, {Plesha}, {Plitzke}, {Pohner}, {Poletis}, {Pollizzi}, {Polster}, {Pontius}, {Pontoppidan}, {Porges}, {Potter}, {Prescott}, {Proffitt}, {Pueyo}, {Quispe Neira}, {Radich}, {Rager}, {Rameau}, {Ramey}, {Ramos Alarcon}, {Rampini}, {Rapp}, {Rashford}, {Rauscher}, {Ravindranath}, {Rawle}, {Rawlings}, {Ray}, {Regan}, {Rehm}, {Rehm}, {Reid}, {Reis}, {Renk}, {Reoch}, {Ressler}, {Rest},
  {Reynolds}, {Richon}, {Richon}, {Ridgaway}, {Riedel}, {Rieke}, {Rieke}, {Rifelli}, {Rigby}, {Riggs}, {Ringel}, {Ritchie}, {Rix}, {Robberto}, {Robinson}, {Robinson}, {Robinson}, {Rock}, {Rodriguez}, {Rodr{\'\i}guez del Pino}, {Roellig}, {Rohrbach}, {Roman}, {Romelfanger}, {Romo}, {Rosales}, {Rose}, {Roteliuk}, {Roth}, {Rothwell}, {Rouzaud}, {Rowe}, {Rowlands}, {Roy}, {Royer}, {Rui}, {Rumler}, {Rumpl}, {Russ}, {Ryan}, {Ryan}, {Saad}, {Sabata}, {Sabatino}, {Sabbi}, {Sabelhaus}, {Sabia}, {Sahu}, {Saif}, {Salvignol}, {Samara-Ratna}, {Samuelson}, {Sanders}, {Sappington}, {Sargent}, {Sauer}, {Savadkin}, {Sawicki}, {Schappell}, {Scheffer}, {Scheithauer}, {Scherer}, {Schiff}, {Schlawin}, {Schmeitzky}, {Schmitz}, {Schmude}, {Schneider}, {Schreiber}, {Schroeven-Deceuninck}, {Schultz}, {Schwab}, {Schwartz}, {Scoccimarro}, {Scott}, {Scott}, {Seaton}, {Seely}, {Seery}, {Seidleck}, {Sembach}, {Shanahan}, {Shaughnessy}, {Shaw}, {Shay}, {Sheehan}, {Sheth}, {Shih}, {Shivaei}, {Siegel}, {Sienkiewicz}, {Simmons}, {Simon},
  {Sirianni}, {Sivaramakrishnan}, {Slade}, {Sloan}, {Slocum}, {Slowinski}, {Smith}, {Smith}, {Smith}, {Smith}, {Smith}, {Smith}, {Smolik}, {Soderblom}, {Sohn}, {Sokol}, {Sonneborn}, {Sontag}, {Sooy}, {Soummer}, {Southwood}, {Spain}, {Sparmo}, {Speer}, {Spencer}, {Sprofera}, {Stallcup}, {Stanley}, {Stansberry}, {Stark}, {Starr}, {Stassi}, {Steck}, {Steeley}, {Stephens}, {Stephenson}, {Stewart}, {Stiavelli}, {}, {Strada}, {Straughn}, {Streetman}, {Strickland}, {Strobele}, {Stuhlinger}, {Stys}, {Such}, {Sukhatme}, {Sullivan}, {Sullivan}, {Sumner}, {Sun}, {Sunnquist}, {Swade}, {Swam}, {Swenton}, {Swoish}, {Tam Litten}, {Tamas}, {Tao}, {Taylor}, {Taylor}, {te Plate}, {Van Tea}, {Teague}, {Telfer}, {Temim}, {Texter}, {Thatte}, {Thompson}, {Thompson}, {Thomson}, {Thronson}, {Tierney}, {Tikkanen}, {Tinnin}, {Tippet}, {Todd}, {Tran}, {Trauger}, {Trejo}, {Vinh Truong}, {Tsukamoto}, {Tufail}, {Tumlinson}, {Tustain}, {Tyra}, {Ubeda}, {Underwood}, {Uzzo}, {Vaclavik}, {Valenduc}, {Valenti}, {Van Campen}, {van de Wetering},
  {Van Der Marel}, {van Haarlem}, {Vandenbussche}, {van Dishoeck}, {Vanterpool}, {Vernoy}, {Vila Costas}, {Volk}, {Voorzaat}, {Voyton}, {Vydra}, {Waddy}, {Waelkens}, {Wahlgren}, {Walker}, {Wander}, {Warfield}, {Warner}, {Wasiak}, {Wasiak}, {Wehner}, {Weiler}, {Weilert}, {Weiss}, {Wells}, {Welty}, {Wheate}, {Wheeler}, {White}, {Whitehouse}, {Whiteleather}, {Whitman}, {Williams}, {Willmer}, {Willott}, {Willoughby}, {Wilson}, {Wilson}, {Wilson}, {Windhorst}, {Wislowski}, {Wolfe}, {Wolfe}, {Wolff}, {Wondel}, {Woo}, {Woods}, {Worden}, {Workman}, {Wright}, {Wu}, {Wu}, {Wun}, {Wymer}, {Yadetie}, {Yan}, {Yang}, {Yates}, {Yeager}, {Yerger}, {Young}, {Young}, {Yu}, {Yu}, {Zak}, {Zeidler}, {Zepp}, {Zhou}, {Zincke}, {Zonak}, \& {Zondag}}]{gardner2023}
{Gardner}, J.~P., {Mather}, J.~C., {Abbott}, R., {et~al.} 2023, \pasp, 135, 068001, \dodoi{10.1088/1538-3873/acd1b5}

\bibitem[{{Gehrels}(1986)}]{gehrels1986}
{Gehrels}, N. 1986, \apj, 303, 336, \dodoi{10.1086/164079}

\bibitem[{{Gelli} {et~al.}(2024){Gelli}, {Mason}, \& {Hayward}}]{gelli2024}
{Gelli}, V., {Mason}, C., \& {Hayward}, C.~C. 2024, arXiv e-prints, arXiv:2405.13108, \dodoi{10.48550/arXiv.2405.13108}

\bibitem[{{Giavalisco} {et~al.}(2004){Giavalisco}, {Ferguson}, {Koekemoer}, {Dickinson}, {Alexander}, {Bauer}, {Bergeron}, {Biagetti}, {Brandt}, {Casertano}, {Cesarsky}, {Chatzichristou}, {Conselice}, {Cristiani}, {Da Costa}, {Dahlen}, {de Mello}, {Eisenhardt}, {Erben}, {Fall}, {Fassnacht}, {Fosbury}, {Fruchter}, {Gardner}, {Grogin}, {Hook}, {Hornschemeier}, {Idzi}, {Jogee}, {Kretchmer}, {Laidler}, {Lee}, {Livio}, {Lucas}, {Madau}, {Mobasher}, {Moustakas}, {Nonino}, {Padovani}, {Papovich}, {Park}, {Ravindranath}, {Renzini}, {Richardson}, {Riess}, {Rosati}, {Schirmer}, {Schreier}, {Somerville}, {Spinrad}, {Stern}, {Stiavelli}, {Strolger}, {Urry}, {Vandame}, {Williams}, \& {Wolf}}]{giavalisco2004}
{Giavalisco}, M., {Ferguson}, H.~C., {Koekemoer}, A.~M., {et~al.} 2004, \apjl, 600, L93, \dodoi{10.1086/379232}

\bibitem[{{Gorce} {et~al.}(2018){Gorce}, {Douspis}, {Aghanim}, \& {Langer}}]{gorce2018}
{Gorce}, A., {Douspis}, M., {Aghanim}, N., \& {Langer}, M. 2018, \aap, 616, A113, \dodoi{10.1051/0004-6361/201629661}

\bibitem[{{Goto} {et~al.}(2021){Goto}, {Shimasaku}, {Yamanaka}, {Momose}, {Ando}, {Harikane}, {Hashimoto}, {Inoue}, \& {Ouchi}}]{goto2021}
{Goto}, H., {Shimasaku}, K., {Yamanaka}, S., {et~al.} 2021, \apj, 923, 229, \dodoi{10.3847/1538-4357/ac308b}

\bibitem[{{Greig} {et~al.}(2019){Greig}, {Mesinger}, \& {Ba{\~n}ados}}]{greig2019}
{Greig}, B., {Mesinger}, A., \& {Ba{\~n}ados}, E. 2019, \mnras, 484, 5094, \dodoi{10.1093/mnras/stz230}

\bibitem[{{Greig} {et~al.}(2017){Greig}, {Mesinger}, {Haiman}, \& {Simcoe}}]{greig2017}
{Greig}, B., {Mesinger}, A., {Haiman}, Z., \& {Simcoe}, R.~A. 2017, \mnras, 466, 4239, \dodoi{10.1093/mnras/stw3351}

\bibitem[{{Greig} {et~al.}(2024){Greig}, {Mesinger}, {Ba{\~n}ados}, {Becker}, {Bosman}, {Chen}, {Davies}, {D'Odorico}, {Eilers}, {Gallerani}, {Haehnelt}, {Keating}, {Lai}, {Qin}, {Ryan-Weber}, {Satyavolu}, {Wang}, {Yang}, \& {Zhu}}]{greig2024}
{Greig}, B., {Mesinger}, A., {Ba{\~n}ados}, E., {et~al.} 2024, \mnras, 530, 3208, \dodoi{10.1093/mnras/stae1080}

\bibitem[{{Gutkin} {et~al.}(2016){Gutkin}, {Charlot}, \& {Bruzual}}]{gutkin2016}
{Gutkin}, J., {Charlot}, S., \& {Bruzual}, G. 2016, \mnras, 462, 1757, \dodoi{10.1093/mnras/stw1716}

\bibitem[{{Hainline} {et~al.}(2024{\natexlab{a}}){Hainline}, {Johnson}, {Robertson}, {Tacchella}, {Helton}, {Sun}, {Eisenstein}, {Simmonds}, {Topping}, {Whitler}, {Willmer}, {Rieke}, {Suess}, {Hviding}, {Cameron}, {Alberts}, {Baker}, {Baum}, {Bhatawdekar}, {Bonaventura}, {Boyett}, {Bunker}, {Carniani}, {Charlot}, {Chevallard}, {Chen}, {Curti}, {Curtis-Lake}, {D'Eugenio}, {Egami}, {Endsley}, {Hausen}, {Ji}, {Looser}, {Lyu}, {Maiolino}, {Nelson}, {Pusk{\'a}s}, {Rawle}, {Sandles}, {Saxena}, {Smit}, {Stark}, {Williams}, {Willott}, \& {Witstok}}]{hainline2024_highz}
{Hainline}, K.~N., {Johnson}, B.~D., {Robertson}, B., {et~al.} 2024{\natexlab{a}}, \apj, 964, 71, \dodoi{10.3847/1538-4357/ad1ee4}

\bibitem[{{Hainline} {et~al.}(2024{\natexlab{b}}){Hainline}, {Helton}, {Johnson}, {Sun}, {Topping}, {Leisenring}, {Baker}, {Eisenstein}, {Hausen}, {Hviding}, {Lyu}, {Robertson}, {Tacchella}, {Williams}, {Willmer}, \& {Roellig}}]{hainline2024_bds}
{Hainline}, K.~N., {Helton}, J.~M., {Johnson}, B.~D., {et~al.} 2024{\natexlab{b}}, \apj, 964, 66, \dodoi{10.3847/1538-4357/ad20d1}

\bibitem[{{Harikane} {et~al.}(2024{\natexlab{a}}){Harikane}, {Nakajima}, {Ouchi}, {Umeda}, {Isobe}, {Ono}, {Xu}, \& {Zhang}}]{harikane2024a}
{Harikane}, Y., {Nakajima}, K., {Ouchi}, M., {et~al.} 2024{\natexlab{a}}, \apj, 960, 56, \dodoi{10.3847/1538-4357/ad0b7e}

\bibitem[{{Harikane} {et~al.}(2023){Harikane}, {Ouchi}, {Oguri}, {Ono}, {Nakajima}, {Isobe}, {Umeda}, {Mawatari}, \& {Zhang}}]{harikane2023}
{Harikane}, Y., {Ouchi}, M., {Oguri}, M., {et~al.} 2023, \apjs, 265, 5, \dodoi{10.3847/1538-4365/acaaa9}

\bibitem[{{Harikane} {et~al.}(2024{\natexlab{b}}){Harikane}, {Inoue}, {Ellis}, {Ouchi}, {Nakazato}, {Yoshida}, {Ono}, {Sun}, {Sato}, {Fujimoto}, {Kashikawa}, {McLeod}, {Perez-Gonzalez}, {Sawicki}, {Sugahara}, {Xu}, {Yamanaka}, {Carnall}, {Cullen}, {Dunlop}, {Egami}, {Grogin}, {Isobe}, {Koekemoer}, {Laporte}, {Lee}, {Magee}, {Matsuo}, {Matsuoka}, {Mawatari}, {Nakajima}, {Nakane}, {Tamura}, {Umeda}, \& {Yanagisawa}}]{harikane2024b}
{Harikane}, Y., {Inoue}, A.~K., {Ellis}, R.~S., {et~al.} 2024{\natexlab{b}}, arXiv e-prints, arXiv:2406.18352, \dodoi{10.48550/arXiv.2406.18352}

\bibitem[{{Harris} {et~al.}(2020){Harris}, {Millman}, {van der Walt}, {Gommers}, {Virtanen}, {Cournapeau}, {Wieser}, {Taylor}, {Berg}, {Smith}, {Kern}, {Picus}, {Hoyer}, {van Kerkwijk}, {Brett}, {Haldane}, {del R{\'\i}o}, {Wiebe}, {Peterson}, {G{\'e}rard-Marchant}, {Sheppard}, {Reddy}, {Weckesser}, {Abbasi}, {Gohlke}, \& {Oliphant}}]{harris2020}
{Harris}, C.~R., {Millman}, K.~J., {van der Walt}, S.~J., {et~al.} 2020, \nat, 585, 357, \dodoi{10.1038/s41586-020-2649-2}

\bibitem[{{Hastings}(1970)}]{hastings1970}
{Hastings}, W.~K. 1970, Biometrika, 57, 97, \dodoi{10.1093/biomet/57.1.97}

\bibitem[{{Hegde} {et~al.}(2024){Hegde}, {Wyatt}, \& {Furlanetto}}]{hegde2024}
{Hegde}, S., {Wyatt}, M.~M., \& {Furlanetto}, S.~R. 2024, \jcap, 2024, 025, \dodoi{10.1088/1475-7516/2024/08/025}

\bibitem[{{Heintz} {et~al.}(2024){Heintz}, {Watson}, {Brammer}, {Vejlgaard}, {Hutter}, {Strait}, {Matthee}, {Oesch}, {Jakobsson}, {Tanvir}, {Laursen}, {Naidu}, {Mason}, {Killi}, {Jung}, {Hsiao}, {Abdurro'uf}, {Coe}, {Arrabal Haro}, {Finkelstein}, \& {Toft}}]{heintz2024}
{Heintz}, K.~E., {Watson}, D., {Brammer}, G., {et~al.} 2024, Science, 384, 890, \dodoi{10.1126/science.adj0343}

\bibitem[{{Hoag} {et~al.}(2019){Hoag}, {Brada{\v{c}}}, {Huang}, {Mason}, {Treu}, {Schmidt}, {Trenti}, {Strait}, {Lemaux}, {Finney}, \& {Paddock}}]{hoag2019}
{Hoag}, A., {Brada{\v{c}}}, M., {Huang}, K., {et~al.} 2019, \apj, 878, 12, \dodoi{10.3847/1538-4357/ab1de7}

\bibitem[{{Hsiao} {et~al.}(2023){Hsiao}, {Abdurro'uf}, {Coe}, {Larson}, {Jung}, {Mingozzi}, {Dayal}, {Kumari}, {Kokorev}, {Vikaeus}, {Brammer}, {Furtak}, {Adamo}, {Andrade-Santos}, {Antwi-Danso}, {Bradac}, {Bradley}, {Broadhurst}, {Carnall}, {Conselice}, {Diego}, {Donahue}, {Eldridge}, {Fujimoto}, {Henry}, {Hernandez}, {Hutchison}, {James}, {Norman}, {Park}, {Pirzkal}, {Postman}, {Ricotti}, {Rigby}, {Vanzella}, {Welch}, {Wilkins}, {Windhorst}, {Xu}, {Zackrisson}, \& {Zitrin}}]{hsiao2023}
{Hsiao}, T. Y.-Y., {Abdurro'uf}, {Coe}, D., {et~al.} 2023, arXiv e-prints, arXiv:2305.03042, \dodoi{10.48550/arXiv.2305.03042}

\bibitem[{{Hunter}(2007)}]{hunter2007}
{Hunter}, J.~D. 2007, Computing in Science and Engineering, 9, 90, \dodoi{10.1109/MCSE.2007.55}

\bibitem[{{Illingworth} {et~al.}(2016){Illingworth}, {Magee}, {Bouwens}, {Oesch}, {Labbe}, {van Dokkum}, {Whitaker}, {Holden}, {Franx}, \& {Gonzalez}}]{illingworth2016}
{Illingworth}, G., {Magee}, D., {Bouwens}, R., {et~al.} 2016, arXiv e-prints, arXiv:1606.00841, \dodoi{10.48550/arXiv.1606.00841}

\bibitem[{{Inoue} {et~al.}(2014){Inoue}, {Shimizu}, {Iwata}, \& {Tanaka}}]{inoue2014}
{Inoue}, A.~K., {Shimizu}, I., {Iwata}, I., \& {Tanaka}, M. 2014, \mnras, 442, 1805, \dodoi{10.1093/mnras/stu936}

\bibitem[{{Inoue} {et~al.}(2018){Inoue}, {Hasegawa}, {Ishiyama}, {Yajima}, {Shimizu}, {Umemura}, {Konno}, {Harikane}, {Shibuya}, {Ouchi}, {Shimasaku}, {Ono}, {Kusakabe}, {Higuchi}, \& {Lee}}]{inoue2018}
{Inoue}, A.~K., {Hasegawa}, K., {Ishiyama}, T., {et~al.} 2018, \pasj, 70, 55, \dodoi{10.1093/pasj/psy048}

\bibitem[{{Izotov} {et~al.}(2021){Izotov}, {Worseck}, {Schaerer}, {Guseva}, {Chisholm}, {Thuan}, {Fricke}, \& {Verhamme}}]{izotov2021}
{Izotov}, Y.~I., {Worseck}, G., {Schaerer}, D., {et~al.} 2021, \mnras, 503, 1734, \dodoi{10.1093/mnras/stab612}

\bibitem[{{Jakobsen} {et~al.}(2022){Jakobsen}, {Ferruit}, {Alves de Oliveira}, {Arribas}, {Bagnasco}, {Barho}, {Beck}, {Birkmann}, {B{\"o}ker}, {Bunker}, {Charlot}, {de Jong}, {de Marchi}, {Ehrenwinkler}, {Falcolini}, {Fels}, {Franx}, {Franz}, {Funke}, {Giardino}, {Gnata}, {Holota}, {Honnen}, {Jensen}, {Jentsch}, {Johnson}, {Jollet}, {Karl}, {Kling}, {K{\"o}hler}, {Kolm}, {Kumari}, {Lander}, {Lemke}, {L{\'o}pez-Caniego}, {L{\"u}tzgendorf}, {Maiolino}, {Manjavacas}, {Marston}, {Maschmann}, {Maurer}, {Messerschmidt}, {Moseley}, {Mosner}, {Mott}, {Muzerolle}, {Pirzkal}, {Pittet}, {Plitzke}, {Posselt}, {Rapp}, {Rauscher}, {Rawle}, {Rix}, {R{\"o}del}, {Rumler}, {Sabbi}, {Salvignol}, {Schmid}, {Sirianni}, {Smith}, {Strada}, {te Plate}, {Valenti}, {Wettemann}, {Wiehe}, {Wiesmayer}, {Willott}, {Wright}, {Zeidler}, \& {Zincke}}]{jakobsen2022}
{Jakobsen}, P., {Ferruit}, P., {Alves de Oliveira}, C., {et~al.} 2022, \aap, 661, A80, \dodoi{10.1051/0004-6361/202142663}

\bibitem[{{Ji} {et~al.}(2023){Ji}, {Williams}, {Tacchella}, {Suess}, {Baker}, {Alberts}, {Bunker}, {Johnson}, {Robertson}, {Sun}, {Eisenstein}, {Rieke}, {Maseda}, {Hainline}, {Hausen}, {Rieke}, {Willmer}, {Egami}, {Shivaei}, {Carniani}, {Charlot}, {Chevallard}, {Curtis-Lake}, {Looser}, {Maiolino}, {Willott}, {Chen}, {Helton}, {Lyu}, {Nelson}, {Bhatawdekar}, {Boyett}, \& {Sandles}}]{ji2023}
{Ji}, Z., {Williams}, C.~C., {Tacchella}, S., {et~al.} 2023, arXiv e-prints, arXiv:2305.18518, \dodoi{10.48550/arXiv.2305.18518}

\bibitem[{{Jin} {et~al.}(2023){Jin}, {Yang}, {Fan}, {Wang}, {Ba{\~n}ados}, {Bian}, {Davies}, {Eilers}, {Farina}, {Hennawi}, {Pacucci}, {Venemans}, \& {Walter}}]{jin2023}
{Jin}, X., {Yang}, J., {Fan}, X., {et~al.} 2023, \apj, 942, 59, \dodoi{10.3847/1538-4357/aca678}

\bibitem[{{Jones} {et~al.}(2024{\natexlab{a}}){Jones}, {Bunker}, {Saxena}, {Witstok}, {Stark}, {Arribas}, {Baker}, {Bhatawdekar}, {Bowler}, {Boyett}, {Cameron}, {Carniani}, {Charlot}, {Chevallard}, {Curti}, {Curtis-Lake}, {Eisenstein}, {Hainline}, {Hausen}, {Ji}, {Johnson}, {Kumari}, {Looser}, {Maiolino}, {Maseda}, {Parlanti}, {Rix}, {Robertson}, {Sandles}, {Scholtz}, {Smit}, {Tacchella}, {{\"U}bler}, {Williams}, \& {Willott}}]{jones2024a}
{Jones}, G.~C., {Bunker}, A.~J., {Saxena}, A., {et~al.} 2024{\natexlab{a}}, \aap, 683, A238, \dodoi{10.1051/0004-6361/202347099}

\bibitem[{{Jones} {et~al.}(2024{\natexlab{b}}){Jones}, {Bunker}, {Saxena}, {Arribas}, {Bhatawdekar}, {Boyett}, {Cameron}, {Carniani}, {Charlot}, {Curtis-Lake}, {Hainline}, {Johnson}, {Kumari}, {Maseda}, {Rix}, {Robertson}, {Tacchella}, {{\"U}bler}, {Williams}, {Willott}, {Witstok}, \& {Zhu}}]{jones2024b}
---. 2024{\natexlab{b}}, arXiv e-prints, arXiv:2409.06405, \dodoi{10.48550/arXiv.2409.06405}

\bibitem[{{Jung} {et~al.}(2020){Jung}, {Finkelstein}, {Dickinson}, {Hutchison}, {Larson}, {Papovich}, {Pentericci}, {Straughn}, {Guo}, {Malhotra}, {Rhoads}, {Song}, {Tilvi}, \& {Wold}}]{jung2020}
{Jung}, I., {Finkelstein}, S.~L., {Dickinson}, M., {et~al.} 2020, \apj, 904, 144, \dodoi{10.3847/1538-4357/abbd44}

\bibitem[{{Kannan} {et~al.}(2022){Kannan}, {Garaldi}, {Smith}, {Pakmor}, {Springel}, {Vogelsberger}, \& {Hernquist}}]{kannan2022}
{Kannan}, R., {Garaldi}, E., {Smith}, A., {et~al.} 2022, \mnras, 511, 4005, \dodoi{10.1093/mnras/stab3710}

\bibitem[{{Kaurov} \& {Gnedin}(2015)}]{kaurov2015}
{Kaurov}, A.~A., \& {Gnedin}, N.~Y. 2015, \apj, 810, 154, \dodoi{10.1088/0004-637X/810/2/154}

\bibitem[{{Kokorev} {et~al.}(2024){Kokorev}, {Atek}, {Chisholm}, {Endsley}, {Chemerynska}, {Mu{\~n}oz}, {Furtak}, {Pan}, {Berg}, {Fujimoto}, {Oesch}, {Weibel}, {Adamo}, {Blaizot}, {Bouwens}, {Dessauges-Zavadsky}, {Khullar}, {Korber}, {Goovaerts}, {Jecmen}, {Labb{\'e}}, {Leclercq}, {Marques-Chaves}, {Mason}, {McQuinn}, {Naidu}, {Natarajan}, {Nelson}, {Rosdahl}, {Saldana-Lopez}, {Schaerer}, {Trebitsch}, {Volonteri}, \& {Zitrin}}]{kokorev2024}
{Kokorev}, V., {Atek}, H., {Chisholm}, J., {et~al.} 2024, arXiv e-prints, arXiv:2411.13640, \dodoi{10.48550/arXiv.2411.13640}

\bibitem[{{Konno} {et~al.}(2014){Konno}, {Ouchi}, {Ono}, {Shimasaku}, {Shibuya}, {Furusawa}, {Nakajima}, {Naito}, {Momose}, {Yuma}, \& {Iye}}]{konno2014}
{Konno}, A., {Ouchi}, M., {Ono}, Y., {et~al.} 2014, \apj, 797, 16, \dodoi{10.1088/0004-637X/797/1/16}

\bibitem[{{Kravtsov} \& {Belokurov}(2024)}]{kravtsov2024}
{Kravtsov}, A., \& {Belokurov}, V. 2024, arXiv e-prints, arXiv:2405.04578, \dodoi{10.48550/arXiv.2405.04578}

\bibitem[{{Kron}(1980)}]{kron1980}
{Kron}, R.~G. 1980, \apjs, 43, 305, \dodoi{10.1086/190669}

\bibitem[{{Lam} {et~al.}(2019){Lam}, {Bouwens}, {Labb{\'e}}, {Schaye}, {Schmidt}, {Maseda}, {Bacon}, {Boogaard}, {Nanayakkara}, {Richard}, {Mahler}, \& {Urrutia}}]{lam2019}
{Lam}, D., {Bouwens}, R.~J., {Labb{\'e}}, I., {et~al.} 2019, \aap, 627, A164, \dodoi{10.1051/0004-6361/201935227}

\bibitem[{{Leethochawalit} {et~al.}(2023){Leethochawalit}, {Roberts-Borsani}, {Morishita}, {Trenti}, \& {Treu}}]{leethochawalit2023_hst_uvlf}
{Leethochawalit}, N., {Roberts-Borsani}, G., {Morishita}, T., {Trenti}, M., \& {Treu}, T. 2023, \mnras, 524, 5454, \dodoi{10.1093/mnras/stad2202}

\bibitem[{{Leung} {et~al.}(2023){Leung}, {Bagley}, {Finkelstein}, {Ferguson}, {Koekemoer}, {P{\'e}rez-Gonz{\'a}lez}, {Morales}, {Kocevski}, {Yang}, {Somerville}, {Wilkins}, {Yung}, {Fujimoto}, {Larson}, {Papovich}, {Pirzkal}, {Berg}, {Lotz}, {Castellano}, {Ch{\'a}vez Ortiz}, {Cheng}, {Dickinson}, {Giavalisco}, {Hathi}, {Hutchison}, {Jung}, {Kartaltepe}, {Natarajan}, \& {Rothberg}}]{leung2023}
{Leung}, G. C.~K., {Bagley}, M.~B., {Finkelstein}, S.~L., {et~al.} 2023, \apjl, 954, L46, \dodoi{10.3847/2041-8213/acf365}

\bibitem[{{Li} {et~al.}(2024){Li}, {Dekel}, {Sarkar}, {Aung}, {Giavalisco}, {Mandelker}, \& {Tacchella}}]{li2024}
{Li}, Z., {Dekel}, A., {Sarkar}, K.~C., {et~al.} 2024, \aap, 690, A108, \dodoi{10.1051/0004-6361/202348727}

\bibitem[{{Livermore} {et~al.}(2017){Livermore}, {Finkelstein}, \& {Lotz}}]{livermore2017}
{Livermore}, R.~C., {Finkelstein}, S.~L., \& {Lotz}, J.~M. 2017, \apj, 835, 113, \dodoi{10.3847/1538-4357/835/2/113}

\bibitem[{{Looser} {et~al.}(2023){Looser}, {D'Eugenio}, {Maiolino}, {Tacchella}, {Curti}, {Arribas}, {Baker}, {Baum}, {Bonaventura}, {Boyett}, {Bunker}, {Carniani}, {Charlot}, {Chevallard}, {Curtis-Lake}, {Danhaive}, {Eisenstein}, {de Graaff}, {Hainline}, {Ji}, {Johnson}, {Kumari}, {Nelson}, {Parlanti}, {Rix}, {Robertson}, {Rodr{\'\i}guez Del Pino}, {Sandles}, {Scholtz}, {Smit}, {Stark}, {{\"U}bler}, {Williams}, {Willott}, \& {Witstok}}]{looser2023}
{Looser}, T.~J., {D'Eugenio}, F., {Maiolino}, R., {et~al.} 2023, arXiv e-prints, arXiv:2306.02470, \dodoi{10.48550/arXiv.2306.02470}

\bibitem[{{Looser} {et~al.}(2024){Looser}, {D'Eugenio}, {Maiolino}, {Witstok}, {Sandles}, {Curtis-Lake}, {Chevallard}, {Tacchella}, {Johnson}, {Baker}, {Suess}, {Carniani}, {Ferruit}, {Arribas}, {Bonaventura}, {Bunker}, {Cameron}, {Charlot}, {Curti}, {de Graaff}, {Maseda}, {Rawle}, {Rix}, {Del Pino}, {Smit}, {{\"U}bler}, {Willott}, {Alberts}, {Egami}, {Eisenstein}, {Endsley}, {Hausen}, {Rieke}, {Robertson}, {Shivaei}, {Williams}, {Boyett}, {Chen}, {Ji}, {Jones}, {Kumari}, {Nelson}, {Perna}, {Saxena}, \& {Scholtz}}]{looser2024}
---. 2024, \nat, 629, 53, \dodoi{10.1038/s41586-024-07227-0}

\bibitem[{{Lovell} {et~al.}(2021){Lovell}, {Vijayan}, {Thomas}, {Wilkins}, {Barnes}, {Irodotou}, \& {Roper}}]{lovell2021}
{Lovell}, C.~C., {Vijayan}, A.~P., {Thomas}, P.~A., {et~al.} 2021, \mnras, 500, 2127, \dodoi{10.1093/mnras/staa3360}

\bibitem[{{Madau} \& {Dickinson}(2014)}]{madau2014}
{Madau}, P., \& {Dickinson}, M. 2014, \araa, 52, 415, \dodoi{10.1146/annurev-astro-081811-125615}

\bibitem[{{Madau} {et~al.}(1999){Madau}, {Haardt}, \& {Rees}}]{madau1999}
{Madau}, P., {Haardt}, F., \& {Rees}, M.~J. 1999, \apj, 514, 648, \dodoi{10.1086/306975}

\bibitem[{{Mascia} {et~al.}(2024){Mascia}, {Pentericci}, {Calabr{\`o}}, {Santini}, {Napolitano}, {Arrabal Haro}, {Castellano}, {Dickinson}, {Ocvirk}, {Lewis}, {Amor{\'\i}n}, {Bagley}, {Bhatawdekar}, {Cleri}, {Costantin}, {Dekel}, {Finkelstein}, {Fontana}, {Giavalisco}, {Grogin}, {Hathi}, {Hirschmann}, {Holwerda}, {Jung}, {Kartaltepe}, {Koekemoer}, {Lucas}, {Papovich}, {P{\'e}rez-Gonz{\'a}lez}, {Pirzkal}, {Trump}, {Wilkins}, \& {Yung}}]{mascia2024}
{Mascia}, S., {Pentericci}, L., {Calabr{\`o}}, A., {et~al.} 2024, \aap, 685, A3, \dodoi{10.1051/0004-6361/202347884}

\bibitem[{{Mason} {et~al.}(2023){Mason}, {Trenti}, \& {Treu}}]{mason2023}
{Mason}, C.~A., {Trenti}, M., \& {Treu}, T. 2023, \mnras, 521, 497, \dodoi{10.1093/mnras/stad035}

\bibitem[{{Mason} {et~al.}(2018){Mason}, {Treu}, {Dijkstra}, {Mesinger}, {Trenti}, {Pentericci}, {de Barros}, \& {Vanzella}}]{mason2018a}
{Mason}, C.~A., {Treu}, T., {Dijkstra}, M., {et~al.} 2018, \apj, 856, 2, \dodoi{10.3847/1538-4357/aab0a7}

\bibitem[{{Mason} {et~al.}(2019){Mason}, {Fontana}, {Treu}, {Schmidt}, {Hoag}, {Abramson}, {Amorin}, {Brada{\v{c}}}, {Guaita}, {Jones}, {Henry}, {Malkan}, {Pentericci}, {Trenti}, \& {Vanzella}}]{mason2019_kmos}
{Mason}, C.~A., {Fontana}, A., {Treu}, T., {et~al.} 2019, \mnras, 485, 3947, \dodoi{10.1093/mnras/stz632}

\bibitem[{{McGreer} {et~al.}(2015){McGreer}, {Mesinger}, \& {D'Odorico}}]{mcgreer2015}
{McGreer}, I.~D., {Mesinger}, A., \& {D'Odorico}, V. 2015, \mnras, 447, 499, \dodoi{10.1093/mnras/stu2449}

\bibitem[{{McLeod} {et~al.}(2016){McLeod}, {McLure}, \& {Dunlop}}]{mcleod2016}
{McLeod}, D.~J., {McLure}, R.~J., \& {Dunlop}, J.~S. 2016, \mnras, 459, 3812, \dodoi{10.1093/mnras/stw904}

\bibitem[{{McLeod} {et~al.}(2024){McLeod}, {Donnan}, {McLure}, {Dunlop}, {Magee}, {Begley}, {Carnall}, {Cullen}, {Ellis}, {Hamadouche}, \& {Stanton}}]{mcleod2024}
{McLeod}, D.~J., {Donnan}, C.~T., {McLure}, R.~J., {et~al.} 2024, \mnras, 527, 5004, \dodoi{10.1093/mnras/stad3471}

\bibitem[{{McLure} {et~al.}(2013){McLure}, {Dunlop}, {Bowler}, {Curtis-Lake}, {Schenker}, {Ellis}, {Robertson}, {Koekemoer}, {Rogers}, {Ono}, {Ouchi}, {Charlot}, {Wild}, {Stark}, {Furlanetto}, {Cirasuolo}, \& {Targett}}]{mclure2013}
{McLure}, R.~J., {Dunlop}, J.~S., {Bowler}, R.~A.~A., {et~al.} 2013, \mnras, 432, 2696, \dodoi{10.1093/mnras/stt627}

\bibitem[{{Metropolis} {et~al.}(1953){Metropolis}, {Rosenbluth}, {Rosenbluth}, {Teller}, \& {Teller}}]{metropolis1953}
{Metropolis}, N., {Rosenbluth}, A.~W., {Rosenbluth}, M.~N., {Teller}, A.~H., \& {Teller}, E. 1953, \jcp, 21, 1087, \dodoi{10.1063/1.1699114}

\bibitem[{{Mirocha} \& {Furlanetto}(2023)}]{mirocha2023}
{Mirocha}, J., \& {Furlanetto}, S.~R. 2023, \mnras, 519, 843, \dodoi{10.1093/mnras/stac3578}

\bibitem[{{Morales} {et~al.}(2021){Morales}, {Mason}, {Bruton}, {Gronke}, {Haardt}, \& {Scarlata}}]{morales2021}
{Morales}, A.~M., {Mason}, C.~A., {Bruton}, S., {et~al.} 2021, \apj, 919, 120, \dodoi{10.3847/1538-4357/ac1104}

\bibitem[{{Morishita} {et~al.}(2023){Morishita}, {Roberts-Borsani}, {Treu}, {Brammer}, {Mason}, {Trenti}, {Vulcani}, {Wang}, {Acebron}, {Bah{\'e}}, {Bergamini}, {Boyett}, {Bradac}, {Calabr{\`o}}, {Castellano}, {Chen}, {De Lucia}, {Filippenko}, {Fontana}, {Glazebrook}, {Grillo}, {Henry}, {Jones}, {Kelly}, {Koekemoer}, {Leethochawalit}, {Lu}, {Marchesini}, {Mascia}, {Mercurio}, {Merlin}, {Metha}, {Nanayakkara}, {Nonino}, {Paris}, {Pentericci}, {Rosati}, {Santini}, {Strait}, {Vanzella}, {Windhorst}, \& {Xie}}]{morishita2023}
{Morishita}, T., {Roberts-Borsani}, G., {Treu}, T., {et~al.} 2023, \apjl, 947, L24, \dodoi{10.3847/2041-8213/acb99e}

\bibitem[{{Morishita} {et~al.}(2024){Morishita}, {Stiavelli}, {Chary}, {Trenti}, {Bergamini}, {Chiaberge}, {Leethochawalit}, {Roberts-Borsani}, {Shen}, \& {Treu}}]{morishita2024}
{Morishita}, T., {Stiavelli}, M., {Chary}, R.-R., {et~al.} 2024, \apj, 963, 9, \dodoi{10.3847/1538-4357/ad1404}

\bibitem[{{Mu{\~n}oz} {et~al.}(2024){Mu{\~n}oz}, {Mirocha}, {Chisholm}, {Furlanetto}, \& {Mason}}]{munoz2024}
{Mu{\~n}oz}, J.~B., {Mirocha}, J., {Chisholm}, J., {Furlanetto}, S.~R., \& {Mason}, C. 2024, arXiv e-prints, arXiv:2404.07250, \dodoi{10.48550/arXiv.2404.07250}

\bibitem[{{Naidu} {et~al.}(2022){Naidu}, {Oesch}, {van Dokkum}, {Nelson}, {Suess}, {Brammer}, {Whitaker}, {Illingworth}, {Bouwens}, {Tacchella}, {Matthee}, {Allen}, {Bezanson}, {Conroy}, {Labbe}, {Leja}, {Leonova}, {Magee}, {Price}, {Setton}, {Strait}, {Stefanon}, {Toft}, {Weaver}, \& {Weibel}}]{naidu2022}
{Naidu}, R.~P., {Oesch}, P.~A., {van Dokkum}, P., {et~al.} 2022, \apjl, 940, L14, \dodoi{10.3847/2041-8213/ac9b22}

\bibitem[{{Nakane} {et~al.}(2024){Nakane}, {Ouchi}, {Nakajima}, {Harikane}, {Ono}, {Umeda}, {Isobe}, {Zhang}, \& {Xu}}]{nakane2024}
{Nakane}, M., {Ouchi}, M., {Nakajima}, K., {et~al.} 2024, \apj, 967, 28, \dodoi{10.3847/1538-4357/ad38c2}

\bibitem[{{Napolitano} {et~al.}(2024{\natexlab{a}}){Napolitano}, {Castellano}, {Pentericci}, {Arrabal Haro}, {Fontana}, {Treu}, {Bergamini}, {Calabro}, {Mascia}, {Morishita}, {Roberts-Borsani}, {Santini}, {Vanzella}, {Vulcani}, {Zakharova}, {Bakx}, {Dickinson}, {Grillo}, {Leethochawalit}, {Llerena}, {Merlin}, {Paris}, {Rojas-Ruiz}, {Rosati}, {Wang}, {Yoon}, \& {Zavala}}]{napolitano2024}
{Napolitano}, L., {Castellano}, M., {Pentericci}, L., {et~al.} 2024{\natexlab{a}}, arXiv e-prints, arXiv:2410.10967, \dodoi{10.48550/arXiv.2410.10967}

\bibitem[{{Napolitano} {et~al.}(2024{\natexlab{b}}){Napolitano}, {Pentericci}, {Santini}, {Calabr{\`o}}, {Mascia}, {Llerena}, {Castellano}, {Dickinson}, {Finkelstein}, {Amor{\'\i}n}, {Arrabal Haro}, {Bagley}, {Bhatawdekar}, {Cleri}, {Davis}, {Gardner}, {Gawiser}, {Giavalisco}, {Hathi}, {Holwerda}, {Hu}, {Jung}, {Kartaltepe}, {Koekemoer}, {Larson}, {Merlin}, {Mobasher}, {Papovich}, {Park}, {Pirzkal}, {Trump}, {Wilkins}, \& {Yung}}]{napolitano2024_lya}
{Napolitano}, L., {Pentericci}, L., {Santini}, P., {et~al.} 2024{\natexlab{b}}, \aap, 688, A106, \dodoi{10.1051/0004-6361/202449644}

\bibitem[{{Ning} {et~al.}(2022){Ning}, {Jiang}, {Zheng}, \& {Wu}}]{ning2022}
{Ning}, Y., {Jiang}, L., {Zheng}, Z.-Y., \& {Wu}, J. 2022, \apj, 926, 230, \dodoi{10.3847/1538-4357/ac4268}

\bibitem[{{Oesch} {et~al.}(2018){Oesch}, {Bouwens}, {Illingworth}, {Labb{\'e}}, \& {Stefanon}}]{oesch2018}
{Oesch}, P.~A., {Bouwens}, R.~J., {Illingworth}, G.~D., {Labb{\'e}}, I., \& {Stefanon}, M. 2018, \apj, 855, 105, \dodoi{10.3847/1538-4357/aab03f}

\bibitem[{{Oesch} {et~al.}(2016){Oesch}, {Brammer}, {van Dokkum}, {Illingworth}, {Bouwens}, {Labb{\'e}}, {Franx}, {Momcheva}, {Ashby}, {Fazio}, {Gonzalez}, {Holden}, {Magee}, {Skelton}, {Smit}, {Spitler}, {Trenti}, \& {Willner}}]{oesch2016}
{Oesch}, P.~A., {Brammer}, G., {van Dokkum}, P.~G., {et~al.} 2016, \apj, 819, 129, \dodoi{10.3847/0004-637X/819/2/129}

\bibitem[{{Oesch} {et~al.}(2023){Oesch}, {Brammer}, {Naidu}, {Bouwens}, {Chisholm}, {Illingworth}, {Matthee}, {Nelson}, {Qin}, {Reddy}, {Shapley}, {Shivaei}, {van Dokkum}, {Weibel}, {Whitaker}, {Wuyts}, {Covelo-Paz}, {Endsley}, {Fudamoto}, {Giovinazzo}, {Herard-Demanche}, {Kerutt}, {Kramarenko}, {Labbe}, {Leonova}, {Lin}, {Magee}, {Marchesini}, {Maseda}, {Mason}, {Matharu}, {Meyer}, {Neufeld}, {Prieto Lyon}, {Schaerer}, {Sharma}, {Shuntov}, {Smit}, {Stefanon}, {Wyithe}, \& {Xiao}}]{oesch2023_fresco}
{Oesch}, P.~A., {Brammer}, G., {Naidu}, R.~P., {et~al.} 2023, \mnras, 525, 2864, \dodoi{10.1093/mnras/stad2411}

\bibitem[{{Oke} \& {Gunn}(1983)}]{oke1983}
{Oke}, J.~B., \& {Gunn}, J.~E. 1983, \apj, 266, 713, \dodoi{10.1086/160817}

\bibitem[{{Ono} {et~al.}(2023){Ono}, {Harikane}, {Ouchi}, {Yajima}, {Abe}, {Isobe}, {Shibuya}, {Wise}, {Zhang}, {Nakajima}, \& {Umeda}}]{ono2023}
{Ono}, Y., {Harikane}, Y., {Ouchi}, M., {et~al.} 2023, \apj, 951, 72, \dodoi{10.3847/1538-4357/acd44a}

\bibitem[{{Ono} {et~al.}(2024){Ono}, {Harikane}, {Ouchi}, {Nakajima}, {Isobe}, {Shibuya}, {Nakane}, {Umeda}, {Xu}, \& {Zhang}}]{ono2024}
---. 2024, \pasj, 76, 219, \dodoi{10.1093/pasj/psae004}

\bibitem[{{Ouchi} {et~al.}(2010){Ouchi}, {Shimasaku}, {Furusawa}, {Saito}, {Yoshida}, {Akiyama}, {Ono}, {Yamada}, {Ota}, {Kashikawa}, {Iye}, {Kodama}, {Okamura}, {Simpson}, \& {Yoshida}}]{ouchi2010}
{Ouchi}, M., {Shimasaku}, K., {Furusawa}, H., {et~al.} 2010, \apj, 723, 869, \dodoi{10.1088/0004-637X/723/1/869}

\bibitem[{{Ouchi} {et~al.}(2018){Ouchi}, {Harikane}, {Shibuya}, {Shimasaku}, {Taniguchi}, {Konno}, {Kobayashi}, {Kajisawa}, {Nagao}, {Ono}, {Inoue}, {Umemura}, {Mori}, {Hasegawa}, {Higuchi}, {Komiyama}, {Matsuda}, {Nakajima}, {Saito}, \& {Wang}}]{ouchi2018}
{Ouchi}, M., {Harikane}, Y., {Shibuya}, T., {et~al.} 2018, \pasj, 70, S13, \dodoi{10.1093/pasj/psx074}

\bibitem[{{Pahl} {et~al.}(2025){Pahl}, {Topping}, {Shapley}, {Sanders}, {Reddy}, {Clarke}, {Kehoe}, {Bento}, \& {Brammer}}]{pahl2025}
{Pahl}, A., {Topping}, M.~W., {Shapley}, A., {et~al.} 2025, \apj, 981, 134, \dodoi{10.3847/1538-4357/adb1ab}

\bibitem[{{Pei}(1992)}]{pei1992}
{Pei}, Y.~C. 1992, \apj, 395, 130, \dodoi{10.1086/171637}

\bibitem[{{P{\'e}rez-Gonz{\'a}lez} {et~al.}(2023){P{\'e}rez-Gonz{\'a}lez}, {Costantin}, {Langeroodi}, {Rinaldi}, {Annunziatella}, {Ilbert}, {Colina}, {N{\o}rgaard-Nielsen}, {Greve}, {{\"O}stlin}, {Wright}, {Alonso-Herrero}, {{\'A}lvarez-M{\'a}rquez}, {Caputi}, {Eckart}, {Le F{\`e}vre}, {Labiano}, {Garc{\'\i}a-Mar{\'\i}n}, {Hjorth}, {Kendrew}, {Pye}, {Tikkanen}, {van der Werf}, {Walter}, {Ward}, {Bik}, {Boogaard}, {Bosman}, {G{\'o}mez}, {Gillman}, {Iani}, {Jermann}, {Melinder}, {Meyer}, {Moutard}, {van Dishoek}, {Henning}, {Lagage}, {Guedel}, {Peissker}, {Ray}, {Vandenbussche}, {Garc{\'\i}a-Argum{\'a}nez}, \& {Mar{\'\i}a M{\'e}rida}}]{perez-gonzalez2023}
{P{\'e}rez-Gonz{\'a}lez}, P.~G., {Costantin}, L., {Langeroodi}, D., {et~al.} 2023, \apjl, 951, L1, \dodoi{10.3847/2041-8213/acd9d0}

\bibitem[{{P{\'e}rez-Gonz{\'a}lez} {et~al.}(2025){P{\'e}rez-Gonz{\'a}lez}, {{\"O}stlin}, {Costantin}, {Melinder}, {Finkelstein}, {Somerville}, {Annunziatella}, {{\'A}lvarez-M{\'a}rquez}, {Colina}, {Dekel}, {Dickinson}, {Ferguson}, {Li}, {Yung}, {Bagley}, {Boogard}, {Burgarella}, {Calabr{\`o}}, {Caputi}, {Cheng}, {Eckart}, {Giavalisco}, {Gillman}, {Greve}, {Hamed}, {Hathi}, {Hjorth}, {Huertas-Company}, {Kartaltepe}, {Koekemoer}, {Kokorev}, {Labiano}, {Langeroodi}, {Leung}, {Natarajan}, {Papovich}, {Peissker}, {Pentericci}, {Pirzkal}, {Rinaldi}, {van der Werf}, \& {Walter}}]{perez-gonzalez2025}
{P{\'e}rez-Gonz{\'a}lez}, P.~G., {{\"O}stlin}, G., {Costantin}, L., {et~al.} 2025, arXiv e-prints, arXiv:2503.15594, \dodoi{10.48550/arXiv.2503.15594}

\bibitem[{{Rieke} {et~al.}(2005){Rieke}, {Kelly}, \& {Horner}}]{rieke2005}
{Rieke}, M.~J., {Kelly}, D., \& {Horner}, S. 2005, in Society of Photo-Optical Instrumentation Engineers (SPIE) Conference Series, Vol. 5904, Cryogenic Optical Systems and Instruments XI, ed. J.~B. {Heaney} \& L.~G. {Burriesci}, 1--8, \dodoi{10.1117/12.615554}

\bibitem[{{Rieke} {et~al.}(2023{\natexlab{a}}){Rieke}, {Kelly}, {Misselt}, {Stansberry}, {Boyer}, {Beatty}, {Egami}, {Florian}, {Greene}, {Hainline}, {Leisenring}, {Roellig}, {Schlawin}, {Sun}, {Tinnin}, {Williams}, {Willmer}, {Wilson}, {Clark}, {Rohrbach}, {Brooks}, {Canipe}, {Correnti}, {DiFelice}, {Gennaro}, {Girard}, {Hartig}, {Hilbert}, {Koekemoer}, {Nikolov}, {Pirzkal}, {Rest}, {Robberto}, {Sunnquist}, {Telfer}, {Wu}, {Ferry}, {Lewis}, {Baum}, {Beichman}, {Doyon}, {Dressler}, {Eisenstein}, {Ferrarese}, {Hodapp}, {Horner}, {Jaffe}, {Johnstone}, {Krist}, {Martin}, {McCarthy}, {Meyer}, {Rieke}, {Trauger}, \& {Young}}]{rieke2023_nircam}
{Rieke}, M.~J., {Kelly}, D.~M., {Misselt}, K., {et~al.} 2023{\natexlab{a}}, \pasp, 135, 028001, \dodoi{10.1088/1538-3873/acac53}

\bibitem[{{Rieke} {et~al.}(2023{\natexlab{b}}){Rieke}, {Robertson}, {Tacchella}, {Hainline}, {Johnson}, {Hausen}, {Ji}, {Willmer}, {Eisenstein}, {Pusk{\'a}s}, {Alberts}, {Arribas}, {Baker}, {Baum}, {Bhatawdekar}, {Bonaventura}, {Boyett}, {Bunker}, {Cameron}, {Carniani}, {Charlot}, {Chevallard}, {Chen}, {Curti}, {Curtis-Lake}, {Danhaive}, {DeCoursey}, {Dressler}, {Egami}, {Endsley}, {Helton}, {Hviding}, {Kumari}, {Looser}, {Lyu}, {Maiolino}, {Maseda}, {Nelson}, {Rieke}, {Rix}, {Sandles}, {Saxena}, {Sharpe}, {Shivaei}, {Skarbinski}, {Smit}, {Stark}, {Stone}, {Suess}, {Sun}, {Topping}, {{\"U}bler}, {Villanueva}, {Wallace}, {Williams}, {Willott}, {Whitler}, {Witstok}, \& {Woodrum}}]{rieke2023_jades}
{Rieke}, M.~J., {Robertson}, B., {Tacchella}, S., {et~al.} 2023{\natexlab{b}}, \apjs, 269, 16, \dodoi{10.3847/1538-4365/acf44d}

\bibitem[{{Rigby} {et~al.}(2023){Rigby}, {Perrin}, {McElwain}, {Kimble}, {Friedman}, {Lallo}, {Doyon}, {Feinberg}, {Ferruit}, {Glasse}, {Rieke}, {Rieke}, {Wright}, {Willott}, {Colon}, {Milam}, {Neff}, {Stark}, {Valenti}, {Abell}, {Abney}, {Abul-Huda}, {Acton}, {Adams}, {Adler}, {Aguilar}, {Ahmed}, {Albert}, {Alberts}, {Aldridge}, {Allen}, {Altenburg}, {{\'A}lvarez-M{\'a}rquez}, {Alves de Oliveira}, {Andersen}, {Anderson}, {Anderson}, {Argyriou}, {Armstrong}, {Arribas}, {Artigau}, {Arvai}, {Atkinson}, {Bacon}, {Bair}, {Banks}, {Barrientes}, {Barringer}, {Bartosik}, {Bast}, {Baudoz}, {Beatty}, {Bechtold}, {Beck}, {Bergeron}, {Bergkoetter}, {Bhatawdekar}, {Birkmann}, {Blazek}, {Blome}, {Boccaletti}, {B{\"o}ker}, {Boia}, {Bonaventura}, {Bond}, {Bosley}, {Boucarut}, {Bourque}, {Bouwman}, {Bower}, {Bowers}, {Boyer}, {Bradley}, {Brady}, {Braun}, {Breda}, {Bresnahan}, {Bright}, {Britt}, {Bromenschenkel}, {Brooks}, {Brooks}, {Brown}, {Brown}, {Brown}, {Bunker}, {Burger}, {Bushouse}, {Cale}, {Cameron}, {Cameron},
  {Canipe}, {Caplinger}, {Caputo}, {Cara}, {Carey}, {Carniani}, {Carrasquilla}, {Carruthers}, {Case}, {Catherine}, {Chance}, {Chapman}, {Charlot}, {Charlow}, {Chayer}, {Chen}, {Cherinka}, {Chichester}, {Chilton}, {Chonis}, {Clampin}, {Clark}, {Clark}, {Coe}, {Coleman}, {Comber}, {Comeau}, {Connolly}, {Cooper}, {Cooper}, {Coppock}, {Correnti}, {Cossou}, {Coulais}, {Coyle}, {Cracraft}, {Curti}, {Cuturic}, {Davis}, {Davis}, {Dean}, {DeLisa}, {deMeester}, {Dencheva}, {Dencheva}, {DePasquale}, {Deschenes}, {Hunor Detre}, {Diaz}, {Dicken}, {DiFelice}, {Dillman}, {Dixon}, {Doggett}, {Donaldson}, {Douglas}, {DuPrie}, {Dupuis}, {Durning}, {Easmin}, {Eck}, {Edeani}, {Egami}, {Ehrenwinkler}, {Eisenhamer}, {Eisenhower}, {Elie}, {Elliott}, {Elliott}, {Ellis}, {Engesser}, {Espinoza}, {Etienne}, {Etxaluze}, {Falini}, {Feeney}, {Ferry}, {Filippazzo}, {Fincham}, {Fix}, {Flagey}, {Florian}, {Flynn}, {Fontanella}, {Ford}, {Forshay}, {Fox}, {Franz}, {Fu}, {Fullerton}, {Galkin}, {Galyer}, {Garc{\'\i}a Mar{\'\i}n}, {Gardner},
  {Gardner}, {Garland}, {Garrett}, {Gasman}, {Gaspar}, {Gaudreau}, {Gauthier}, {Geers}, {Geithner}, {Gennaro}, {Giardino}, {Girard}, {Giuliano}, {Glassmire}, {Glauser}, {Glazer}, {Godfrey}, {Golimowski}, {Gollnitz}, {Gong}, {Gonzaga}, {Gordon}, {Gordon}, {Goudfrooij}, {Greene}, {Greenhouse}, {Grimaldi}, {Groebner}, {Grundy}, {Guillard}, {Gutman}, {Ha}, {Haderlein}, {Hagedorn}, {Hainline}, {Haley}, {Hami}, {Hamilton}, {Hammel}, {Hansen}, {Harkins}, {Harr}, {Hart}, {Hart}, {Hartig}, {Hashimoto}, {Haskins}, {Hathaway}, {Havey}, {Hayden}, {Hecht}, {Heller-Boyer}, {Henriques}, {Henry}, {Hermann}, {Hernandez}, {Hesman}, {Hicks}, {Hilbert}, {Hines}, {Hoffman}, {Holfeltz}, {Holler}, {Hoppa}, {Hott}, {Howard}, {Howard}, {Hunter}, {Hunter}, {Hurst}, {Husemann}, {Hustak}, {Ilinca Ignat}, {Illingworth}, {Irish}, {Jackson}, {Jahromi}, {Jakobsen}, {James}, {James}, {Januszewski}, {Jenkins}, {Jirdeh}, {Johnson}, {Johnson}, {Jones}, {Jones}, {Jones}, {Jones}, {Jordan}, {Jordan}, {Jurczyk}, {Jurling}, {Kaleida}, {Kalmanson},
  {Kammerer}, {Kang}, {Kao}, {Karakla}, {Kavanagh}, {Kelly}, {Kendrew}, {Kennedy}, {Kenny}, {Keski-kuha}, {Keyes}, {Kidwell}, {Kinzel}, {Kirk}, {Kirkpatrick}, {Kirshenblat}, {Klaassen}, {Knapp}, {Knight}, {Knollenberg}, {Koehler}, {Koekemoer}, {Kovacs}, {Kulp}, {Kumari}, {Kyprianou}, {La Massa}, {Labador}, {Labiano}, {Lagage}, {Lajoie}, {Lallo}, {Lam}, {Lamb}, {Lambros}, {Lampenfield}, {Langston}, {Larson}, {Law}, {Lawrence}, {Lee}, {Leisenring}, {Lepo}, {Leveille}, {Levenson}, {Levine}, {Levy}, {Lewis}, {Lewis}, {Libralato}, {Lightsey}, {Link}, {Liu}, {Lo}, {Lockwood}, {Logue}, {Long}, {Long}, {Loomis}, {Lopez-Caniego}, {Lorenzo Alvarez}, {Love-Pruitt}, {Lucy}, {Luetzgendorf}, {Maghami}, {Maiolino}, {Major}, {Malla}, {Malumuth}, {Manjavacas}, {Mannfolk}, {Marrione}, {Marston}, {Martel}, {Maschmann}, {Masci}, {Masciarelli}, {Maszkiewicz}, {Mather}, {McKenzie}, {McLean}, {McMaster}, {Melbourne}, {Mel{\'e}ndez}, {Menzel}, {Merz}, {Meyett}, {Meza}, {Miskey}, {Misselt}, {Moller}, {Morrison}, {Morse}, {Moseley},
  {Mosier}, {Mountain}, {Mueckay}, {Mueller}, {Mullally}, {Murphy}, {Murray}, {Murray}, {Mustelier}, {Muzerolle}, {Mycroft}, {Myers}, {Myrick}, {Nanavati}, {Nance}, {Nayak}, {Naylor}, {Nelan}, {Nickson}, {Nielson}, {Nieto-Santisteban}, {Nikolov}, {Noriega-Crespo}, {O'Shaughnessy}, {O'Sullivan}, {Ochs}, {Ogle}, {Oleszczuk}, {Olmsted}, {Osborne}, {Ottens}, {Owens}, {Pacifici}, {Pagan}, {Page}, {Park}, {Parrish}, {Patapis}, {Paul}, {Pauly}, {Pavlovsky}, {Pedder}, {Peek}, {Pena-Guerrero}, {Penanen}, {Perez}, {Perna}, {Perriello}, {Phillips}, {Pietraszkiewicz}, {Pinaud}, {Pirzkal}, {Pitman}, {Piwowar}, {Platais}, {Player}, {Plesha}, {Pollizi}, {Polster}, {Pontoppidan}, {Porterfield}, {Proffitt}, {Pueyo}, {Pulliam}, {Quirt}, {Quispe Neira}, {Ramos Alarcon}, {Ramsay}, {Rapp}, {Rapp}, {Rauscher}, {Ravindranath}, {Rawle}, {Regan}, {Reichard}, {Reis}, {Ressler}, {Rest}, {Reynolds}, {Rhue}, {Richon}, {Rickman}, {Ridgaway}, {Ritchie}, {Rix}, {Robberto}, {Robinson}, {Robinson}, {Robinson}, {Rock}, {Rodriguez}, {Rodriguez
  Del Pino}, {Roellig}, {Rohrbach}, {Roman}, {Romelfanger}, {Rose}, {Roteliuk}, {Roth}, {Rothwell}, {Rowlands}, {Roy}, {Royer}, {Royle}, {Rui}, {Rumler}, {Runnels}, {Russ}, {Rustamkulov}, {Ryden}, {Ryer}, {Sabata}, {Sabatke}, {Sabbi}, {Samuelson}, {Sapp}, {Sappington}, {Sargent}, {Sauer}, {Scheithauer}, {Schlawin}, {Schlitz}, {Schmitz}, {Schneider}, {Schreiber}, {Schulze}, {Schwab}, {Scott}, {Sembach}, {Shanahan}, {Shaughnessy}, {Shaw}, {Shawger}, {Shay}, {Sheehan}, {Shen}, {Sherman}, {Shiao}, {Shih}, {Shivaei}, {Sienkiewicz}, {Sing}, {Sirianni}, {Sivaramakrishnan}, {Skipper}, {Sloan}, {Slocum}, {Slowinski}, {Smith}, {Smith}, {Smith}, {Smith}, {Snyder}, {Soh}, {Sohn}, {Soto}, {Spencer}, {Stallcup}, {Stansberry}, {Starr}, {Starr}, {Stewart}, {Stiavelli}, {Straughn}, {Strickland}, {Stys}, {Summers}, {Sun}, {Sunnquist}, {Swade}, {Swam}, {Swaters}, {Swoish}, {Taylor}, {Taylor}, {Te Plate}, {Tea}, {Teague}, {Telfer}, {Temim}, {Thatte}, {Thompson}, {Thompson}, {Thomson}, {Tikkanen}, {Tippet}, {Todd}, {Toolan},
  {Tran}, {Trejo}, {Truong}, {Tsukamoto}, {Tustain}, {Tyra}, {Ubeda}, {Underwood}, {Uzzo}, {Van Campen}, {Vandal}, {Vandenbussche}, {Vila}, {Volk}, {Wahlgren}, {Waldman}, {Walker}, {Wander}, {Warfield}, {Warner}, {Wasiak}, {Watkins}, {Weaver}, {Weilert}, {Weiser}, {Weiss}, {Weissman}, {Welty}, {West}, {Wheate}, {Wheatley}, {Wheeler}, {White}, {Whiteaker}, {Whitehouse}, {Whiteleather}, {Whitman}, {Williams}, {Willmer}, {Willoughby}, {Wilson}, {Wirth}, {Wislowski}, {Wolf}, {Wolfe}, {Wolff}, {Workman}, {Wright}, {Wu}, {Wu}, {Wymer}, {Yates}, {Yeager}, {Yeates}, {Yerger}, {Yoon}, {Young}, {Yu}, {Zak}, {Zeidler}, {Zhou}, {Zielinski}, {Zincke}, \& {Zonak}}]{rigby2023}
{Rigby}, J., {Perrin}, M., {McElwain}, M., {et~al.} 2023, \pasp, 135, 048001, \dodoi{10.1088/1538-3873/acb293}

\bibitem[{{Rinaldi} {et~al.}(2024){Rinaldi}, {Caputi}, {Iani}, {Costantin}, {Gillman}, {Perez Gonzalez}, {{\"O}stlin}, {Colina}, {Greve}, {N{\o}rgard-Nielsen}, {Wright}, {{\'A}lvarez-M{\'a}rquez}, {Eckart}, {Garc{\'\i}a-Mar{\'\i}n}, {Hjorth}, {Ilbert}, {Kendrew}, {Labiano}, {Le F{\`e}vre}, {Pye}, {Tikkanen}, {Walter}, {van der Werf}, {Ward}, {Annunziatella}, {Azzollini}, {Bik}, {Boogaard}, {Bosman}, {Crespo G{\'o}mez}, {Jermann}, {Langeroodi}, {Melinder}, {Meyer}, {Moutard}, {Peissker}, {van Dishoeck}, {G{\"u}del}, {Henning}, {Lagage}, {Ray}, {Vandenbussche}, {Waelkens}, \& {Dayal}}]{rinaldi2024}
{Rinaldi}, P., {Caputi}, K.~I., {Iani}, E., {et~al.} 2024, \apj, 969, 12, \dodoi{10.3847/1538-4357/ad4147}

\bibitem[{{Robertson} {et~al.}(2024){Robertson}, {Johnson}, {Tacchella}, {Eisenstein}, {Hainline}, {Arribas}, {Baker}, {Bunker}, {Carniani}, {Cargile}, {Carreira}, {Charlot}, {Chevallard}, {Curti}, {Curtis-Lake}, {D'Eugenio}, {Egami}, {Hausen}, {Helton}, {Jakobsen}, {Ji}, {Jones}, {Maiolino}, {Maseda}, {Nelson}, {P{\'e}rez-Gonz{\'a}lez}, {Pusk{\'a}s}, {Rieke}, {Smit}, {Sun}, {{\"U}bler}, {Whitler}, {Williams}, {Willmer}, {Willott}, \& {Witstok}}]{robertson2024_jof}
{Robertson}, B., {Johnson}, B.~D., {Tacchella}, S., {et~al.} 2024, \apj, 970, 31, \dodoi{10.3847/1538-4357/ad463d}

\bibitem[{{Robertson}(2022)}]{robertson2022}
{Robertson}, B.~E. 2022, \araa, 60, 121, \dodoi{10.1146/annurev-astro-120221-044656}

\bibitem[{{Robertson} {et~al.}(2015){Robertson}, {Ellis}, {Furlanetto}, \& {Dunlop}}]{robertson2015}
{Robertson}, B.~E., {Ellis}, R.~S., {Furlanetto}, S.~R., \& {Dunlop}, J.~S. 2015, \apjl, 802, L19, \dodoi{10.1088/2041-8205/802/2/L19}

\bibitem[{{Robertson} {et~al.}(2013){Robertson}, {Furlanetto}, {Schneider}, {Charlot}, {Ellis}, {Stark}, {McLure}, {Dunlop}, {Koekemoer}, {Schenker}, {Ouchi}, {Ono}, {Curtis-Lake}, {Rogers}, {Bowler}, \& {Cirasuolo}}]{robertson2013}
{Robertson}, B.~E., {Furlanetto}, S.~R., {Schneider}, E., {et~al.} 2013, \apj, 768, 71, \dodoi{10.1088/0004-637X/768/1/71}

\bibitem[{{Robertson} {et~al.}(2023){Robertson}, {Tacchella}, {Johnson}, {Hainline}, {Whitler}, {Eisenstein}, {Endsley}, {Rieke}, {Stark}, {Alberts}, {Dressler}, {Egami}, {Hausen}, {Rieke}, {Shivaei}, {Williams}, {Willmer}, {Arribas}, {Bonaventura}, {Bunker}, {Cameron}, {Carniani}, {Charlot}, {Chevallard}, {Curti}, {Curtis-Lake}, {D'Eugenio}, {Jakobsen}, {Looser}, {L{\"u}tzgendorf}, {Maiolino}, {Maseda}, {Rawle}, {Rix}, {Smit}, {{\"U}bler}, {Willott}, {Witstok}, {Baum}, {Bhatawdekar}, {Boyett}, {Chen}, {de Graaff}, {Florian}, {Helton}, {Hviding}, {Ji}, {Kumari}, {Lyu}, {Nelson}, {Sandles}, {Saxena}, {Suess}, {Sun}, {Topping}, \& {Wallace}}]{robertson2023_jades_highz}
{Robertson}, B.~E., {Tacchella}, S., {Johnson}, B.~D., {et~al.} 2023, Nature Astronomy, 7, 611, \dodoi{10.1038/s41550-023-01921-1}

\bibitem[{{Rojas-Ruiz} {et~al.}(2020){Rojas-Ruiz}, {Finkelstein}, {Bagley}, {Stevans}, {Finkelstein}, {Larson}, {Mechtley}, \& {Diekmann}}]{rojas-ruiz2020}
{Rojas-Ruiz}, S., {Finkelstein}, S.~L., {Bagley}, M.~B., {et~al.} 2020, \apj, 891, 146, \dodoi{10.3847/1538-4357/ab7659}

\bibitem[{{Rosdahl} {et~al.}(2022){Rosdahl}, {Blaizot}, {Katz}, {Kimm}, {Garel}, {Haehnelt}, {Keating}, {Martin-Alvarez}, {Michel-Dansac}, \& {Ocvirk}}]{rosdahl2022}
{Rosdahl}, J., {Blaizot}, J., {Katz}, H., {et~al.} 2022, \mnras, 515, 2386, \dodoi{10.1093/mnras/stac1942}

\bibitem[{{Rowe} {et~al.}(2015){Rowe}, {Jarvis}, {Mandelbaum}, {Bernstein}, {Bosch}, {Simet}, {Meyers}, {Kacprzak}, {Nakajima}, {Zuntz}, {Miyatake}, {Dietrich}, {Armstrong}, {Melchior}, \& {Gill}}]{rowe2015}
{Rowe}, B.~T.~P., {Jarvis}, M., {Mandelbaum}, R., {et~al.} 2015, Astronomy and Computing, 10, 121, \dodoi{10.1016/j.ascom.2015.02.002}

\bibitem[{{Salpeter}(1955)}]{salpeter1955}
{Salpeter}, E.~E. 1955, \apj, 121, 161, \dodoi{10.1086/145971}

\bibitem[{{Saxena} {et~al.}(2024){Saxena}, {Cameron}, {Katz}, {Bunker}, {Chevallard}, {D'Eugenio}, {Arribas}, {Bhatawdekar}, {Boyett}, {Cargile}, {Carniani}, {Charlot}, {Curti}, {Curtis-Lake}, {Hainline}, {Ji}, {Johnson}, {Jones}, {Kumari}, {Laseter}, {Maseda}, {Robertson}, {Simmonds}, {Tacchella}, {Ubler}, {Williams}, {Willott}, {Witstok}, \& {Zhu}}]{saxena2024}
{Saxena}, A., {Cameron}, A.~J., {Katz}, H., {et~al.} 2024, arXiv e-prints, arXiv:2411.14532, \dodoi{10.48550/arXiv.2411.14532}

\bibitem[{{Schechter}(1976)}]{schechter1976}
{Schechter}, P. 1976, \apj, 203, 297, \dodoi{10.1086/154079}

\bibitem[{{Schouws} {et~al.}(2024){Schouws}, {Bouwens}, {Ormerod}, {Smit}, {Algera}, {Sommovigo}, {Hodge}, {Ferrara}, {Oesch}, {Rowland}, {van Leeuwen}, {Stefanon}, {Herard-Demanche}, {Fudamoto}, {R{\"o}ttgering}, \& {van der Werf}}]{schouws2024}
{Schouws}, S., {Bouwens}, R.~J., {Ormerod}, K., {et~al.} 2024, arXiv e-prints, arXiv:2409.20549, \dodoi{10.48550/arXiv.2409.20549}

\bibitem[{{S{\'e}rsic}(1963)}]{sersic1963}
{S{\'e}rsic}, J.~L. 1963, Boletin de la Asociacion Argentina de Astronomia La Plata Argentina, 6, 41

\bibitem[{{Shen} {et~al.}(2023){Shen}, {Vogelsberger}, {Boylan-Kolchin}, {Tacchella}, \& {Kannan}}]{shen2023}
{Shen}, X., {Vogelsberger}, M., {Boylan-Kolchin}, M., {Tacchella}, S., \& {Kannan}, R. 2023, \mnras, 525, 3254, \dodoi{10.1093/mnras/stad2508}

\bibitem[{{Shibuya} {et~al.}(2015){Shibuya}, {Ouchi}, \& {Harikane}}]{shibuya2015}
{Shibuya}, T., {Ouchi}, M., \& {Harikane}, Y. 2015, \apjs, 219, 15, \dodoi{10.1088/0067-0049/219/2/15}

\bibitem[{{Shull} {et~al.}(2012){Shull}, {Harness}, {Trenti}, \& {Smith}}]{shull2012}
{Shull}, J.~M., {Harness}, A., {Trenti}, M., \& {Smith}, B.~D. 2012, \apj, 747, 100, \dodoi{10.1088/0004-637X/747/2/100}

\bibitem[{{Simmonds} {et~al.}(2024){Simmonds}, {Tacchella}, {Hainline}, {Johnson}, {Pusk{\'a}s}, {Robertson}, {Baker}, {Bhatawdekar}, {Boyett}, {Bunker}, {Cargile}, {Carniani}, {Chevallard}, {Curti}, {Curtis-Lake}, {Ji}, {Jones}, {Kumari}, {Laseter}, {Maiolino}, {Maseda}, {Rinaldi}, {Stoffers}, {{\"U}bler}, {Villanueva}, {Williams}, {Willot}, {Witstok}, \& {Zhu}}]{simmonds2024}
{Simmonds}, C., {Tacchella}, S., {Hainline}, K., {et~al.} 2024, arXiv e-prints, arXiv:2409.01286, \dodoi{10.48550/arXiv.2409.01286}

\bibitem[{{Stanway} {et~al.}(2003){Stanway}, {Bunker}, \& {McMahon}}]{stanway2003}
{Stanway}, E.~R., {Bunker}, A.~J., \& {McMahon}, R.~G. 2003, \mnras, 342, 439, \dodoi{10.1046/j.1365-8711.2003.06546.x}

\bibitem[{{Stark}(2016)}]{stark2016}
{Stark}, D.~P. 2016, \araa, 54, 761, \dodoi{10.1146/annurev-astro-081915-023417}

\bibitem[{{Sun} {et~al.}(2023){Sun}, {Faucher-Gigu{\`e}re}, {Hayward}, {Shen}, {Wetzel}, \& {Cochrane}}]{sun2023}
{Sun}, G., {Faucher-Gigu{\`e}re}, C.-A., {Hayward}, C.~C., {et~al.} 2023, \apjl, 955, L35, \dodoi{10.3847/2041-8213/acf85a}

\bibitem[{{Tacchella} {et~al.}(2023){Tacchella}, {Johnson}, {Robertson}, {Carniani}, {D'Eugenio}, {Kumari}, {Maiolino}, {Nelson}, {Suess}, {{\"U}bler}, {Williams}, {Adebusola}, {Alberts}, {Arribas}, {Bhatawdekar}, {Bonaventura}, {Bowler}, {Bunker}, {Cameron}, {Curti}, {Egami}, {Eisenstein}, {Frye}, {Hainline}, {Helton}, {Ji}, {Looser}, {Lyu}, {Perna}, {Rawle}, {Rieke}, {Rieke}, {Saxena}, {Sandles}, {Shivaei}, {Simmonds}, {Sun}, {Willmer}, {Willott}, \& {Witstok}}]{tacchella2023}
{Tacchella}, S., {Johnson}, B.~D., {Robertson}, B.~E., {et~al.} 2023, \mnras, 522, 6236, \dodoi{10.1093/mnras/stad1408}

\bibitem[{{Tang} {et~al.}(2024){Tang}, {Stark}, {Topping}, {Mason}, \& {Ellis}}]{tang2024_lya}
{Tang}, M., {Stark}, D.~P., {Topping}, M.~W., {Mason}, C., \& {Ellis}, R.~S. 2024, arXiv e-prints, arXiv:2408.01507, \dodoi{10.48550/arXiv.2408.01507}

\bibitem[{{Topping} {et~al.}(2024){Topping}, {Stark}, {Endsley}, {Whitler}, {Hainline}, {Johnson}, {Robertson}, {Tacchella}, {Chen}, {Alberts}, {Baker}, {Bunker}, {Carniani}, {Charlot}, {Chevallard}, {Curtis-Lake}, {DeCoursey}, {Egami}, {Eisenstein}, {Ji}, {Maiolino}, {Williams}, {Willmer}, {Willott}, \& {Witstok}}]{topping2024}
{Topping}, M.~W., {Stark}, D.~P., {Endsley}, R., {et~al.} 2024, \mnras, 529, 4087, \dodoi{10.1093/mnras/stae800}

\bibitem[{{Umeda} {et~al.}(2024){Umeda}, {Ouchi}, {Nakajima}, {Harikane}, {Ono}, {Xu}, {Isobe}, \& {Zhang}}]{umeda2024}
{Umeda}, H., {Ouchi}, M., {Nakajima}, K., {et~al.} 2024, \apj, 971, 124, \dodoi{10.3847/1538-4357/ad554e}

\bibitem[{{{\v{D}}urov{\v{c}}{\'\i}kov{\'a}} {et~al.}(2024){{\v{D}}urov{\v{c}}{\'\i}kov{\'a}}, {Eilers}, {Chen}, {Satyavolu}, {Kulkarni}, {Simcoe}, {Keating}, {Haehnelt}, \& {Ba{\~n}ados}}]{durovcikova24}
{{\v{D}}urov{\v{c}}{\'\i}kov{\'a}}, D., {Eilers}, A.-C., {Chen}, H., {et~al.} 2024, \apj, 969, 162, \dodoi{10.3847/1538-4357/ad4888}

\bibitem[{{Vidal-Garc{\'\i}a} {et~al.}(2017){Vidal-Garc{\'\i}a}, {Charlot}, {Bruzual}, \& {Hubeny}}]{vidal-garcia2017}
{Vidal-Garc{\'\i}a}, A., {Charlot}, S., {Bruzual}, G., \& {Hubeny}, I. 2017, \mnras, 470, 3532, \dodoi{10.1093/mnras/stx1324}

\bibitem[{{Vijayan} {et~al.}(2021){Vijayan}, {Lovell}, {Wilkins}, {Thomas}, {Barnes}, {Irodotou}, {Kuusisto}, \& {Roper}}]{vijayan2021}
{Vijayan}, A.~P., {Lovell}, C.~C., {Wilkins}, S.~M., {et~al.} 2021, \mnras, 501, 3289, \dodoi{10.1093/mnras/staa3715}

\bibitem[{{Wang} {et~al.}(2023){Wang}, {Fujimoto}, {Labb{\'e}}, {Furtak}, {Miller}, {Setton}, {Zitrin}, {Atek}, {Bezanson}, {Brammer}, {Leja}, {Oesch}, {Price}, {Chemerynska}, {Cutler}, {Dayal}, {van Dokkum}, {Goulding}, {Greene}, {Fudamoto}, {Khullar}, {Kokorev}, {Marchesini}, {Pan}, {Weaver}, {Whitaker}, \& {Williams}}]{wang2023}
{Wang}, B., {Fujimoto}, S., {Labb{\'e}}, I., {et~al.} 2023, \apjl, 957, L34, \dodoi{10.3847/2041-8213/acfe07}

\bibitem[{{Wang} {et~al.}(2020){Wang}, {Davies}, {Yang}, {Hennawi}, {Fan}, {Barth}, {Jiang}, {Wu}, {Mudd}, {Ba{\~n}ados}, {Bian}, {Decarli}, {Eilers}, {Farina}, {Venemans}, {Walter}, \& {Yue}}]{wang2020}
{Wang}, F., {Davies}, F.~B., {Yang}, J., {et~al.} 2020, \apj, 896, 23, \dodoi{10.3847/1538-4357/ab8c45}

\bibitem[{{Weibel} {et~al.}(2025){Weibel}, {Oesch}, {Williams}, {Jespersen}, {Shuntov}, {Whitaker}, {Atek}, {Bezanson}, {Brammer}, {Chemerynska}, {Cloonan}, {Dayal}, {Furtak}, {Hutter}, {Ji}, {Maseda}, \& {Xiao}}]{weibel2025}
{Weibel}, A., {Oesch}, P.~A., {Williams}, C.~C., {et~al.} 2025, arXiv e-prints, arXiv:2507.06292, \dodoi{10.48550/arXiv.2507.06292}

\bibitem[{{Whitaker} {et~al.}(2019){Whitaker}, {Ashas}, {Illingworth}, {Magee}, {Leja}, {Oesch}, {van Dokkum}, {Mowla}, {Bouwens}, {Franx}, {Holden}, {Labb{\'e}}, {Rafelski}, {Teplitz}, \& {Gonzalez}}]{whitaker2019}
{Whitaker}, K.~E., {Ashas}, M., {Illingworth}, G., {et~al.} 2019, \apjs, 244, 16, \dodoi{10.3847/1538-4365/ab3853}

\bibitem[{{Whitler} {et~al.}(2023){Whitler}, {Endsley}, {Stark}, {Topping}, {Chen}, \& {Charlot}}]{whitler2023_ceers}
{Whitler}, L., {Endsley}, R., {Stark}, D.~P., {et~al.} 2023, \mnras, 519, 157, \dodoi{10.1093/mnras/stac3535}

\bibitem[{{Whitler} {et~al.}(2020){Whitler}, {Mason}, {Ren}, {Dijkstra}, {Mesinger}, {Pentericci}, {Trenti}, \& {Treu}}]{whitler2020}
{Whitler}, L.~R., {Mason}, C.~A., {Ren}, K., {et~al.} 2020, \mnras, 495, 3602, \dodoi{10.1093/mnras/staa1178}

\bibitem[{{Wilkins} {et~al.}(2011){Wilkins}, {Bunker}, {Stanway}, {Lorenzoni}, \& {Caruana}}]{wilkins2011}
{Wilkins}, S.~M., {Bunker}, A.~J., {Stanway}, E., {Lorenzoni}, S., \& {Caruana}, J. 2011, \mnras, 417, 717, \dodoi{10.1111/j.1365-2966.2011.19315.x}

\bibitem[{{Wilkins} {et~al.}(2023){Wilkins}, {Vijayan}, {Lovell}, {Roper}, {Irodotou}, {Caruana}, {Seeyave}, {Kuusisto}, {Thomas}, \& {Parris}}]{wilkins2023}
{Wilkins}, S.~M., {Vijayan}, A.~P., {Lovell}, C.~C., {et~al.} 2023, \mnras, 519, 3118, \dodoi{10.1093/mnras/stac3280}

\bibitem[{{Williams} {et~al.}(2023){Williams}, {Tacchella}, {Maseda}, {Robertson}, {Johnson}, {Willott}, {Eisenstein}, {Willmer}, {Ji}, {Hainline}, {Helton}, {Alberts}, {Baum}, {Bhatawdekar}, {Boyett}, {Bunker}, {Carniani}, {Charlot}, {Chevallard}, {Curtis-Lake}, {de Graaff}, {Egami}, {Franx}, {Kumari}, {Maiolino}, {Nelson}, {Rieke}, {Sandles}, {Shivaei}, {Simmonds}, {Smit}, {Suess}, {Sun}, {{\"U}bler}, \& {Witstok}}]{williams2023_jems}
{Williams}, C.~C., {Tacchella}, S., {Maseda}, M.~V., {et~al.} 2023, \apjs, 268, 64, \dodoi{10.3847/1538-4365/acf130}

\bibitem[{{Willott} {et~al.}(2024){Willott}, {Desprez}, {Asada}, {Sarrouh}, {Abraham}, {Brada{\v{c}}}, {Brammer}, {Estrada-Carpenter}, {Iyer}, {Martis}, {Matharu}, {Mowla}, {Muzzin}, {Noirot}, {Sawicki}, {Strait}, {Rihtar{\v{s}}i{\v{c}}}, \& {Withers}}]{willott2024}
{Willott}, C.~J., {Desprez}, G., {Asada}, Y., {et~al.} 2024, \apj, 966, 74, \dodoi{10.3847/1538-4357/ad35bc}

\bibitem[{{Witstok} {et~al.}(2024){Witstok}, {Jakobsen}, {Maiolino}, {Helton}, {Johnson}, {Robertson}, {Tacchella}, {Cameron}, {Smit}, {Bunker}, {Saxena}, {Sun}, {Arribas}, {Baker}, {Bhatawdekar}, {Boyett}, {Cargile}, {Carniani}, {Charlot}, {Chevallard}, {Curti}, {Curtis-Lake}, {D'Eugenio}, {Eisenstein}, {Hainline}, {Jones}, {Kumari}, {Maseda}, {P{\'e}rez-Gonz{\'a}lez}, {Rinaldi}, {Scholtz}, {{\"U}bler}, {Williams}, {Willmer}, {Willott}, \& {Zhu}}]{witstok2024}
{Witstok}, J., {Jakobsen}, P., {Maiolino}, R., {et~al.} 2024, arXiv e-prints, arXiv:2408.16608, \dodoi{10.48550/arXiv.2408.16608}

\bibitem[{{Witten} {et~al.}(2024){Witten}, {McClymont}, {Laporte}, {Roberts-Borsani}, {Sijacki}, {Tacchella}, {Simmonds}, {Katz}, {Ellis}, {Witstok}, {Maiolino}, {Ji}, {Hayes}, {Looser}, \& {D'Eugenio}}]{witten2024}
{Witten}, C., {McClymont}, W., {Laporte}, N., {et~al.} 2024, arXiv e-prints, arXiv:2407.07937, \dodoi{10.48550/arXiv.2407.07937}

\bibitem[{{Yang} {et~al.}(2020{\natexlab{a}}){Yang}, {Wang}, {Fan}, {Hennawi}, {Davies}, {Yue}, {Banados}, {Wu}, {Venemans}, {Barth}, {Bian}, {Boutsia}, {Decarli}, {Farina}, {Green}, {Jiang}, {Li}, {Mazzucchelli}, \& {Walter}}]{yang2020_damping_wing}
{Yang}, J., {Wang}, F., {Fan}, X., {et~al.} 2020{\natexlab{a}}, \apjl, 897, L14, \dodoi{10.3847/2041-8213/ab9c26}

\bibitem[{{Yang} {et~al.}(2020{\natexlab{b}}){Yang}, {Wang}, {Fan}, {Hennawi}, {Davies}, {Yue}, {Eilers}, {Farina}, {Wu}, {Bian}, {Pacucci}, \& {Lee}}]{yang2020_tau_eff}
---. 2020{\natexlab{b}}, \apj, 904, 26, \dodoi{10.3847/1538-4357/abbc1b}

\bibitem[{{Yang} {et~al.}(2022){Yang}, {Morishita}, {Leethochawalit}, {Castellano}, {Calabr{\`o}}, {Treu}, {Bonchi}, {Fontana}, {Mason}, {Merlin}, {Paris}, {Trenti}, {Roberts-Borsani}, {Bradac}, {Vanzella}, {Vulcani}, {Marchesini}, {Ding}, {Nanayakkara}, {Birrer}, {Glazebrook}, {Jones}, {Boyett}, {Santini}, {Strait}, \& {Wang}}]{yang2022}
{Yang}, L., {Morishita}, T., {Leethochawalit}, N., {et~al.} 2022, \apjl, 938, L17, \dodoi{10.3847/2041-8213/ac8803}

\bibitem[{{Yung} {et~al.}(2019){Yung}, {Somerville}, {Finkelstein}, {Popping}, \& {Dav{\'e}}}]{yung2019}
{Yung}, L.~Y.~A., {Somerville}, R.~S., {Finkelstein}, S.~L., {Popping}, G., \& {Dav{\'e}}, R. 2019, \mnras, 483, 2983, \dodoi{10.1093/mnras/sty3241}

\bibitem[{{Yung} {et~al.}(2024){Yung}, {Somerville}, {Finkelstein}, {Wilkins}, \& {Gardner}}]{yung2024}
{Yung}, L.~Y.~A., {Somerville}, R.~S., {Finkelstein}, S.~L., {Wilkins}, S.~M., \& {Gardner}, J.~P. 2024, \mnras, 527, 5929, \dodoi{10.1093/mnras/stad3484}

\bibitem[{{Zhu} {et~al.}(2022){Zhu}, {Becker}, {Bosman}, {Keating}, {D'Odorico}, {Davies}, {Christenson}, {Ba{\~n}ados}, {Bian}, {Bischetti}, {Chen}, {Davies}, {Eilers}, {Fan}, {Gaikwad}, {Greig}, {Haehnelt}, {Kulkarni}, {Lai}, {Pallottini}, {Qin}, {Ryan-Weber}, {Walter}, {Wang}, \& {Yang}}]{zhu2022}
{Zhu}, Y., {Becker}, G.~D., {Bosman}, S. E.~I., {et~al.} 2022, \apj, 932, 76, \dodoi{10.3847/1538-4357/ac6e60}

\bibitem[{{Zhu} {et~al.}(2024){Zhu}, {Becker}, {Bosman}, {Cain}, {Keating}, {Nasir}, {D'Odorico}, {Ba{\~n}ados}, {Bian}, {Bischetti}, {Bolton}, {Chen}, {D'Aloisio}, {Davies}, {Davies}, {Eilers}, {Fan}, {Gaikwad}, {Greig}, {Haehnelt}, {Kulkarni}, {Lai}, {Puchwein}, {Qin}, {Ryan-Weber}, {Satyavolu}, {Spina}, {Walter}, {Wang}, {Wolfson}, \& {Yang}}]{zhu2024}
---. 2024, \mnras, 533, L49, \dodoi{10.1093/mnrasl/slae061}

\end{thebibliography}
\bibliographystyle{aasjournal}



\end{document}